\documentclass[%
onecolumn,
nofootinbib,
amsmath,amssymb,
aps,
prd,
]{revtex4}%
\numberwithin{equation}{section}%numbering in terms of sections
\usepackage{graphicx}% Include figure files
\usepackage{dcolumn}% Align table columns on decimal point
\usepackage{bm}% bold math
\usepackage{multirow}%
\usepackage{amsmath,amssymb,amsfonts}%
\usepackage{amsthm}%
\usepackage{mathrsfs}%
\usepackage[title]{appendix}%
\usepackage{xcolor}%
\usepackage{textcomp}%
\usepackage{color} 

\usepackage[hidelinks]{hyperref}%link to the section

\newcommand{\be}{\begin{equation}}
\newcommand{\ee}{\end{equation}}
\newcommand\bea{\begin{eqnarray}}
\newcommand\eea{\end{eqnarray}}
\newcommand{\dd}{\mbox{d}}

\begin{document}
%\preprint{APS/123-QED}

%\preprint{}


\title{Charged particle interference in Kerr-Newman spacetime based on teleparallel gravity}                      



\author{Zhongyou Mo}
\email[]{mozhy7@mail.sysu.edu.cn}

%\thanks{}

\affiliation{School of Science, Sun Yat-Sen University, Shenzhen, 518107, China}

\affiliation{Department of Physics, Southern University of Science and Technology, Shenzhen, 518055, China}

\begin{abstract}
The quantum interference of charged particles in Kerr-Newman spacetime is studied based on teleparallel gravity. We calculate the gravitational phase difference, electromagnetic phase difference, and fringe shifts. We find that the gravitational phase difference contains three parts, respectively corresponding to the result in Kerr spacetime, the effect of Kerr-Newman black hole's charge on spacetime, and a mixed contribution from gravity and electromagnetic interaction. As for the electromagnetic phase difference, it contains two parts, respectively representing a pure electromagnetic contribution and a contribution in which electromagnetism mixes with gravitation. We compare the magnitudes of these phase differences. Afterwards, we extend the results to the case of dyonic Kerr-Newman spacetime. We discuss the effects of the black hole's magnetic charge and rotation, and propose two gedanken experiments to test the Dirac quantization condition. We also study the case that the universality of gravity breaks down, and show that the deviation from the weak equivalence principle can be determined from the fringe shifts. Additionally, we compare the results in teleparallel gravity with those in general relativity. For a generic weak gravitational field, we show that the gravitational phases in these two theories are consistent at the first order of the gravitational gauge potential but differ at higher orders.
\end{abstract}

%%%%%%%%%%%%%%%%%%%%%%%%%%%%%%%%%%
%%%%%%%

%\keywords{}

\maketitle


\tableofcontents%%%%%%

%%%%%

%%%%%

%%%%%
\section{Introduction\label{Introduction}}

The quantum mechanical phase in the gravitational field has been a topic of interest for a long time. As for such phase, Stodolsky argued that in the semiclassical limit, it is proportional to the proper time, for a massive particle~\cite{Stodolsky:1978ks}
\be 
\phi=\frac{m}{\hbar} \int_A^B \dd s,
\label{qmphase1}
\ee
where the path $AB$ is along the classical trajectory of the particle. In the weak field approximation, this phase can be separated into two parts. The first part corresponds to a proper time defined by the Minkowski metric, while the second part is called the gravitationally induced phase defined by the deviation from the Minkowski metric~\cite{Stodolsky:1978ks}. By this way, Stodolsky showed that the phase \eqref{qmphase1} correctly produces the result of the interference experiment of neutrons~\cite{Overhauser:1974,Colella:1975dq,Greenberger:1979zz}. In addition to the above experiment, this theory was applied to the phenomenon of neutrino oscillations. For example, in Ref.~\cite{Ahluwalia} Ahluwalia and Burgard showed that in the gravitational field, every part of the neutrino mass eigenstates gains a gravitationally induced phase during the propagation, which eventually affects the oscillation of the neutrino. Stodolsky's theory was also applied in Grossman and Lipkin's paper \cite{Grossman:1996eh} to neutrinos which travel in a gravitational field.  

As shown above, it is convenient to separate \eqref{qmphase1} into a ``free part'' and a ``gravitationally induced part'' for the cases of weak gravitational fields. But for a general gravitational field, it is a hard task (recall $\dd s=(g_{\mu\nu}\dd x^\mu \dd x^\nu)^{1/2}$). Fortunately, such difficulty can be conquered in the frame of Teleparallel Gravity (TG), the teleparallel equivalent of general relativity~\cite{Aldrovandi:2013wha}. In TG, there is a phase equal to~\eqref{qmphase1}, given by~\cite{Aldrovandi:2013wha}
\be
\phi=\frac{m}{\hbar}\int_A^B  (u_a \dd x^a+u_a \dot{A}{^a}_{b\beta}x^b \dd x^\beta +u_a B{^a}_\beta\dd x^\beta),
\label{phi three}
\ee
where $u_a=\eta_{ab}u^b$, with $u^a={h^a}_\mu \dd x^\mu /\dd s$ the four-velocity defined in tangent-space (which is a Minkowski spacetime attached to each point of spacetime), and ${h^a}_\mu$ is the tetrad. Here the indices related to spacetime are denoted by Greek letters, while the indices for tangent-space are denoted by the first letters of the Latin alphabet. According to Ref.~\cite{Aldrovandi:2013wha}, the first term in \eqref{phi three} stands for a free particle, the second term corresponds to the interaction with the inertial effects, with $\dot{A}{^a}_{b\beta}$ a Lorentz connection, and the last term with the gauge potential $B{^a}_\beta$ represents the gravitational interaction. In practical computations, it is convenient to choose a ``proper frame''~\cite{Krssak:2018ywd}, such that the Lorentz connection vanishes (note that the proper frames reduce to the inertial frames of special relativity when gravity is removed~\cite{Krssak:2018ywd}). In Ref.~\cite{Aldrovandi:2013wha} the gravitational phase is defined by the last two terms of \eqref{phi three}, while in Ref.~\cite{Aldrovandi:2003pd} it is defined by the last term. Here we adopt the second option, i.e., the gravitational phase is defined as $\hbar^{-1}m\int u_a B{^a}_\beta\dd x^\beta$, considering that the fundamental field in TG is represented by the gauge potential~\cite{Aldrovandi:2006cy}. By such definition, the gravitational phase resembles the Aharonov-Bohm phase $\hbar^{-1}q\int A_\mu\dd x^\mu$ in electromagnetism~\cite{Aharonov:1959fk}. In the previous work~\cite{Mo}, the authors found a general method to calculate the gravitational phase for a given spacetime (i.e., expressing the gauge potential by the tetrad), and applied it to Kerr spacetime.

So far, we have only talked about the pure gravitational case. If we consider the quantum interference of charged particles in the presence of both a gravitational field and an electromagnetic field, the contribution from electromagnetic interaction should be also included. A relevant scenario with the earth's gravitational field and an electric field was studied by Anandan and Stodolsky in Ref.~\cite{Anandan:1983ry}, where they found that the effect of gravity is modified by a term relating to the charge, the gravitational acceleration, the speed of light, and the Planck constant. Another scenario with a magnetic field and the earth's gravity was discussed by Anandan, who showed that the Aharonov-Bohm phase shift is modified by a factor related to the acceleration of gravity~\cite{Anandan:1983fe}. Besides, Kagramanova {\em et al.} studied the interference of charged particles in  Pleba\'nski–Demia\'nski spacetime~\cite{Kagramanova:2008bv}, involving both the gravitational field and the electromagnetic field. Although these works are interesting, all of them are in the framework of general relativity. Instead, we are interested in the calculations in terms of TG, for the reason mentioned in the previous paragraph.

In this paper, we study the interference of charged particles in Kerr-Newman spacetime, based on the method developed in Ref.~\cite{Mo}. Differently from Ref.~\cite{Mo}, here we need to consider the contributions from charges for the gravitational phase $\phi_g$, and also need to compute the electromagnetic phase $\phi_e$. As extensions, we also consider a dyonic black hole and dyonic particles, and study the case which violates the universality of gravitation. We emphasize that this work is not just a generalization of Ref.~\cite{Mo} to include more black hole parameters. Instead, there is something genuinely new. Firstly, compared with that in Ref.~\cite{Mo}, the gravitational phase difference and the corresponding fringe shifts here are no longer some quantities only related to the gravitational interaction. Instead, we show that they also contain mixed gravitational-electromagnetic terms. We also prove the local Lorentz invariance of the gravitational phase (see Sec.~\ref{QMphase}), which was not done in Ref.~\cite{Mo}. Besides, we find a simpler way to compute the fringe shift to avoid repeating tedious calculations when the setup is flipped (see the last third paragraph in Sec.~\ref{interference}). New derivations and discussions are presented in this work compared with the Kerr case in Ref.~\cite{Mo}. For example, the phase formula and the conserved equations for energy and angular momentum are modified taking the electromagnetic interaction into account. For dyonic Kerr-Newman spacetime, we also discuss the effect of the magnetic charge in the quantum interference, the Dirac strings, and the Dirac quantization condition. In the subject that the weak equivalence principle is violated, we modify the formulas in terms of the premise $m_g\ne m_i$, where $m_g$ and $m_i$ are the gravitational mass and inertial mass, e.g., we re-derive the equation of motion and the conserved equations (see Appendix~\ref{without}). Consequently, such violation brings an extra term to the gravitational phase difference, compared with the case in universal gravity. Another difference is that we use a proper tetrad for which the Lorentz connection is vanishing (we derive such tetrad based on the one given by Ref.~\cite{Arcos:2004tzt}), while the tetrad used in Ref.~\cite{Mo} actually has a non-vanishing Lorentz connection (see the sentences after \eqref{chichi}). We also point out a minor error in Ref.~\cite{Mo} (read the sentences following \eqref{t varphiNew2}). In addition to including electromagnetic interaction in the quantum interference, a motivation of this paper is to compare the results in TG with those in general relativity. Though TG and general relativity are equivalent in classical level~\cite{Aldrovandi:2013wha}, they may be inconsistent in quantum aspects. For this reason, it is essential to make such comparison. In this paper, we use the formulation of teleparallel gravity in Ref.~\cite{Aldrovandi:2013wha} where such theory is local Lorentz invariant. In such formulation, the tetrad is related to both the Lorentz connection and gravitational gauge potential. In the practical calculations of the interference experiment, we choose a proper tetrad for which $\dot{A}{^a}_{b\mu}=0$. The teleparallel gravity theory we use is equivalent to general relativity except that in Sec.~\ref{Non-Universal} (and Appendix~\ref{without}) we discuss the version which violates the weak equivalence principle.

The contents of this paper are arranged as follows. In Sec.~\ref{GEphase} we show briefly how to compute the gravitational phase and the electromagnetic phase, then apply them to Kerr-Newman spacetime. After that, we study an interference experiment in such spacetime in Sec.~\ref{interference0}, in which we calculate the phase differences and the fringe shifts. In Sec.~\ref{dyonic black hole}, we extend the above results to the cases of dyonic Kerr-Newman black hole and dyonic particles, then analyze the effect of the black hole's magnetic charge and discuss the ways to test the Dirac quantization condition. Afterwards, in Sec.~\ref{Non-Universal} we study the case when the universality of gravity is broken down. Then in Sec.~\ref{comparison}, we compare the phase difference and the gravitational phase in TG with those in general relativity. Finally, we make a summary in Sec.~\ref{summary}, in which we also analyze the effect of the black hole's rotation, talk about the observational aspects of the quantum interference, and discuss the gravitoelectromagnetic analogies. Throughout this paper, the units $c=G=1$ and the metric signature $(+,-,-,-)$ are used (unless we explicitly specify). 

\section{Gravitational phase and electromagnetic phase\label{GEphase}}

\subsection{Phases in generic spacetime\label{QMphase}}
As mentioned in Sec.~\ref{Introduction}, the gravitational phase takes the following form~\cite{Aldrovandi:2013wha}
\be
\phi_g=\frac{m}{\hbar}\int_A^B  u_a B{^a}_\beta\dd x^\beta.
\label{phi B}
\ee
In the presence of electromagnetic field, we also need to consider the electromagnetic phase, which is given by~\cite{Aharonov:1959fk,Wu:1975es}\footnote{The phase \eqref{EMphase} corresponds to the electromagnetic interaction term of the action for an electric charge (as \eqref{actionS} shows).}
\be
\phi_e=\frac{q}{\hbar} \int_A^B  A_\beta \dd x^\beta,
\label{EMphase}
\ee
with $A_\beta$ the electromagnetic potential and $q$ the charge of the particle. We would like to mention that the free and inertial parts $\phi_{\rm f}$ and $\phi_{\rm i}$ of the phase \eqref{phi three} (i.e., the first two terms) could contribute non-vanishing phase differences in the interference experiment, considering that the path is along the trajectory of the charged particle, which is deflected by both gravity and electromagnetic interaction. According to \eqref{phi three}, we have
\bea
\phi_{\rm f}+\phi_{\rm i}
&=&\frac{m}{\hbar}\int  u_a (\partial_\beta x^a+\dot{A}{^a}_{b\beta}x^b ) \dd x^\beta
 \nonumber\\
&=&\frac{m}{\hbar}\int  u_a ({h^a}_\beta-{B^a}_\beta ) \dd x^\beta
 \nonumber\\
&=&\frac{m}{\hbar}\int  u_\beta \dd x^\beta
-\phi_{\rm g},
\label{free phase0}
\eea
where $\partial_\beta$ denotes the partial derivative with respect to $x^\beta$.

For simplicity, in this paper we always choose a proper frame, such that the Lorentz connection vanishes. Then the expression \eqref{phi B} can be rewritten as~\cite{Mo}
\be
\phi_g=\frac{1}{\hbar}\int_A^B S_\beta \dd x^\beta,
\qquad
S_\beta=P_\nu {B^\nu}_\beta,
\label{phi_g}
\ee
with $P_\nu$ the covariant four-momentum, and
\be
B{^\nu}_\beta=h_a{^\nu}B{^a}_\beta,
\qquad
B{^a}_\beta=h{^a}_\beta -\delta{^a}_{\mu'}\frac{\partial x^{\mu'}}{\partial x^\beta},
\label{B field}
\ee
where $x^{\mu'}$ are the cartesian coordinates. Note that the second equation in \eqref{B field} holds only for proper frames. In short, to calculate the gravitational phase, we need to know the covariant four-momentum, the tetrad, and the Jacobi matrix of the transformation between the cartesian coordinates and the coordinates we consider. As for $\phi_e$, the computation is direct, provided the electromagnetic potential is given.

%%%%%%%%%%%

\subsection{Local Lorentz invariance\label{Local Lorentz}}

We hope that the observable results in the interference experiment (e.g., the fringe shifts) are local Lorentz invariant. Hence, it is essential to discuss whether the teleparallel theory we use (especially the gravitational phase \eqref{phi B} and electromagnetic phase \eqref{EMphase}) are invariant under local Lorentz transformations.  

Firstly, we mention that the teleparallel gravity theory we adopt from Ref.~\cite{Aldrovandi:2013wha} is local Lorentz invariant, as shown by Ref.~\cite{Krssak:2018ywd}. In this theory, the tetrad and torsion are related to the Lorentz connection by
\bea
{h^a}_\mu
&=& \partial_\mu x^a+\dot{A}{^a}_{b\mu}x^b +B{^a}_\mu,
\label{hdotAB}
\\
\dot{T}{^a}_{\nu\rho}
&=&\partial_\nu {h^a}_\rho -\partial_\rho {h^a}_\nu +\dot{A}{^a}_{e\nu} {h^e}_\rho -\dot{A}{^a}_{e\rho} {h^e}_\nu,
\label{hT expression}
\eea
and ${h^a}_\mu$ and $\dot{T}{^a}_{\nu\rho}$ are Lorentz covariant:
\be
{h'^a}_\mu={\Lambda^a}_b\, {h^b}_\mu,
\qquad
\dot{T}'{^a}_{\nu\rho} ={\Lambda^a}_b\, \dot{T}{^b}_{\nu\rho},
\label{hT}
\ee
where ${\Lambda^a}_b={\Lambda^a}_b(x)$ is a local Lorentz transform matrix. In Ref.~\cite{Aldrovandi:2013wha}, the first equation of \eqref{hT} is derived by the relations $g_{\mu\nu}=\eta_{cd} {h^c}_\mu {h^d}_\nu$ and $\eta_{cd}{\Lambda^c}_a {\Lambda^d}_b =\eta_{ab}$, while the second equation is given without detailed derivations. However, we can check that the latter also holds by applying~\cite{Aldrovandi:2013wha}
\be
\dot{A}'{^a}_{b\mu}
={\Lambda^a}_c \, \dot{A}{^c}_{d\mu}  ({\Lambda^{-1})^d }_b
+{\Lambda^a}_c \,\partial_\mu 
({\Lambda^{-1})^c }_b ,
\label{connection transfrom}
\ee
when we transform $\dot{T}{^a}_{\nu\rho}$. As claimed in Ref.~\cite{Aldrovandi:2013wha}, the Lagrangian density of the gravitational field\footnote{In SI units, the right hand side of \eqref{dotL} should be multiplied by $c^4/G$.}
\be
\dot{\mathcal{L}}
=\frac{h}{16\pi}\Bigl(
\frac{1}{4} \dot{T}{^a}_{bc} \dot{T}{_a}^{bc}
+\frac{1}{2} \dot{T}{^a}_{bc} \dot{T}{^{cb}}_a
-\dot{T}{^a}_{ba} \dot{T}{^{cb}}_c
\Bigr),
\label{dotL}
\ee
is local Lorentz invariant, where $h={\rm det}({h^a}_\mu)=\sqrt{-g}$ and $\dot{T}{^a}_{bc}=\dot{T}{^a}_{\nu\rho} {h_b}^\nu {h_c}^\rho$. One can check this conclusion by applying \eqref{hT} to \eqref{dotL}. Though the Lagrangian looks dependent of the Lorentz connection (because of \eqref{hT expression}), it can be separated as two parts where the first part is independent of such connection and the second part is a divergence including such connection~\cite{Krssak:2018ywd}. Thus the field equation derived from the Lagrangian \eqref{dotL} is actually independent of the Lorentz connection. Such conclusion coincides with the interpretation that the Lorentz connection stands for inertial effects only~\cite{Krssak:2018ywd}. By the way, the Lagrangian \eqref{dotL} is different from the Einstein–Hilbert Lagrangian only by a divergence~\cite{Aldrovandi:2013wha}, i.e., $\dot{\mathcal{L}}=\mathcal{L}_{\rm EH}-\partial_\mu (c^4 h (8\pi G)^{-1}\dot{T}{^{\nu\mu}}_\nu )$, hence these two Lagrangians are actually equivalent (Ref.~\cite{Aldrovandi:2013wha} also shows that the field equation in teleparallel gravity is equivalent to the Einstein field equation). This is the reason why this theory is usually
called the teleparallel equivalent of general relativity~\cite{Krssak:2018ywd}. Finally, we would like to mention that there is a version of teleparallel theory called ``pure tetrad teleparallel gravity'' which lacks local Lorentz covariance~\cite{Cho:1975dh}, where the torsion is defined as $\dot{T}{^a}_{\nu\rho}=\partial_\nu {h^a}_\rho -\partial_\rho {h^a}_\nu$ rather than \eqref{hT expression}~\cite{Krssak:2015rqa}. Indeed, one can check that the torsion in a ``pure tetrad teleparallel gravity'' theory is not local Lorentz covariant.  

Now we discuss the local Lorentz invariance about the particle dynamics in the presence of gravity and electromagnetic field. The action is
\be
\mathcal{S}
=-m\int \dd s -\int q A_\beta \dd x^\beta.
\label{S particle action}
\ee
In teleparallel gravity, the spacetime interval is expressed as~\cite{Aldrovandi:2013wha}
\be
\dd s
=-u_a {h^a}_\beta \dd x^\beta
=-u_a {h^a}_\beta {h_b}^\beta {h^b},
\label{s three}
\ee
where in the last step we used $
\dd x^\mu={h_a}^\mu {h^a}$ derived from the relations $h^a={h^a}_\nu \dd x^\nu$ and ${h_a}^\mu {h^a}_\nu =\delta^\mu_\nu$. Since the tetrad is local Lorentz covariant, the spacetime internal \eqref{s three} is local Lorentz invariant (one can check it using the first equation in \eqref{hT}). The last term in \eqref{S particle action} is also local Lorentz invariant because $A_\mu \dd x^\mu={h^a}_\mu A_a \dd x^\mu$ and
\be
{h'^a}_\mu A'_a \dd x^\mu
=({\Lambda^a}_b \, {h^b}_\mu ) \bigl[ ({\Lambda^{-1})^c }_a \, A_c \bigr]\,  
\dd x^\mu
={h^c}_\mu A_c \dd x^\mu,
\label{Adx invariance}
\ee
where we used ${\Lambda^a}_b ({\Lambda^{-1})^c }_a=\delta_c^b$. Therefore, the action \eqref{S particle action} is local Lorentz invariant. Following the derivation in Ref.~\cite{Aldrovandi:2013wha}, the variation of the action \eqref{S particle action} gives the equation of motion
\be
\frac{\dd u_a}{\dd s} -\dot{A}{^b}_{a\rho} u_b u^\rho 
=\dot{T}{^b}_{a\rho} u_b u^\rho
+\frac{q}{m_{\rm i} }  F_{a\nu} u^\nu,
\label{EoM teleparallel}
\ee
where $\dot{T}{^b}_{a\rho}={h_a}^\nu\,\dot{T}{^b}_{\nu\rho}$, and $F_{a\nu}={h_a}^\mu F_{\mu\nu}$ with $F_{\mu\nu}$ the electromagnetic tensor. Using \eqref{hT}, \eqref{connection transfrom}, and $F'_{a\nu}={(\Lambda^{-1})^b}_a F_{b\nu}$, one can verify that the equation \eqref{EoM teleparallel} is local Lorentz covariant.

Now we prove the local Lorentz invariances of the phases \eqref{phi B} and \eqref{EMphase}. The electromagnetic phase \eqref{EMphase} is obviously local Lorentz invariant because of \eqref{Adx invariance}. As for the former, firstly we know that gravitational gauge potential is local Lorentz covariant~\cite{Aldrovandi:2013wha}
\be
B'{^a}_\mu={\Lambda^a}_b \, B{^b}_\mu,
\label{BprimeB}
\ee
such that 
\be
u'_a B'{^a}_\mu\dd x^\mu
=({(\Lambda^{-1})^g}_a \, u_g)
 ({\Lambda^a}_c \, B{^c}_\mu)\,  
 \dd x^\mu 
=u_c B{^c}_\mu\dd x^\mu.
\ee
Therefore, the gravitational phase \eqref{phi B} is also local Lorentz invariant. We would like to mention that \eqref{BprimeB} is implied by \eqref{hT} considering the relation \eqref{hdotAB}. Indeed, we have $B{^a}_\beta=h{^a}_\mu-e{^a}_\mu$ with
\be
e{^a}_\mu=\partial_\mu x^a+\dot{A}{^a}_{b\mu}x^b
={h^p}_\mu h_p\, x^a+\dot{A}{^a}_{b\mu}x^b,
\ee
where we use the relation $\partial_\mu={h^p}_\mu h_p$~\cite{Aldrovandi:2013wha}. Then we have
\bea
e'{^a}_\mu
&=&({\Lambda^p}_d {h^d}_\mu) (  {(\Lambda^{-1})^q}_p  h_q ) ({\Lambda^a}_c x^c)
+({\Lambda^a}_c \, \dot{A}{^c}_{d\mu}  {(\Lambda^{-1} )^d}_b 
+{\Lambda^a}_c \,\partial_\mu {(\Lambda^{-1} )^c}_b )
({\Lambda^b}_p x^p)
\nonumber\\
&=&{\Lambda^a}_c (\partial_\mu x^c +\dot{A}{^c}_{p\mu}x^p)
+(\partial_\mu {\Lambda^a}_c) x^c
+{\Lambda^a}_c {\Lambda^b}_p (\partial_\mu  {(\Lambda^{-1} )^c}_b ) x^p
\nonumber\\
&=&{\Lambda^a}_c (\partial_\mu x^c +\dot{A}{^c}_{p\mu}x^p)
\nonumber\\
&=&{\Lambda^a}_c\, e{^c}_\mu,
\eea
where we used ${h^p}_\mu h_p=\partial_\mu$ in the second step and ${\Lambda^a}_c \partial_\mu {(\Lambda^{-1})^c}_b=- {(\Lambda^{-1})^c}_b \partial_\mu {\Lambda^a}_c$ in the third step because of ${\Lambda^a}_c{(\Lambda^{-1})^c}_b=\delta^a_b$.

In summary, the teleparallel gravity formalism we use from Ref.~\cite{Aldrovandi:2013wha} is local Lorentz invariant. Specially, the gravitational Lagrangian \eqref{dotL}, particle action \eqref{S particle action}, gravitational phase \eqref{phi B}, and electromagnetic phase \eqref{EMphase} are local Lorentz invariant, and the tetrad \eqref{hdotAB}, torsion \eqref{hT expression}, and equation of motion \eqref{EoM teleparallel} are local Lorentz covariant. Since the gravitational phase differences, electromagnetic phase differences, and fringe shifts in this paper are computed by applying the formulas \eqref{phi B} and \eqref{EMphase}, they are also local Lorentz invariant. 

Finally, we would like to mention that though the gauge potential ${B^a}_\mu$ is local Lorentz covariant (recall \eqref{BprimeB}), the specific expression in the right hand side of the second equation in \eqref{B field} is not local Lorentz covariant, recalling that such expression holds only in proper frames. In other words, in a non-proper frame, one should not use the expression ${B^a}_\mu={h^a}_\mu-{\delta^a}_{\mu'} \partial_\beta x^{\mu'}$ to calculate the phase \eqref{phi_g}, because in general the Lorentz connection does not vanish (recall \eqref{connection transfrom}). In a generic frame, such expression should be replaced by ${B^a}_\mu={h^a}_\mu-{e^a}_\mu$, which is local Lorentz covariant, i.e., ${B'^a}_\mu={h'^a}_\mu-{e'^a}_\mu$.

%%%%%%%%%%%
\subsection{Phases in Kerr-Newman spacetime\label{KN}}

In this section we show how to calculate the gravitational phase and the electromagnetic phase in Kerr-Newman spacetime, in terms of \eqref{phi_g} and \eqref{EMphase} respectively.

Discuss the gravitational phase firstly. Using Boyer-Lindquist coordinates $K (t, r, \theta, \varphi)$, the metric of Kerr-Newman spacetime is written as~\cite{Arcos:2004tzt}
\be
(g_{\mu\nu})=
\begin{pmatrix}
g_{00}&{}&{}&g_{03}\\
{}&g_{11}&{}&{}\\
{}&{}&g_{22}&{}\\
g_{30}&{}&{}&g_{33}
\end{pmatrix},
\label{metric}
\ee
where
\bea
&&g_{00}=1-\frac{R_0 r}{\rho ^2},
\qquad
g_{11}=-\frac{\rho ^2}{\Delta},
\qquad
g_{22}=-\rho ^2,
\nonumber\\
&&g_{33}=-\left(r^2+a^2+\frac{R_0 r a^2 \sin^2\theta}{\rho^2}\right)\sin^2\theta,
\qquad
g_{03}=g_{30}=\frac{R_0 r a \sin^2\theta}{\rho^2},
\nonumber\\
&&R_0=2M-\frac{Q^2}{r}, 
\qquad
\rho^2=r^2+a^2 \cos^2\theta,
 \qquad
\Delta=r^2-R_0 r +a^2.
\label{rgrho}
\eea
Here $M$ is the mass of the black hole, $Q$ is its charge, and $a$ is  its angular momentum per unit mass.\footnote{Notice that $a=J/(Mc)$ in SI units, with $J$ the angular momentum of the black hole.} Similar to the inverse Kerr metric~\cite{Chandrasekhar}, the inverse of \eqref{metric} is 
\be
(g^{\mu\nu})=
\begin{pmatrix}
g^{00}&{}&{}&g^{03}\\
{}&g^{11}&{}&{}\\
{}&{}&g^{22}&{}\\
g^{30}&{}&{}&g^{33}
\end{pmatrix},
\label{inverse metric}
\ee
where
\bea
&&g^{00}=\frac{\Sigma^2}{\rho^2\Delta},
\qquad
g^{11}=-\frac{\Delta}{\rho^2},
\qquad
g^{22}=-\frac{1}{\rho^2},
\qquad
g^{33}=-\frac{\Delta-a^2\sin^2\theta}{\rho^2\sin^2\theta\,\Delta},
\nonumber\\
&&g^{03}=g^{30}=\frac{R_0 r a}{\rho^2\Delta},
\qquad
\Sigma =\sqrt{\left(a^2+r^2\right)^2-a^2 \sin^2\theta\,\Delta}\, .
\eea

A tetrad in Kerr-Newman spacetime with a vanishing Lorentz connection is (see Appendix~\ref{derive tetrad} for how to find it)
\be
(h{^a}_\mu)
=\begin{pmatrix}
\frac{\chi}{\rho} & 0 & 0 &\frac{a(2Mr-Q^2)}{\rho\chi} \sin^2\theta \\
0& \frac{r}{\chi_0}\sin\theta\cos\varphi
\, &\rho_0 \cos\theta \cos\varphi &
-\frac{\chi_0}{\chi}\rho\sin\theta\sin\varphi\\
0&\frac{r}{\chi_0}\sin\theta\sin\varphi 
& \rho_0 \cos\theta \sin\varphi
 &\frac{\chi_0}{\chi}\rho \sin\theta\cos\varphi\\
0& \frac{\rho_0}{\chi_0}\cos\theta & -r\sin\theta &0
\end{pmatrix},
\label{b matrix}
\ee
where
\bea
\rho_0&=&\sqrt{r^2+a^2},
\label{rho0}
\\
\chi_0 &=& \sqrt{r^2+a^2+Q^2-2Mr},
\\
\chi &=& \sqrt{r^2+a^2\cos^2\theta +Q^2-2Mr}.
\label{chichi}
\eea
In Ref.~\cite{Arcos:2004tzt}, the authors also found a tetrad in such spacetime (see \eqref{tetrad Kerr}). However, as shown in Appendix~\ref{derive tetrad}, the Lorentz connection of \eqref{tetrad Kerr} is non-vanishing, which is the reason why we do not directly adopt \eqref{tetrad Kerr}. Besides, we mention that the tetrad used in Ref.~\cite{Mo} for Kerr spacetime actually has a non-vanishing Lorentz connection (with the same argument in Appendix~\ref{derive tetrad}, one can prove that it is the same as the Lorentz connection of \eqref{tetrad Kerr}). In terms of \eqref{b matrix} we get
\be
(h_a{^\mu})
=\begin{pmatrix}
\frac{\rho}{\chi} & 0 & 0 &0 \\
\frac{a(2Mr-Q^2)}{\rho\chi\chi_0}\sin\theta\sin\varphi
&\frac{r\chi_0}{\rho^2} \sin\theta \cos\varphi &\frac{\rho_0}{\rho^2} \cos\theta \cos\varphi& 
-\frac{\chi}{\rho\chi_0}\csc\theta\sin\varphi\\
-\frac{a(2Mr-Q^2)}{\rho\chi\chi_0}\sin\theta\cos\varphi
&\frac{r\chi_0}{\rho^2} \sin\theta \sin\varphi 
&\frac{\rho_0}{\rho^2} \cos\theta \sin\varphi
& \frac{\chi}{\rho\chi_0}\csc\theta\cos\varphi\\
0& \frac{\rho_0\chi_0}{\rho^2}\cos\theta & -\frac{r}{\rho^2}\sin\theta &0
\end{pmatrix}.
\label{inverse tetrad}
\ee
In deriving \eqref{inverse tetrad}, we used the relation ${h^a}_\mu {h_c}^\mu=\delta^a_c$ for the tetrad \cite{Aldrovandi:2013wha}. By the way, in \eqref{inverse tetrad} one can see ${h_0}^i=0$, hence such tetrad corresponds to a stationary observer~\cite{Ulhoa:2019ibd} (whose four-velocity satisfies $U^i=0$). Note that a tetrad which satisfies ${h_0}^i=U^i=0$ is called ``stationary reference frame''~\cite{Ulhoa:2019ibd} (some people call it ``static reference frame''~\cite{Maluf:2023rwe}).

According to \eqref{B field}, to find the gravitational gauge potential, we also need to know the transformation between the cartesian coordinates $K'$ and the Boyer-Lindquist coordinates $K$. The relation between them is~\cite{Landau:1975pou}
\be
\begin{cases}
t'=t,\\
x'=\sqrt{r^2+a^2}\, \sin\theta\cos\varphi,\\
y'=\sqrt{r^2+a^2}\, \sin\theta\sin\varphi,\\
z'=r\, \cos\theta,
\end{cases}
\label{transformationKK}
\ee
with the transformation matrix
\be
({J^{\mu'}}_\mu)=\Bigl(\frac{\partial x^{\mu'}}{\partial x^\mu}\Bigr)=
\begin{pmatrix}
1 & 0 & 0 &0\\
0& \frac{r}{\rho_0}\sin\theta \cos\varphi &\rho_0 \cos\theta\cos \varphi &-\rho_0\sin\theta\sin\varphi\\
0& \frac{r}{\rho_0}\sin\theta \sin\varphi &\rho_0 \cos\theta\sin \varphi &\rho_0\sin\theta\cos\varphi\\
0&\cos\theta &-r\sin\theta &0
\end{pmatrix}.
\label{Jacobi}
\ee
Plugging \eqref{b matrix} and \eqref{Jacobi} into the second formula of \eqref{B field}, we get the gauge potential
\be
(B{^a}_\beta)=
\begin{pmatrix}
\frac{\chi}{\rho}-1 & 0 & 0 &\frac{a(2Mr-Q^2)}{\rho\chi} \sin^2\theta \\
0& (\frac{1}{\chi_0}-\frac{1}{\rho_0} )r\sin\theta\cos\varphi
&0 &
( \rho_0-\frac{\chi_0}{\chi}\rho )\sin\theta\sin\varphi\\
0&(\frac{1}{\chi_0}-\frac{1}{\rho_0} )r\sin\theta\sin\varphi 
&0
 &(\frac{\chi_0}{\chi}\rho -\rho_0 ) \sin\theta\cos\varphi\\
0&(\frac{\rho_0}{\chi_0}-1 )\cos\theta & 0 &0
\end{pmatrix}.
\label{B Kerr}
\ee
Then replacing \eqref{B Kerr} and \eqref{inverse tetrad} into the first equation of \eqref{B field}, we obtain
\be
(B{^\nu}_\beta)
=\begin{pmatrix}
1-\frac{\rho}{\chi}  & 0 & 0 & \frac{a(2Mr-Q^2)\rho_0}{\rho\chi\chi_0} \sin^2\theta \\
0& 1-\frac{\chi_0}{\rho_0} & 0 &0\\
0& 0 & 0 &0\\
0&0 &0 & 1-\frac{\rho_0\chi}{\rho\chi_0}
\end{pmatrix}.
\label{B expression}
\ee

Afterwards, inserting \eqref{B expression} into the second equation of \eqref{phi_g}, implies
\be
(S_\beta)
=\begin{pmatrix}
P_0(1-\frac{\rho}{\chi})\\
P_1(1-\frac{\chi_0}{\rho_0})\\
0\\
P_0 \frac{a(2Mr-Q^2)\rho_0}{\rho\chi\chi_0} \sin^2\theta +P_3(1-\frac{\rho_0\chi}{\rho\chi_0})
\end{pmatrix}.
\label{S beta}
\ee
As for the four-momentum in \eqref{S beta}, let us focus on the equation of motion of the particle. For a charged particle in Kerr-Newman spacetime, there are two conserved quantities $\mathcal{E}$ and $\mathcal{L}$ relating to the $t$ and $\varphi$ components of the momentum~\cite{Misner1973}
\bea
\mathcal{E}&=& P_0+qA_0, 
\label{conserved quantities1}
\\
-\mathcal{L}&=&P_3+qA_3,
\label{conserved quantities}
\eea
with $A_\mu$ the electromagnetic potential given by
\be
A_\mu\dd x^\mu
=\frac{Qr}{\rho^2}\bigl(\dd t -a\sin^2\theta\, \dd\varphi\bigr).
\label{dmudxmu}
\ee
We would like to mention that \eqref{conserved quantities1} and \eqref{conserved quantities} look different from (33.31a) and (33.31b) of Ref.~\cite{Misner1973} (but $\mathcal{E}$ and $\mathcal{L}$ are equivalent to those of them). This is because the metric signature $(-,+,+,+)$ in~\cite{Misner1973} is different from ours. Correspondingly, the sign of the electromagnetic potential here is opposite to the sign in~\cite{Misner1973}.

As shown in~\cite{Misner1973}, the quantity $\mathcal{E}$ is the energy of the particle at infinity, while $\mathcal{L}$ corresponds to its angular momentum along the axis of the black hole. Plugging \eqref{conserved quantities1} and \eqref{conserved quantities} into \eqref{S beta}, and using $P_\mu=g_{\mu\nu}P^\nu$ for $P_1$, we get
\be
(S_\beta)
=\begin{pmatrix}
(\mathcal{E}-qA_0)(1-\frac{\rho}{\chi})\\
g_{11}P^1(1-\frac{\chi_0}{\rho_0})\\
0\\
(\mathcal{E}-qA_0) \frac{a(2Mr-Q^2)\rho_0}{\rho\chi\chi_0} \sin^2\theta -(\mathcal{L}+qA_3)(1-\frac{\rho_0\chi}{\rho\chi_0})
\end{pmatrix}.
\label{S beta2}
\ee
The expression of $P^1$ can be found from the equation~\cite{Hackmann:2013pva}
\be
\Bigl(\frac{d\bar{r}}{d\gamma}\Bigr)^2=R(\bar{r}),
\label{EoMtwo}
\ee
where $\gamma$ is the Mino time~\cite{Mino:2003yg} defined by $d\gamma=\rho^{-2}M d\tau$ with $\tau$ the eigentime, and the symbol bar denotes the normalization to $M$ (namely, $\bar{x}=x/M$ for $x=r$, $a$, $L$, $Q$, while $\bar{K}=K/M^2$ and $\bar{\Delta}(\bar{r}) =\bar{r}^2-2\bar{r} +\bar{a}^2 +\bar{Q}^2$), and $R(\bar{r})$ is defined as
\be
R(\bar{r})=\bigl[(\bar{r}^2+\bar{a}^2)E -\bar{a}\bar{L} -\frac{q}{m}\bar{Q}\bar{r}\bigr]^2 
-(\epsilon \bar{r}^2 +\bar{K})\bar{\Delta}(\bar{r}),
\ee
where $K$ is the Carter constant~\cite{Carter:1968rr}, with $E=\mathcal{E}/m$ and $L=\mathcal{L}/m$, and $\epsilon=1$ for timelike geodesics while $\epsilon=0$ for null geodesics. Here we take $\epsilon=1$, because we only consider massive particles. Using \eqref{EoMtwo}, we get the expression
\be
P^1=m\frac{\dd r}{\dd\tau}
=\pm \frac{1}{\rho^2} \sqrt{ \! \bigl[ ( r^2 \! + \! a^2 )\mathcal{E}  \! - \! a\mathcal{L}  \! - \! q Qr \bigr]^2  \! - \! m^2( r^2  \! + \! K)(r^2  \! - \! 2Mr  \! + \! a^2  \! + \! Q^2) }.
\label{P1}
\ee
By the way, according to Ref.~\cite{Hackmann:2013pva}, we also have the following equation
\be
P^2=m\frac{d\theta}{d\tau}
=\pm \frac{1}{\rho^2} \sqrt{m^2 K -m^2 a^2\cos^2\theta -\frac{1}{\sin^2\theta}(a\mathcal{E}\sin^2\theta -\mathcal{L} )^2}.
\label{PP2}
\ee

Then recalling \eqref{phi_g}, the gravitational phase in Kerr-Newman spacetime is
\be
\phi_g=\frac{1}{\hbar}\int_A^B S_\beta \dd x^\beta,
\label{phi_gKN}
\ee
with $S_\beta$ given in \eqref{S beta2}. As for the electromagnetic phase, we have from \eqref{EMphase} and  \eqref{dmudxmu} the following result 
\be
\phi_e 
=\frac{1}{\hbar}\int_A^B T_\beta \dd x^\beta,
\label{phie}
\ee
where
\be
T_\beta
=\begin{pmatrix}
q\frac{Qr}{\rho^2}
\\
0
\\
0
\\
-q\frac{Q r}{\rho^2} a\sin^2\theta
\end{pmatrix}.
\label{Qbeta}
\ee

Besides, because in this paper we only consider proper frames ($\dot{A}{^a}_{b\beta}=0$), the inertial phase $\phi_{\rm i}$ in \eqref{free phase0} vanishes, such that the free phase takes the form
\be
\phi_{\rm f}
=\frac{m}{\hbar}\int  u_\beta \dd x^\beta -\phi_{\rm g}
=\frac{1}{\hbar} \int F_\beta \dd x^\beta,
\label{free phase}
\ee
where
\bea
(F_\beta)
&=&m \begin{pmatrix}
u_0 \frac{\rho}{\chi}\\
u_1 \frac{\chi_0}{\rho_0} \\
u_2\\
u_0 \frac{a(Q^2-2Mr)\rho_0}{\rho\chi\chi_0} \sin^2\theta +u_3 \frac{\rho_0\chi}{\rho\chi_0}
\end{pmatrix}
=
\begin{pmatrix}
(\mathcal{E}-q A_0)\frac{\rho}{\chi}\\
g_{11} P^1 \frac{\chi_0}{\rho_0}\\
g_{22} P^2\\
(\mathcal{E}-q A_0) \frac{a(Q^2-2Mr)\rho_0}{\rho\chi\chi_0} \sin^2\theta 
-(\mathcal{L}+ q A_3 )\frac{\rho_0\chi}{\rho\chi_0}
\end{pmatrix}.
\label{F beta020}
\eea
Finally, the total quantum phase is
\be
\phi_{\rm total}
=\phi_g+\phi_e+\phi_{\rm f}
=\frac{1}{\hbar}\int_A^B S_\beta \dd x^\beta
+\frac{1}{\hbar}\int_A^B T_\beta dx^\beta
+\frac{1}{\hbar}\int_A^B F_\beta dx^\beta.
\label{phige2}
\ee

%%%%%
%%%%%%%%%%%
\section{Quantum interference in Kerr-Newman spacetime\label{interference0}}

\subsection{Interference experiment\label{interference}}

\begin{figure}[h]
\centering
\includegraphics[width=0.7\textwidth]{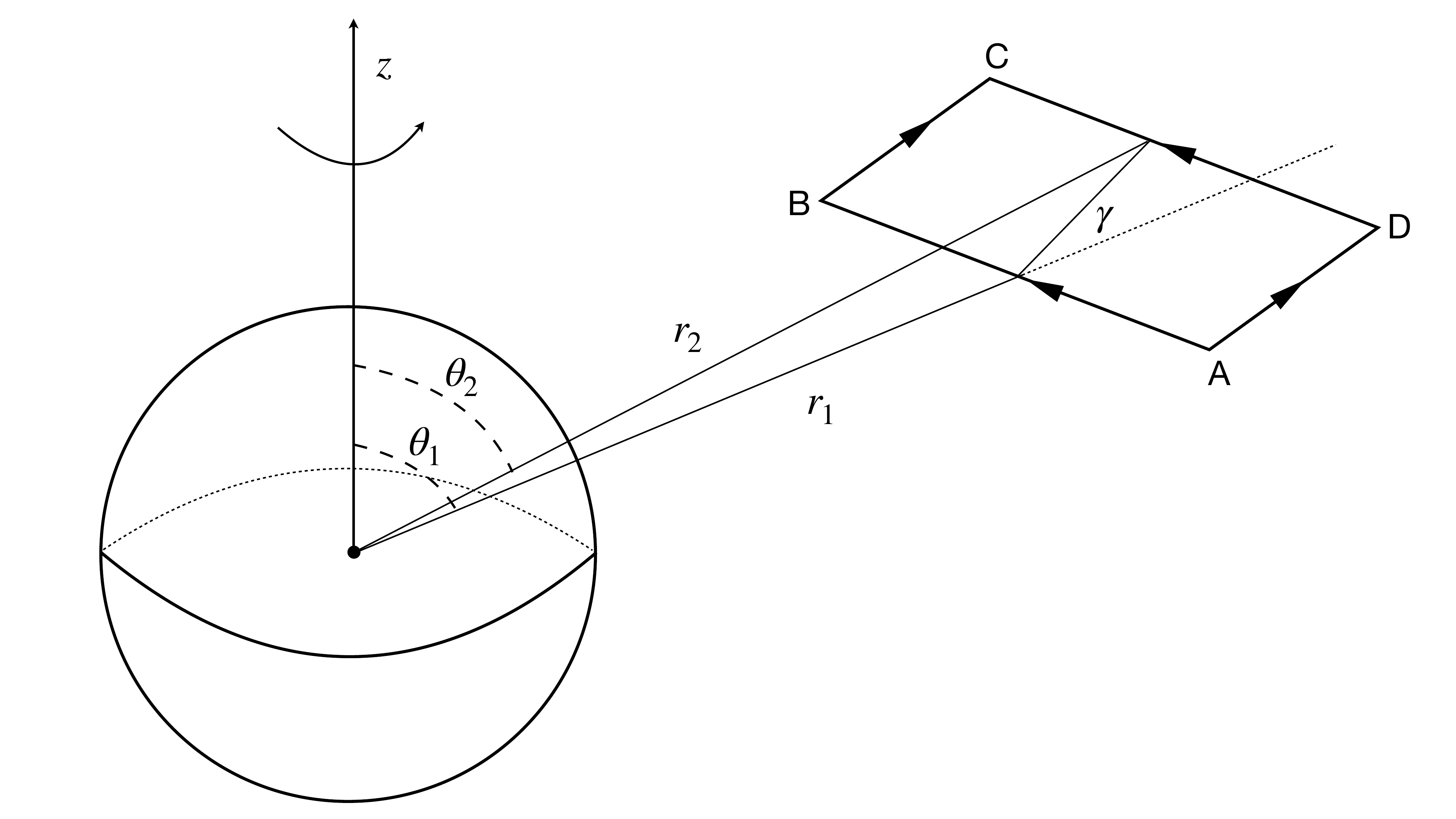}
\caption{An interference experiment in Kerr-Newman spacetime, where the setup is put in the region $r\gg r_g$. The beams start from the point A and interfere at the point C. The curved arrow represents the direction of the black hole's rotation, and $z$ denotes the rotation axis. The coordinates $(r_1,\theta_1)$ and $(r_2,\theta_2)$ correspond to the points on AB and DC respectively. The height of the rectangle is $l$, and the length of the base AB is $s$. The angle between the rectangle and $r_1$ is $\gamma$. Here AB is perpendicular to $\vec{r}_1$, $\vec{r}_2$, and the axis $z$. This figure is plotted based on the second figure of Ref.~\cite{Mo}.}
\label{experimentKN}
\end{figure}

In Ref.~\cite{Mo}, a gravitationally induced interference experiment was discussed in Kerr spacetime, where the authors considered two paths along a parallelogram in the asymptotic region and computed the phase shifts. Now we study this experiment again, but consider charged particles in Kerr-Newman spacetime, with a rectangle as the interferometer, as FIG.~\ref{experimentKN} shows. The coherent beams start from A and interfere at C. Note that the situation of FIG.~\ref{experimentKN} corresponds to $\varphi_{AB}>0$. What we measure are the fringe shifts when the setup is rotated from $\gamma=0$ to $\gamma=\pi$ or is flipped such that A and B swap. 

The setup is put in the region satisfying $r_g/r\ll 1$, where $r_g=2M$ is the Schwarzschild radius, and we take the assumption
\be
O\Bigl(\frac{r_g}{r}\Bigr)\sim O\Bigl(\frac{a}{r}\Bigr)\sim O\Bigl(\frac{Q}{r}\Bigr).
\label{O three}
\ee
Besides, we adopt the following assumptions same as those in Ref.~\cite{Mo}:\\
(a) The size of the setup is much smaller than its distance from the black hole, such that  approximately $r$ and $\theta$ keep constant along AB and DC; 
\\
(b) The energy $\mathcal{E}$ of the particle keeps constant even when it turns direction at B and D (hence the magnitude of the velocity (see \eqref{vL}) does not change at B and D).

In the following contents we calculate the gravitational phase difference firstly, then the electromagnetic phase difference, finally the free phase difference. Same as the formula in Ref.~\cite{Mo}, the gravitational phase difference between the paths ADC and ABC can be written as (see Appendix~\ref{formula derivation})
\be
\delta\phi^g
=(\phi_{AD}^g+\phi_{DC}^g)-(\phi_{AB}^g+\phi_{BC}^g)
=\delta\phi_a+\delta\phi_b +\delta\phi_c,
\label{delta phi0g}
\ee
where
\bea
\delta\phi_a
&=& 
\frac{1}{\hbar}\bigl[ (S_0^D)_{DC} t_{AB}+ (S_3^D)_{DC}\varphi_{AB}\bigr] 
-\phi_{AB}^g,
\label{delta phi ag}
\\
\delta\phi_b
&=& 
\frac{1}{\hbar}
\bigl[(S_0^D)_{DC} -(S_0^A)_{AB} \bigr] (t_{DC}-t_{AB}),
\label{delta phi bg}
\\
\delta\phi_c
&=& 
\frac{1}{\hbar}\bigl[(S_3^D)_{DC} -(S_3^B)_{BC} \bigr] (\varphi_{DC}-\varphi_{AB}).
\label{delta phi cg}
\eea
Here the superscripts A, B, C, and D denote the positions, while the subscripts AB, BC, AD, and DC denote the paths. As \eqref{delta phi ag}, \eqref{delta phi bg}, and \eqref{delta phi cg} show, the phase difference depends on the components $S_0$ and $S_3$. Expand them at the third-order of \eqref{O three} (see \eqref{S beta2} for $S_\beta$)
\bea
S_0
&\approx&
\mathcal{E}\Bigl(-\frac{r_g}{2r}
-\frac{3 r_g^2}{8 r^2}
+\frac{Q^2}{2r^2}
-\frac{5r_g^3}{16r^3}+ \frac{a^2 r_g}{2r^3} \cos^2\theta
+\frac{3 r_g Q^2}{4r^3}
\Bigr)
+q \Bigl(\frac{r_g Q}{2 r^2}+\frac{3 r_g^2 Q}{8 r^3}-\frac{Q^3}{2 r^3}\Bigr),
\label{S beta3}
\\
S_3
&\approx&
\mathcal{E} r \sin^2\theta\, \Bigl(\frac{a r_g}{r^2}
  +\frac{a r_g^2}{r^3}
 -\frac{a Q^2}{r^3}
 \Bigr) 
 -\mathcal{L}  \sin^2\theta\, \frac{a^2 r_g}{2r^3}
-q r \sin^2\theta\, \frac{a r_g Q}{r^3}.
\label{S beta3-2}
\eea
Before calculating the phase difference \eqref{delta phi0g}, let us discuss $\phi^g_{AB}$ firstly. Applying \eqref{phi_gKN}, we get the gravitational phase along the path AB as follows
\be
\phi^g_{AB} \approx
\frac{1}{\hbar}
(S_0 t_{AB} + S_3 \varphi_{AB}),
 \label{phiAB0}
\ee
where $t_{AB}=t_B-t_A$ and $\varphi_{AB}=\varphi_B-\varphi_A$. In deriving \eqref{phiAB0}, we used the assumption~(a) and the fact that $S_\beta$ does not depend on $t$ and $\varphi$ (see \eqref{S beta2}). Substituting \eqref{S beta3} and \eqref{S beta3-2} into \eqref{phiAB0}, yields
\bea
\phi^g_{AB} &\approx&
\frac{t_{AB}}{\hbar}  \mathcal{E} 
\Bigl(
-\frac{r_g}{2 r_1}-\frac{3 r_g^2}{8 r_1^2} 
+\frac{Q^2}{2 r_1^2}
-\frac{5 r_g^3}{16 r_1^3}+ \frac{a^2 r_g}{2 r_1^3} \cos^2\theta_1 
+\frac{3 r_g Q^2}{4 r_1^3}
\Bigr)
\nonumber\\
&&+\frac{\varphi_{AB}}{\hbar} 
\sin^2\theta_1 \Bigl[
\mathcal{E} r_1 
\Bigl(
 \frac{a r_g}{ r_1^2}
 +\frac{a r_g^2}{ r_1^3}
 -\frac{a Q^2}{ r_1^3}
  \Bigr)
 -\frac{1}{2} \mathcal{L}_{AB} \frac{a^2 r_g}{ r_1^3}
 \Bigr]
 \nonumber\\
&&
+q\Bigl[
\frac{t_{AB}}{\hbar}
  \Bigl(\frac{r_g Q}{2 r_1^2}+\frac{3 r_g^2 Q}{8 r_1^3}-\frac{Q^3}{2 r_1^3}
 \Bigr)
-\frac{\varphi_{AB}}{\hbar}  r_1 \sin^2\theta_1\, \frac{a r_g Q}{ r_1^3}
\Bigr].
\label{phiAB}
\eea
Similar to those in Ref.~\cite{Mo}, the quantities relating to observations are given by (see Appendix~\ref{observation})
\bea
&&t_{AB} \approx s\Bigl(
\frac{1}{v\sqrt{g_{00} }}-\frac{g_{03} }{g_{00}\sqrt{\Gamma_{33}} }
\Bigr),
\label{t varphi}
\\
&&\varphi_{AB} \approx \frac{s}{\sqrt{\Gamma_{33} }},
\label{t varphi2}
\\
&&\mathcal{E}=
\frac{m \sqrt{g_{00}} }{\sqrt{1-v^2}} +q A_0,
\label{mathcalE}
\\
&&\mathcal{L} = \frac{1}{g_{00}}\bigl(
-g_{03} +v^\varphi \Gamma_{33}  \sqrt{g_{00}}
\bigr) 
(\mathcal{E}-q A_0)
-q A_3,
\label{L given}
\eea
where $v$ is the magnitude of the velocity defined in \eqref{vL}, with the three-velocity and the three-dimensional metric tensor defined by~\cite{Landau:1975pou}
\be
v^k=\frac{\dd x^k}{\sqrt{g_{00}} (\dd x^0+\frac{g_{0i}}{g_{00}}
 \dd x^i)},\qquad
\Gamma_{ij} =-g_{ij}+\frac{g_{0i}g_{0j}}{g_{00}}.
\label{v components}
\ee
Note that \eqref{t varphi2} holds only when $\varphi_{AB}\ll 1$ (namely $s\ll \sqrt{\Gamma_{33}}$) is satisfied. One can check such condition for the case $M=a=Q=0$, in which we have $\tan(\varphi_{AB}/2)=s/(2 r_1\sin\theta_1)$. In such case (flat spacetime), only when $\varphi_{AB}\ll 1$ holds, the approximation $\varphi_{AB}\approx s/(r_1\sin\theta_1)$ holds (which implies $s\ll r_1\sin\theta_1$). As for $t_{DC}$ and $\varphi_{DC}$, we only need to make the replacements $r_1\rightarrow r_2$, $\theta_1\rightarrow\theta_2$, and $v\rightarrow v_2$ in \eqref{t varphi} and \eqref{t varphi2}, where $v_2$ is the magnitude of the velocity on DC.

Now we return to the phase difference $\delta\phi^g$. We can compute it following the same procedure in Ref.~\cite{Mo}. Namely, plugging \eqref{S beta3}, \eqref{S beta3-2}, and \eqref{phiAB} into \eqref{delta phi0g}, and using \eqref{t varphi}, \eqref{t varphi2}, \eqref{mathcalE}, \eqref{L given}, and the relations~\cite{Mo}\footnote{Though the two relations in \eqref{r2 theta2} are derived in Ref.~\cite{Mo} for Kerr spacetime, they still hold in Kerr-Newman spacetime.}
\be
r_2\approx r_1 +\frac{l\cos\gamma}{\sqrt{-g_{11}(r_1,\theta_1)} },
\qquad
\theta_2\approx \Bigl| \theta_1-
\frac{l\sin\gamma }{\sqrt{-g_{22}(r_1,\theta_1)}} \Bigr|,
\label{r2 theta2}
\ee
we expand $\delta\phi^g$ in the neighborhoods of $r_1$ and $\theta_1$ to the first-order of $l/r_1$. Finally we derive 
\be
\delta\phi^g
\approx
\delta\phi^g_0
+\delta\phi^g_Q
+\delta\phi^g_q,
\label{delta phi g}
\ee
where
\bea
\delta\phi^g_0
&=&\frac{m l s}{ \hbar r_1 \sqrt{1-v^2}} \Bigl\{
\frac{1}{v}\cos\gamma\,
\Bigl(\frac{r_g}{2r_1}+\frac{r_g^2}{2r_1^2}\Bigr)
-\frac{a r_g}{r_1^2}\Bigl(2\cos\theta_1\, \sin\gamma+\cos\gamma\,\sin\theta_1
\Bigr)
\nonumber\\
&&
+\frac{1}{v}\cos\gamma\,\frac{r_g^3}{2r_1^3}
-\frac{a r_g^2}{r_1^3}
\Bigl(
  \cos\theta_1\,\sin\gamma
 +\frac{3}{2}\cos\gamma\,\sin\theta_1
 \Bigr)
\nonumber\\
&&
+\frac{a^2 r_g}{r_1^3}
\Bigl[ \frac{1}{v}\Bigl(\frac{1}{4}\cos\gamma\,-\frac{7}{4}\cos\gamma\,\cos^2\theta_1
+\frac{1}{2}\sin(2\theta_1)
\sin\gamma
\Bigr)
\nonumber\\
&&
+ v\sin\theta_1\,\Bigl(
\frac{3}{2} \cos\gamma\, \sin\theta_1
+\sin\gamma\, \cos\theta_1
\Bigr)
\Bigr]
\Bigr\},
\label{delta g0}
\\
\delta\phi^g_Q
&=&\frac{m l s}{ \hbar r_1 \sqrt{1-v^2}} \Bigl[
-\frac{1}{v}\cos\gamma\, \Bigl(
\frac{Q^2}{r_1^2} +\frac{3r_g Q^2}{2 r_1^3}
\Bigr)
+2\sin(\theta_1+\gamma)\frac{a Q^2}{r_1^3}
\Bigr],
\label{delta gQ}
\\
\delta\phi^g_q
&=&\frac{q l s}{ \hbar r_1} \Bigl[
\frac{1}{v}\cos\gamma\, \Bigl(
-\frac{r_g Q}{2r_1^2} 
-\frac{3r_g^2 Q}{8 r_1^3}
+\frac{Q^3}{2r_1^3} 
\Bigr)
+\cos\gamma\,\sin\theta_1\, \frac{a r_g Q}{r_1^3}
\Bigr].
\label{delta gqq}
\eea
To restore SI units, one can multiply the right hand sides of \eqref{delta g0}, \eqref{delta gQ}, and \eqref{delta gqq} by $c$, $c$, and $c/\sqrt{4\pi \varepsilon_0 G}$ respectively, and simultaneously need to make the following replacements (note that in SI units, the right hand side of the first equation in \eqref{v components} should be multiplied by $c$)
\be
v\rightarrow \frac{v}{c},
\qquad 
v^r\rightarrow \frac{v^r}{c},
\qquad 
v^\theta\rightarrow \frac{v^\theta}{c}, 
\qquad r_g\rightarrow \frac{2GM}{c^2}, 
\qquad 
Q\rightarrow \frac{Q}{c^2} \sqrt{\frac{G}{4\pi\varepsilon_0}},
\label{replacements vrQ}
\ee
in these equations.

If we let $Q=0$, the gravitational phase difference is reduced to $\delta\phi_0^g$,  corresponding to the gravitational phase difference in Kerr spacetime. We notice that the term \eqref{delta g0} coincides with the result (75) of Ref.~\cite{Mo} (except that the last term in \eqref{delta g0} looks different from that in Ref.~\cite{Mo}), though ${h^a}_\mu|_{Q=0}$ is different from the  tetrad of the latter. The reason is that when $Q=0$, the expressions of $S_0$ and $S_3$ are the same as those in Ref.~\cite{Mo} (one can check this claim by setting $Q=0$ in \eqref{S beta2} and comparing the result with (64) of Ref.~\cite{Mo}), and that only these two components are used to calculate the gravitational phase difference (see \eqref{delta phi ag}, \eqref{delta phi bg}, and \eqref{delta phi cg}). The term $\delta\phi^g_Q$ represents the contribution from the charge of the black hole, by means of the gravitation rather than the electromagnetic interaction. It is not surprising, because in Kerr-Newman spacetime, the charge of the black hole also plays a role in the metric. As for $\delta\phi^g_q$, it is caused by both gravitation and electromagnetic interaction. This term comes from the relations \eqref{mathcalE} and \eqref{L given}. 

As for the electromagnetic phase difference, similar to \eqref{delta phi0g}, it is calculated by the following formula (Appendix~\ref{formula derivation})
\be
\delta\phi^e
=\delta\phi^e_a+\delta\phi^e_b +\delta\phi^e_c,
\label{delta phi0EM}
\ee
where
\bea
\delta\phi^e_a
&=& 
\frac{q}{\hbar}\bigl[ (A_0^D)_{DC} t_{AB}+ (A_3^D)_{DC}\varphi_{AB}\bigr]
-\frac{q}{\hbar}\bigl[ (A_0^A)_{AB} t_{AB}+ (A_3^A)_{AB}\varphi_{AB}\bigr] ,
\label{delta phi agEM}
\\
\delta\phi^e_b
&=& 
\frac{q}{\hbar}
\bigl[(A_0^D)_{DC} -(A_0^A)_{AB} \bigr] (t_{DC}-t_{AB}),
\label{delta phi bgEM}
\\
\delta\phi^e_c
&=& 
\frac{q}{\hbar}\bigl[(A_3^D)_{DC} -(A_3^B)_{BC} \bigr] (\varphi_{DC}-\varphi_{AB}).
\label{delta phi cgEM}
\eea
Then repeating the above calculations, we derive
\be
\delta\phi^e
\approx
\delta\phi^e_0 +\delta\phi^e_G,
\label{delta e}
\ee
where
\bea
\delta\phi^e_0
&=&
\frac{q l s}{ \hbar r_1} \Bigl\{
- \frac{1}{v}
\frac{Q}{r_1} \cos\gamma
+\frac{a Q}{r_1^2}
\bigl(2\sin\gamma\,\cos\theta_1 +\cos\gamma\,\sin\theta_1\bigr)
\nonumber\\
&&
+\frac{1}{v} \frac{a^2 Q}{4 r_1^3}
\Bigl[
\cos\gamma\, \bigl(5+7\cos(2\theta_1) \bigr)
-4\sin\gamma\,\sin(2\theta_1)
\Bigr]
\Bigr\},
\label{deltaPhie0}
\\
\delta\phi^e_G
&=&
\frac{q l s}{ \hbar r_1}
\frac{a r_g Q}{2 r_1^3}\cos\gamma\,\sin\theta_1.
\label{deltaPhieG}
\eea
The term $\delta\phi^e_0$ corresponds to a pure electromagnetic interaction, while $\delta\phi^e_G$ is a mixed gravitational–electromagnetic term. In SI units, the former is independent of the gravitational constant $G$, while the latter is proportional to it, as follows
\bea
(\delta\phi^e_0)_{\rm SI}
&=&
\frac{1}{4\pi\varepsilon_0} \frac{q l s}{ \hbar r_1} \Bigl\{
- \frac{1}{v}
\frac{Q}{r_1} \cos\gamma
+\frac{1}{ c}\frac{a Q}{r_1^2}
\bigl(2\sin\gamma\,\cos\theta_1 +\cos\gamma\,\sin\theta_1\bigr)
\nonumber\\
&&
+\frac{1}{v} \frac{a^2 Q}{4 r_1^3}
\Bigl[
\cos\gamma\, \bigl(5+7\cos(2\theta_1) \bigr)
-4\sin\gamma\,\sin(2\theta_1)
\Bigr]
\Bigr\},
\label{deltaPhie0SI}
\\
(\delta\phi^e_G)_{\rm SI} 
&=&
\frac{1}{4\pi\varepsilon_0}
\frac{G }{c^3} \frac{q l s}{ \hbar r_1}
\frac{a M Q}{r_1^3}\cos\gamma\,\sin\theta_1.
\label{deltaPhieGSI}
\eea
Indeed, by letting $G=0$ in the metric (SI units), and still using the potential \eqref{dmudxmu}, then repeating the above calculations, we obtain only the term $\delta\phi^e_0$ for the electromagnetic phase difference. We would like to mention that the electromagnetic potential \eqref{dmudxmu} is independent of the gravitational constant (one can confirm this conclusion in SI units). It is not surprising, because the electromagnetic field of Kerr-Newman solution is equivalent to the field of a rotating charged perfect conductor in flat space~\cite{Tiomno:1973ku} (such conductor has the susceptibility of
the vacuum or has infinite magnetic susceptibility). This is the reason why we use the potential \eqref{dmudxmu} even at $G=0$.

In \eqref{delta gqq} and \eqref{deltaPhieG} we notice that both $\delta\phi^g_q$ and $\delta\phi^e_G$ are proportional to $q$ and contain the gravitational constant (after restore SI units). However, they have different origins. The term $\delta\phi^g_q$ comes from the conservation equations \eqref{mathcalE} and \eqref{L given} which show that the motion of the particle is affected by both the electromagnetic field and the gravitational field. As for $\delta\phi^e_G$, the gravitational constant comes from \eqref{t varphi} and the first equation of \eqref{r2 theta2} (though \eqref{t varphi2} also contains $G$ in SI units, it does not contribute to the third and lower terms which have $G$). Equations \eqref{t varphi} and \eqref{t varphi2} correspond to the gravitational modifications to the time and angle respectively, while \eqref{r2 theta2} relates to the gravitational modification to the length. Note that \eqref{r2 theta2} is derived using \eqref{vL2} (see Ref.~\cite{Mo}).

For the free phase difference, similar to \eqref{delta phi0g}, we can compute it by the formula (Appendix~\ref{formula derivation})
\bea
\phi^f
&=&\delta\phi_a^f +\delta\phi_b^f +\delta\phi_c^f,
\label{ADCminusABCf}
\eea
where 
\bea
\delta\phi_a^f
&=&\frac{1}{\hbar} \bigl[F_0^D t_{AB} +F_3^D \varphi_{AB}\bigr]_{DC}
-(\phi_{\rm f})_{AB},
\label{deltaPhiaf}
\\
\delta\phi_b^f
&=&\frac{1}{\hbar}[(F_0^D)_{DC}-(F_0^B)_{BC}] (t_{DC}-t_{AB}),
\label{deltaPhibf}
\\
\delta\phi_c^f
&=&\frac{1}{\hbar}[(F_3^D)_{DC}-(F_3^B)_{BC}] (\varphi_{DC}-\varphi_{AB}).
\label{deltaPhicf}
\eea
Then we repeat the above calculations to find
\bea
\delta\phi^f
&=&\frac{l s}{ \hbar r_1}
\Bigl[
\frac{1}{v}
\Bigl(
 -\frac{m \cos\gamma}{4\sqrt{1-v^2}}\frac{r_g^2}{r_1^2}
 +\frac{q\cos\gamma}{2} \frac{r_g Q}{r_1^2}
\Bigr)
+\frac{m}{16\sqrt{1-v^2}}
\Bigl( 12\cos\gamma \frac{r_g Q^2}{ v r_1^3}
- 5\cos\gamma \frac{r_g^3}{ v r_1^3}
\nonumber\\
&&
+ 8\cos\gamma \sin\theta_1  \frac{a r_g^2}{r_1^3}
- 8v \sin\theta_1\,(2\cos\theta_1\sin\gamma +3\sin\theta_1\cos\gamma)
\frac{a^2 r_g}{r_1^3}
\Bigr)
\nonumber\\
&&
+\frac{q}{8 v}\Bigl(
-4 \cos\gamma\frac{Q^3}{r_1^3}
+ 3\cos\gamma\frac{Q r_g^2}{r_1^3}
- 8 v \cos\gamma\sin\theta_1\frac{a r_g Q}{r_1^3}
\Bigr)
\Bigr].
\label{deltaphiFF0}
\eea
We can see that the free phase difference $\delta\phi^f$ only provides the second order and higher order modifications. 
Finally, the total quantum phase difference between the paths ABC and ADC is
\be
\delta\phi_{\rm total}
=\delta\phi^g +\delta\phi^e +\delta\phi^f
=\delta\phi^g_0 +\delta\phi^g_Q +\delta\phi^g_q 
+\delta\phi^e_0
+\delta\phi^e_G
+\delta\phi^f,
\label{deltaphige}
\ee
with $\delta\phi^g_0$,  $\delta\phi^g_Q$,  $\delta\phi^g_q$, $\delta\phi^e_0$, $\delta\phi^e_G$, and $\delta\phi^f$ given in \eqref{delta g0}, \eqref{delta gQ}, \eqref{delta gqq}, \eqref{deltaPhie0}, \eqref{deltaPhieG}, and \eqref{deltaphiFF0} respectively.

Similar to Ref.~\cite{Mo}, with the height $l$ as the axis, we flip the rectangle such that A and B swap, to get a new phase difference $\delta\phi_{\rm ge}'$. After this operation, the $\varphi$ component of the velocity is reversed, namely $(v^\varphi)'=-v^\varphi$, while the magnitude of the velocity is still $v$. Here the symbol ``prime'' is used to denote the quantities after the operation. On the path AB we have
\be
(v^\varphi)'_1=-\sqrt{\frac{v^2}{\Gamma_{33} } },
\label{v varphi prime}
\ee
opposite to \eqref{vphiGamma}. Correspondingly, the angle $\varphi_{AB}$ turns to
\be
\varphi'_{AB}
\approx
-\frac{s}{\sqrt{\Gamma_{33} }}.
\label{t varphiNew2}
\ee
In Ref.~\cite{Mo}, the authors claimed that after flipping the interferometer, $t_{AB}$ does not change while $\mathcal{L}$ changes to $-\mathcal{L}$. However, it is not the case for Kerr-Newman spacetime (it even fails in Kerr spacetime which Ref.~\cite{Mo} studies), though such claim indeed holds in flat spacetime. Instead, according to \eqref{vts} and \eqref{v varphi prime}, the time $t_{AB}$ should be changed to
\be
t'_{AB} 
\approx 
s\Bigl(
\frac{1}{v\sqrt{g_{00} }}+\frac{g_{03} }{g_{00}\sqrt{\Gamma_{33}} }
\Bigr).
\label{t varphiNew}
\ee
The energy $\mathcal{E}'$ is still given by \eqref{mathcalE}, while the angular momentum $\mathcal{L}'$ is obtained by replacing $v^\varphi$ by $(v^\varphi)'$ in \eqref{L given}. With these changes, we can derive a new gravitational phase difference ${\delta\phi'}^g$ and a new electromagnetic phase difference ${\delta\phi'}^e$ as follows
\bea
\delta{\phi}'^g
&\approx&
\delta{\phi}'^g_0
+\delta{\phi}'^g_Q
+\delta{\phi}'^g_q,
\label{delta gFlip}
\\
\delta{\phi}'^e
&\approx&
\delta{\phi}'^e_0 +\delta{\phi}'^e_G,
\label{delta eFlip}
\eea
where
\bea
\delta{\phi}'^g_0
&=&\frac{m l s}{ \hbar r_1 \sqrt{1-v^2}} \Bigl\{
\frac{1}{v}
\cos\gamma\,\Bigl(\frac{r_g}{2r_1}+\frac{r_g^2}{2r_1^2}\Bigr)
+\frac{a r_g}{r_1^2}\Bigl(2\cos\theta_1\, \sin\gamma+\cos\gamma\,\sin\theta_1
\Bigr)
\nonumber\\
&&
+\frac{1}{v}\cos\gamma\,\frac{r_g^3}{2r_1^3}
+\frac{a r_g^2}{r_1^3}
\Bigl(
  \cos\theta_1\,\sin\gamma
 +\frac{3}{2}\cos\gamma\,\sin\theta_1
 \Bigr)
\nonumber\\
&&
+\frac{a^2 r_g}{r_1^3}
\Bigl[ \frac{1}{v}\Bigl(\frac{1}{4}\cos\gamma\,-\frac{7}{4}\cos\gamma\,\cos^2\theta_1
+\frac{1}{2}\sin(2\theta_1)
\sin\gamma
\Bigr)
\nonumber\\
&&
+v\sin\theta_1\,\Bigl(
\frac{3}{2} \cos\gamma\, \sin\theta_1
+\sin\gamma\, \cos\theta_1
\Bigr)
\Bigr]
\Bigr\},
\label{delta g0Flip}
\\
\delta{\phi}'^g_Q
&=&\frac{m l s}{ \hbar r_1 \sqrt{1-v^2}} \Bigl[
-\frac{1}{v}\cos\gamma\, \Bigl(
\frac{Q^2}{r_1^2} +\frac{3r_g Q^2}{2 r_1^3}
\Bigr)
-2\sin(\theta_1+\gamma)\frac{a Q^2}{r_1^3}
\Bigr],
\label{delta gQFlip}
\\
\delta{\phi}'^g_q
&=&\frac{q l s}{ \hbar r_1} \Bigl[
\frac{1}{v}\cos\gamma\, \Bigl(
-\frac{r_g Q}{2r_1^2} 
-\frac{3r_g^2 Q}{8 r_1^3}
+\frac{Q^3}{2r_1^3} 
\Bigr)
-\cos\gamma\,\sin\theta_1\, \frac{a r_g Q}{r_1^3}
\Bigr],
\label{delta gqqFlip}
\\
\delta{\phi}'^e_0
&=&
\frac{q l s}{ \hbar r_1} \Bigl\{
- \frac{1}{v}
\frac{Q}{r_1} \cos\gamma
-\frac{a Q}{r_1^2}
\bigl(2\sin\gamma\,\cos\theta_1 +\cos\gamma\,\sin\theta_1\bigr)
\nonumber\\
&&
+\frac{1}{v} \frac{a^2 Q}{4 r_1^3}
\Bigl[
\cos\gamma\, \bigl(5+7\cos(2\theta_1) \bigr)
-4\sin\gamma\,\sin(2\theta_1)
\Bigr]
\Bigr\},
\label{deltaPhie0Flip}
\\
\delta{\phi}'^e_G
&=&
-\frac{q l s}{ \hbar r_1}
\frac{a r_g Q}{2 r_1^3}\cos\gamma\,\sin\theta_1,
\label{deltaPhieGFlip}
\eea
and 
\bea
\delta\phi'^f
&=&\frac{l s}{ \hbar r_1}
\Bigl[
\frac{1}{v}
\Bigl(
 -\frac{m \cos\gamma}{4\sqrt{1-v^2}}\frac{r_g^2}{r_1^2}
 +\frac{q\cos\gamma}{2} \frac{r_g Q}{r_1^2}
\Bigr)
+\frac{m}{16\sqrt{1-v^2}}
\Bigl( 12\cos\gamma \frac{r_g Q^2}{ v r_1^3}
- 5\cos\gamma \frac{r_g^3}{ v r_1^3}
\nonumber\\
&&
- 8\cos\gamma \sin\theta_1  \frac{a r_g^2}{r_1^3}
- 8v \sin\theta_1\,(2\cos\theta_1\sin\gamma +3\sin\theta_1\cos\gamma)
\frac{a^2 r_g}{r_1^3}
\Bigr)
\nonumber\\
&&
+\frac{q}{8 v}\Bigl(
-4 \cos\gamma\frac{Q^3}{r_1^3}
+ 3\cos\gamma\frac{Q r_g^2}{r_1^3}
+ 8 v \cos\gamma\sin\theta_1\frac{a r_g Q}{r_1^3}
\Bigr)
\Bigr].
\label{deltaphiFF00}
\eea
The new total quantum phase difference is $\delta\phi'_{\rm total}={\delta\phi'}^g+{\delta\phi'}^e+{\delta\phi'}^f$.

\begin{figure}[h]
\centering
\includegraphics[width=0.8\textwidth]{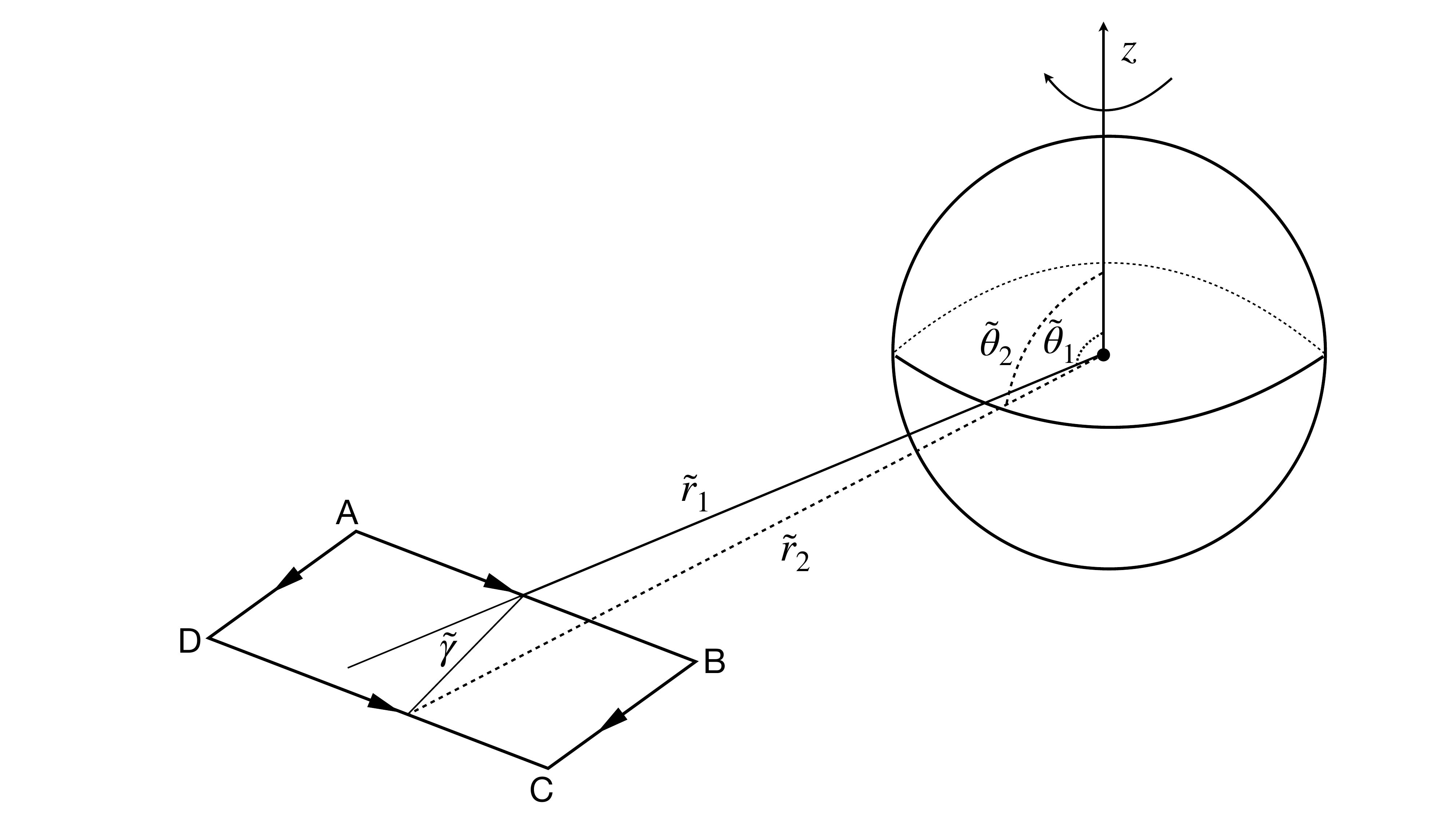}
\caption{The situation after rotating FIG.~\ref{experimentKN} by $\pi$ and flipping the rectangle along its height. The quantities $\tilde{r}_1$, $\tilde{r}_2$, $\tilde{a}$, $\tilde{\gamma}$, $\tilde{\theta}_1$, and $\tilde{\theta}_2$ are the radial coordinates, the spin parameter, and the angles respectively in the perspective of this situation. Now the direction of the axis $z$ is opposite to the direction of the black hole's angular momentum.}
\label{experimentKNFlip}
\end{figure}

Actually we have a simpler way to get the phase differences  \eqref{delta gFlip}, \eqref{delta eFlip}, and \eqref{deltaphiFF00}. For this purpose, we firstly rotate FIG.~\ref{experimentKN} by $\pi$ (around an axis which is perpendicular to $\vec{r}_1$ and $\vec{r}_2$ and passes through the origin) such that the angular momentum of the black hole is reversed, then flip the rectangle such that A and B swap, as FIG.~\ref{experimentKNFlip} shows. Note that we do not rotate the coordinate axes, i.e., we do not change the coordinate system.
We can find that the situation of the rectangle in FIG.~\ref{experimentKNFlip} is the same as the situation in FIG.~\ref{experimentKN}, except that we need to use a new spin parameter $\tilde{a}$, a new angle $\tilde{\gamma}$, and new coordinates $(\tilde{t}_p, \tilde{r}_p, \tilde{\theta}_p, \tilde{\varphi}_p)$ for the point on the rectangle. Here we use $(t_p, r_p, \theta_p, \varphi_p)$ to denote the old coordinates of such point. Evidently we have
\bea
&&\tilde{a} =-a,
\quad
\tilde{\gamma}=-\gamma,
\quad
\tilde{t}_p=t_p,
\quad
\tilde{r}_p=r_p,
\quad
\tilde{\theta}_p=\pi-\theta_p,
\label{coordinates flip}
\\
&&\tilde{\varphi}_p=
\begin{cases}
\varphi_p+\pi, & \text{if $0\le \varphi_p <\pi$},\\
\varphi_p-\pi, & \text{if $\pi\le \varphi_p< 2\pi$}.
\end{cases}
\eea
Therefore, to obtain the new phase differences $\delta{\phi}'^g$, $\delta{\phi}'^e$, and $\delta{\phi}'^f$, we only need to use the above new parameters and new coordinates to replace with those in the old phase differences. Making such replacements in \eqref{delta phi g}, \eqref{delta e}, and \eqref{ADCminusABCf}, we get
\bea
\delta{\phi}'^g &=&\delta{\phi}^g (a\rightarrow \tilde{a}, \gamma\rightarrow \tilde{\gamma}, r_1\rightarrow \tilde{r}_1, \theta_1\rightarrow \tilde{\theta}_1),
\\
\delta{\phi}'^e &=&\delta{\phi}^e (a\rightarrow \tilde{a}, \gamma\rightarrow \tilde{\gamma}, r_1\rightarrow \tilde{r}_1, \theta_1\rightarrow \tilde{\theta}_1),
\\
\delta{\phi}'^f &=&
\delta{\phi}^f (a\rightarrow \tilde{a}, \gamma\rightarrow \tilde{\gamma}, r_1\rightarrow \tilde{r}_1, \theta_1\rightarrow \tilde{\theta}_1).
\eea
Accordingly, recalling \eqref{coordinates flip}, we can plug the replacements
 \be
a\rightarrow -a,
\qquad
\gamma \rightarrow -\gamma,
\qquad
\theta_1\rightarrow \pi -\theta_1,
\label{replacements flip}
\ee
into the expressions of the old phase differences $\delta\phi^g$, $\delta\phi^e$, and $\delta\phi^f$ to get the new phase differences $\delta{\phi}'^g$, $\delta{\phi}'^e$, and $\delta{\phi}'^f$.  Indeed, one can obtain \eqref{delta g0Flip}, \eqref{delta gQFlip}, \eqref{delta gqqFlip}, \eqref{deltaPhie0Flip}, \eqref{deltaPhieGFlip}, and \eqref{deltaphiFF00} by applying the replacements \eqref{replacements flip} to \eqref{delta g0}, \eqref{delta gQ}, \eqref{delta gqq}, \eqref{deltaPhie0}, \eqref{deltaPhieG}, and \eqref{deltaphiFF0} respectively. In Sec.~\ref{dyonic black hole} and Sec.~\ref{Non-Universal} we will use the replacements \eqref{replacements flip} to avoid repeating tedious calculations for the phase differences after flipping the rectangle.

The operation of flipping the rectangle produces a fringe shift\footnote{Notice that in Ref.~\cite{Mo} the fringe shift is defined as the absolute value of $n$. Here we discard the absolute value symbol because we want to keep the sign.}
\be
n=\frac{\delta\phi_{\rm ge}'-\delta\phi_{\rm ge}}{2\pi}
=n^g_0 +n^g_{Q}+n^g_{q} +n^e_0 +n^e_G +n^f,
\label{fringe shift n}
\ee
where
\bea
n^g_0 &=&
\frac{m l s}{\pi\hbar r_1\sqrt{1-v^2} }\Bigl[
\frac{a r_g}{r_1^2}\bigl(
2\sin\gamma\,\cos\theta_1 +\cos\gamma\,\sin\theta_1
 \bigr)
 \nonumber\\
&&
 +\frac{a r_g^2}{r_1^3}
 \Bigl(\sin(\gamma+\theta_1)
 +\frac{1}{2}\cos\gamma\,\sin\theta_1
 \Bigr)
\Bigr],
\label{ng0KN}
\\
n^g_{Q} &=& 
-\frac{2 m l s}{\pi\hbar r_1\sqrt{1-v^2} } \sin(\gamma+\theta_1)
\frac{a Q^2}{r_1^3},
\\
n^g_{q} &=&
-\frac{q l s}{\pi\hbar r_1} \cos\gamma\, \sin\theta_1
\frac{a r_g Q}{r_1^3},
\\
n^e_0 &=&
-\frac{q l s}{\pi\hbar r_1}
\frac{a Q}{r_1^2}
(2\sin\gamma\cos\theta_1
+\cos\gamma\sin\theta_1),
\\
n^e_G &=&
-\frac{q l s}{2\pi\hbar r_1}
\cos\gamma\, \sin\theta_1\, 
\frac{a r_g Q}{r_1^3},
\label{neGequals}
\\
n^f &=&\frac{l s}{\pi \hbar r_1  }
\Bigl[ 
-\frac{m }{2\sqrt{1-v^2}} \cos\gamma\sin\theta_1 
\frac{a r_g^2}{r_1^3}
 +q\cos\gamma\sin\theta_1\frac{a r_g Q}{r_1^3}
\Bigr].
\label{nff0}
\eea
As we mentioned in the sentences after \eqref{replacements vrQ}, when $Q=0$, the expressions of $S_0$ and $S_3$ are the same as those in Ref.~\cite{Mo}. Hence one may expect $n_0^g$ here coincides with the fringe shift derived in Ref.~\cite{Mo}. However, we find that in \eqref{ng0KN}, there is an additional term in the coefficient of the cubic term, compared with the result in Ref.~\cite{Mo}. 
This is not surprising, because as mentioned in the sentences after \eqref{t varphiNew2}, Ref.~\cite{Mo} claimed that after flipping the interferometer, $t_{AB}$ does not change while $\mathcal{L}$ turns to $-\mathcal{L}$, but we think they should change to $t'_{AB}$ and $\mathcal{L}'$ respectively (see \eqref{t varphiNew} and the sentence after it). Nevertheless, the additional term in \eqref{ng0KN} is not from the difference between $\mathcal{L}'$ and $-\mathcal{L}$, because $\mathcal{L}$ does not appear in $S_0$ (see \eqref{S beta3}) and only appears in the coefficients of the third-order and higher-order terms of $S_3$ (see \eqref{S beta3-2}). Actually, such additional term is caused by the difference between $t'_{AB}$ and $t_{AB}$ (see \eqref{t varphiNew} and \eqref{t varphi}).

Finally, we can rotate the rectangle around the axis AB from the angle $\gamma=0$ to $\gamma=\pi$, producing a fringe shift
\be
N
=\frac{\delta\phi_{\rm total}(\gamma=\pi)-\delta\phi_{\rm total}(\gamma=0) }{2\pi}
=-\frac{\delta\phi_{\rm total}(\gamma=0) }{\pi},
\label{fringe shift NN}
\ee
where $\delta\phi_{\rm total}(\gamma=\pi) =-\delta\phi_{\rm total}(\gamma=0)$ is used in the last step according to \eqref{deltaphige}.

%%%%%%%%%
%%%%%%%%%
\subsection{A brief discussion\label{brief discussion}}
Here we make a brief discussion for the phase differences. Talk about the angle dependences firstly. In \eqref{delta g0} and \eqref{deltaPhie0} we can see that both the first-order terms of $\delta\phi^g$ and of $\delta\phi^e$ are proportional to $\cos\gamma$ but independent of $\theta$. When $\cos\gamma=0$ but $\cos\theta_1\ne 0$, both $\delta\phi^g$ and $\delta\phi^e$ are dominated by the second-order terms, which are proportional to $a$, while $\delta\phi^f$ is dominated by the third-order terms (in such case we have $\delta\phi^g|_{a=0}=\delta\phi^e|_{a=0}=\delta\phi^f|_{a=0}=0$). When $\cos\gamma=\cos\theta_1=0$ (e.g., $\gamma=\theta_1=\pi/2$), the term $\delta\phi^g$, $\delta\phi^e$, and $\delta\phi^f$ vanish. In what follows we compare the magnitude of the gravitational phase difference $\delta\phi^g$ in \eqref{delta phi g} with that of the electromagnetic phase difference $\delta\phi^e$ in \eqref{delta e}. 

Discuss the case $\cos\gamma\ne 0$ firstly. For the gravitational phase difference, obviously $|\delta\phi^g_Q|\ll |\delta\phi^g_0|$, and (SI units)
\be
 \biggl|\frac{\delta\phi^g_q}{\delta\phi^g_0} \biggr|
\approx 
\biggl| 
-\frac{qQ }{4\pi\varepsilon_0  r_1}\Bigl/
\frac{m c^2}{\sqrt{1-(v/c)^2} }
\biggr|.
\label{qM}
\ee
It is just the ratio between the Coulomb potential energy and the particle's relativistic energy. Though $\delta\phi^g_q$ is a correction to the pure gravitational phase difference by electromagnetic interaction, its magnitude can be larger than that of the latter, depending on the factor $q/m$. In short, both $|\delta\phi^g_q|\ll |\delta\phi^g_0|$ and $|\delta\phi^g_q| \gg |\delta\phi^g_0|$ are possible. When $|\delta\phi^g_q| \ll |\delta\phi^g_0|$, we have
\be
 \biggl|\frac{\delta\phi^e}{\delta\phi^g} \biggr|
\approx
 \biggl| \frac{\delta\phi^e_0}{\delta\phi^g_0} \biggr|
\approx
\biggl|-\frac{qQ}{ 4\pi\varepsilon_0}\Bigl/
\frac{GMm}{ \sqrt{1-(v/c)^2} } \biggr|
=\biggl|-\frac{1}{ 4\pi\varepsilon_0 G}\frac{q}{m}\frac{Q}{M}\sqrt{1-\frac{v^2}{c^2}} \biggr|
 \ll
 \frac{2r_1}{r_g},
 \quad
\text{with}
\quad
\frac{2r_1}{r_g}\gg 1.
\label{ratioge}
\ee 
In the non-relativistic limit, it is the ratio between the Coulomb potential energy and the Newtonian potential energy. Then we make some numerical evaluations, taking the values of physical constants from Ref.~\cite{allen2000allen}. As an example, consider non-relativistic electrons that $|q/m|\approx 1.76\times 10^{11} {\rm C/kg}$. To let $|\delta\phi^e/\delta\phi^g|\approx 1$, the charge-to-mass ratio of the gravitational source should satisfy $ |Q/M|\approx 4.22\times 10^{-32} {\rm C/kg}$. For earth as the gravitational source, such ratio requires $ |Q|\approx  2.51\times 10^{-7}  {\rm C}$, which is far less than the extreme charge $ |Q|=\sqrt{4\pi\varepsilon_0 G}M\approx 5.15\times 10^{14}  {\rm C}$. Instead, when $|\delta\phi^g_q|\gg |\delta\phi^g_0|$, we have
\be
\biggl|\frac{\delta\phi^e}{\delta\phi^g}\biggr|
\approx
\biggl|\frac{\delta\phi^e_0}{\delta\phi^g_q}\biggr|
\approx
\frac{r_1 c^2}{GM}
=\frac{2 r_1}{r_g},
\quad
\text{with}
\quad
1\ll \frac{2 r_1}{r_g} \ll \biggl|-\frac{1}{ 4\pi\varepsilon_0 G}\frac{q}{m}\frac{Q}{M}\sqrt{1-\frac{v^2}{c^2}}\biggr|.
\label{ratioge2}
\ee
For example, letting $r_1$ equals to earth's equatorial radius and $M$ equals to earth's mass, we get $|\delta\phi^e/\delta\phi^g|\approx 1.44\times 10^9$. 

Now we consider the case $\cos\gamma=0$ but $\cos\theta_1\ne 0$. In this case, $\delta\phi^g$ is dominated by the second term of \eqref{delta g0}. While for $\delta\phi^e$, we assume $a/r_1\ll v$ so that it is dominated by the second term of \eqref{deltaPhie0}. Then we get (SI units)
\be
\biggl|\frac{\delta\phi^e}{\delta\phi^g}\biggr|
\approx
\biggl|-\frac{1}{ 8\pi\varepsilon_0 G}\frac{q}{m}\frac{Q}{M}\sqrt{1-\frac{v^2}{c^2}}\biggr|,
\quad
\text{with}
\quad
\frac{a}{r_1}\ll \frac{v}{c}.
\ee

%%%%%%%%%%%
\section{Extension to dyonic Kerr-Newman spacetime\label{dyonic black hole}}

\subsection{Massive particles\label{massive particles}}

In this section we extend the phase differences \eqref{delta phi g} and \eqref{delta e} to the case of a dyonic Kerr-Newman black hole which has an electric charge $Q$ and a magnetic charge $P$. For the metric of such black hole, we only need to replace the expression of $R_0$ in \eqref{rgrho} by~\cite{Hackmann:2013pva}
\be
R_0=2M-\frac{Q^2+P^2}{r},
\label{R0}
\ee
and \eqref{dmudxmu} should be changed to~\cite{Kasuya:1981ef}
\be
A_\mu\dd x^\mu
=\frac{Qr}{\rho^2}\bigl(\dd t -a\sin^2\theta\dd\varphi\bigr)
-P\Bigl[ \frac{a}{\rho^2}\cos\theta\, \dd t +\Bigl( 1-\frac{r^2+a^2}{\rho^2}\cos\theta\Bigr)\dd\varphi
\Bigr].
\label{dmudxmu2}
\ee
Therefore, for the tetrad and the gravitational gauge potential, we only need to make the replacement $Q^2\rightarrow Q^2+P^2$ in the corresponding quantities of Kerr-Newman spacetime (see Sec.~\ref{KN}). As for the function $S_\beta$ (see \eqref{S beta2}), we also need such replacement, and need to replace the new metric, the new electromagnetic potential, and the new four-momentum into it, where the new expression of $P^1$ is~\cite{Hackmann:2013pva}
\be
P^1
=
\pm \frac{1}{\rho^2} \sqrt{\bigl[(r^2 \! + \! a^2)\mathcal{E} -a\mathcal{L} -q Qr\bigr]^2  \! - \! m^2( r^2  \! + \! K)(r^2  \! - \! 2Mr  \! + \! a^2  \! + \! Q^2  \! + \! P^2) }.
\label{P1 new}
\ee
Then repeating the calculations in Sec.~\ref{interference}, we find the gravitational phase difference for the interference experiment
\be
\delta\phi^{\rm g}
\approx
\delta\phi^{\rm g}_0
+\delta\phi^{\rm g}_{QP}
+\delta\phi^{\rm g}_q,
\label{g phase difference}
\ee
where
\bea 
\delta\phi^{\rm g}_0
&=&\frac{m l s}{ \hbar r_1 \sqrt{1-v^2}} \Bigl\{
\frac{1}{v}
\cos\gamma\,\Bigl(\frac{r_g}{2r_1}+\frac{r_g^2}{2r_1^2}\Bigr)
-\frac{a r_g}{r_1^2}\Bigl(2\cos\theta_1\, \sin\gamma+\cos\gamma\,\sin\theta_1
\Bigr)
\nonumber\\
&&
+\frac{1}{v}\cos\gamma\,\frac{r_g^3}{2r_1^3}
-\frac{a r_g^2}{r_1^3}
\Bigl(
  \cos\theta_1\,\sin\gamma
 +\frac{3}{2}\cos\gamma\,\sin\theta_1
 \Bigr)
\nonumber\\
&&
+\frac{a^2 r_g}{r_1^3}
\Bigl[ \frac{1}{v}\Bigl(\frac{1}{4}\cos\gamma\,-\frac{7}{4}\cos\gamma\,\cos^2\theta_1
+\frac{1}{2}\sin(2\theta_1)
\sin\gamma
\Bigr)
\nonumber\\
&&
+v\sin\theta_1\,\Bigl(
\frac{3}{2} \cos\gamma\, \sin\theta_1
+\sin\gamma\, \cos\theta_1
\Bigr)
\Bigr]
\Bigr\},
\label{delta 0 dyon}
\\
\delta\phi^{\rm g}_{QP}
&=&\frac{m l s}{ \hbar r_1 \sqrt{1-v^2}} \Bigl[
-\frac{1}{v}\cos\gamma\, \Bigl(
\frac{Q^2+P^2}{r_1^2} +\frac{3r_g (Q^2+P^2)}{2 r_1^3}
\Bigr)
\nonumber\\
&&
+2\sin(\theta_1+\gamma)\frac{a (Q^2+P^2)}{r_1^3}
\Bigr],
\label{delta Q dyon}
\\
\delta\phi^{\rm g}_q
&=&\frac{q l s}{ \hbar r_1} \Bigl[
\frac{1}{v}\cos\gamma\, \Bigl(
-\frac{r_g Q}{2r_1^2} 
-\frac{3r_g^2 Q}{8 r_1^3}
+\frac{Q^3}{2r_1^3} 
\Bigr)
+\cos\gamma\,\sin\theta_1\, \frac{a r_g Q}{r_1^3}
\nonumber\\
&&+\frac{\cos\gamma}{2v} 
\frac{P^2 Q}{r_1^3}
+
\frac{1}{4v}\Bigl(\cos(\gamma-\theta_1)+3\cos(\gamma+\theta_1)\Bigr)\frac{P a r_g}{r_1^3}
\Bigr],
\label{delta q0 dyon}
\eea
and the electromagnetic phase difference
\bea
\delta\phi^{\rm e}
&\approx&
\frac{q l s}{ \hbar r_1} \Bigl\{
-\frac{1}{v}\frac{Q}{r_1} \cos\gamma
+\frac{a Q}{r_1^2}
\bigl(2\sin\gamma\,\cos\theta_1 +\cos\gamma\,\sin\theta_1\bigr)
\nonumber\\
&&
+\frac{1}{v}\frac{a^2 Q}{4 r_1^3}
\Bigl[
\cos\gamma\, \bigl(5+7\cos(2\theta_1)\bigr)
-4\sin\gamma\,\sin(2\theta_1)
\Bigr]
\nonumber\\
&&
+\frac{a r_g Q}{2r_1^3}\cos\gamma\,\sin\theta_1
+\frac{P}{r_1}\sin\gamma
+\frac{1}{v}
\frac{P a}{r_1^2}\bigl(2\cos\gamma\, \cos\theta_1 
- \sin\gamma\, \sin\theta_1 \bigr)
\nonumber\\
&&
-\frac{P a^2}{8r_1^3}
\Bigl( 10 \sin\gamma \! + \! 3\sin(\gamma \! - \! 2 \theta_1) 
 \! + \! 11\sin(\gamma 
 \! + \! 2\theta_1)
\Bigr)
 \! - \! \frac{1}{v}\frac{P a r_g}{2r_1^3}\sin\gamma\, \sin\theta_1
\Bigr\},
\label{delta e dyon}
\eea
and the free phase difference
\bea 
\delta\phi^{\rm f}
&=&\frac{l s}{ \hbar r_1}
\Bigl\{
\frac{1}{v}
\Bigl[
 -\frac{m\cos\gamma}{4\sqrt{1-v^2}}\frac{r_g^2}{r_1^2}
 +\frac{q\cos\gamma}{2} \frac{r_g Q}{r_1^2}
\Bigr]
+\frac{m }{16\sqrt{1-v^2}}
\Bigl[12\cos\gamma \frac{r_g (Q^2+P^2)}{ v r_1^3}
\nonumber\\
&&
- 5\cos\gamma \frac{r_g^3}{ v r_1^3}
+ 8\cos\gamma \sin\theta_1  \frac{a r_g^2}{r_1^3}
- 8v \sin\theta_1\,(2\cos\theta_1\sin\gamma +3\sin\theta_1\cos\gamma)
\frac{a^2 r_g}{r_1^3}
\Bigr]
+\frac{q}{8 v}\Bigl[
-4 \cos\gamma\frac{Q^3}{r_1^3}
\nonumber\\
&&
-4 \cos\gamma\frac{Q P^2}{r_1^3}
+ 3\cos\gamma\frac{Q r_g^2}{r_1^3}
- 8 v \cos\gamma\sin\theta_1\frac{a r_g Q}{r_1^3}
+ 4(\sin\gamma\sin\theta_1-2\cos\gamma\cos\theta_1)\frac{a r_g P}{r_1^3}
\Bigr]
\Bigr\}.
\label{deltaphiFF2}
\eea
Finally, the total quantum phase difference is $\delta\phi^{\rm total}=\delta\phi^{\rm g}+\delta\phi^{\rm e}+\delta\phi^{\rm f}$. 

The phase differences $\delta{\phi}'^{\rm g}$, $\delta{\phi}'^{\rm e}$, and $\delta{\phi}'^{\rm f}$, corresponding to flipping the rectangle (with its height as the axis), can be found by implementing the replacements \eqref{replacements flip} in \eqref{g phase difference}, \eqref{delta e dyon}, and \eqref{deltaphiFF2} respectively. Finally, we get the fringe shift of this process
\be
n=\frac{\delta{\phi}'^{\rm total}-\delta\phi^{\rm total}}{2\pi}
= n^{\rm g}_0 +n^{\rm g}_{QP}+n^{\rm g}_{q} +n^{\rm e}_0 +n^{\rm e}_G +n^{\rm f},
\label{fringe shift nDyon}
\ee
where $\delta{\phi}'^{\rm total}=\delta{\phi}'^{\rm g}+\delta{\phi}'^{\rm e}+\delta{\phi}'^{\rm f}$ and
\bea
n^{\rm g}_0 &=&
\frac{m l s}{\pi\hbar r_1\sqrt{1-v^2} }\Bigl[
\frac{a r_g}{r_1^2}\bigl(
2\sin\gamma\,\cos\theta_1 +\cos\gamma\,\sin\theta_1
 \bigr)
 \nonumber\\
&&
+\frac{a r_g^2}{r_1^3}\Bigl(\sin(\gamma+\theta_1)
+\frac{1}{2}\cos\gamma\,\sin\theta_1
\Bigr)
\Bigr],
\\
n^{\rm g}_{QP} &=& 
-\frac{2 m l s}{\pi\hbar r_1\sqrt{1-v^2} } \sin(\gamma+\theta_1)
\frac{a (Q^2+P^2)}{r_1^3},
\\
n^{\rm g}_{q} &=&
-\frac{q l s}{\pi\hbar r_1} \cos\gamma\, \sin\theta_1
\frac{a r_g Q}{r_1^3},
\\
n^{\rm e}_0 &=&
\frac{q l s}{\pi\hbar r_1}
\Bigl[
-\frac{P}{r_1}\sin\gamma
-\frac{a Q}{r_1^2}
(2\sin\gamma\cos\theta_1
 \! + \!\cos\gamma\sin\theta_1)
 \nonumber\\
&&
+\frac{P a^2}{8 r_1^2}\Bigl( 10 \sin\gamma+ 3\sin(\gamma \! -\! 2 \theta_1) 
 \! + \! 11\sin(\gamma 
\! +\! 2\theta_1)
\Bigr)
 \!\Bigr],
\\
n^{\rm e}_G &=&
-\frac{q l s}{2\pi\hbar r_1}
\cos\gamma\, \sin\theta_1\, 
\frac{a r_g Q}{r_1^3},
\\
n^{\rm f} &=&
\frac{l s}{\pi \hbar r_1  }
\Bigl[ 
-\frac{m }{2\sqrt{1-v^2}} \cos\gamma\sin\theta_1
\frac{a r_g^2}{r_1^3}
 +q\cos\gamma\sin\theta_1\frac{a r_g Q}{r_1^3}
\Bigr].
\label{nff2}
\eea
Besides, the fringe shift of rotating the rectangle around the axis AB from $\gamma=0$ to $\gamma=\pi$, is
\be
N=-\frac{\delta\phi^{\rm total}(\gamma=0) }{\pi}.
\label{N dyon}
\ee

We would like to mention that the potential \eqref{dmudxmu2} is singular at $\theta=\pi$. (One can check this statement in the flat spacetime (e.g., let $G=0$ in the metric), by finding out the $\varphi$ component of the vector potential.) Nonetheless, it does not affect the above results, because we have adopted the approximation $s\ll \sqrt{\Gamma_{33}}$ (recall the sentences after \eqref{v components}) such that the paths on the rectangle never touch the singularities. Anyway, there is a method to resolve the above difficulty, i.e., introducing the Dirac string for a magnetic monopole and redefining the electromagnetic field as~\cite{Dirac:1931kp,Dirac:1948um}
\be
F_{\mu\nu}=\partial_\mu A_\nu-\partial_\nu A_\mu +S_{\mu\nu},
\label{field string}
\ee
where $S_{\mu\nu}$ is a quantity vanishing everywhere except on the sheets which are traced out by Dirac strings. In our case, the Dirac string is the negative semi-axis $z$ ($\theta=\pi$). One may worry whether such string affects the measurement if the paths enclose it. For example, consider two paths (``$l_1$'' and ``$l_2$'') which together enclose the negative semi-axis $z$. The gravitational phase difference and the electromagnetic phase difference between them are respectively
\bea
\delta\phi^g&=& \frac{1}{\hbar} \Bigl(\int_{l_2} S_\mu \dd x^\mu-\int_{l_1} S_\mu \dd x^\mu
\Bigr),
\label{gravitational phase shift}
\\
\delta\phi^e&=&\frac{q}{\hbar}\Bigl(\int_{l_2} A_\mu \dd x^\mu-\int_{l_1} A_\mu \dd x^\mu\Bigr).
\label{electromagnetic phase shift}
\eea
We then let the loop shrinks to a point. Obviously, only the terms of $\dd\varphi$ are left, i.e.
\bea
\delta\phi^g|_{\theta\rightarrow \pi}
&=& \frac{1}{\hbar} \Bigl(\int_{l_2} S_\varphi \dd \varphi-\int_{l_1} S_\varphi \dd \varphi\Bigr)\Bigr|_{\theta\rightarrow\pi},
\label{S In}
\\
\delta\phi^e|_{\theta\rightarrow \pi}
&=&\frac{q}{\hbar}\Bigl(\int_{l_2} A_\varphi \dd \varphi-\int_{l_1} A_\varphi \dd \varphi\Bigr)\Bigr|_{\theta\rightarrow\pi}.
\label{A In}
\eea
On the other hand, we have $S_\varphi|_{\theta\rightarrow\pi}=0$ and $A_\varphi|_{\theta\rightarrow\pi}=-2P$ (see \eqref{S beta2} and \eqref{dmudxmu2}). Hence \eqref{S In} and \eqref{A In} become
\bea
\delta\phi^g|_{\theta\rightarrow \pi}&=&0,
\label{gravitational phase shift2}
\\
\delta\phi^e|_{\theta\rightarrow \pi}&=&-\frac{4\pi q P}{\hbar}.
\label{electromagnetic phase shift2}
\eea
To ensure the phase difference \eqref{electromagnetic phase shift2} has no physical effect, it should satisfy $\delta\phi^e|_{\theta\rightarrow\pi}=-2\pi j$ (where $j$ is an integer), which implies the Dirac quantization condition \cite{Dirac:1931kp}
\be
\frac{2 q P}{\hbar} =j.
\label{Dirac quantization}
\ee
In summary, the Dirac string cannot be detected through the gravitational phase difference, and if the Dirac quantization condition holds, such string is also unobservable by the electromagnetic phase difference. 

Besides, one can use the potential \cite{Semiz:1990fm}
\be
A_\mu\dd x^\mu
=\frac{Qr}{\rho^2}\bigl(\dd t -a\sin^2\theta\dd\varphi\bigr)
-P\Bigl[ \frac{a}{\rho^2}\cos\theta\, \dd t +\Bigl(\pm 1-\frac{r^2+a^2}{\rho^2}\cos\theta\Bigr)\dd\varphi
\Bigr],
\label{dmudxmu3}
\ee
to avoid introducing Dirac strings, where the upper sign  is used for the region $0\le \theta \le\pi/2$, while the lower sign is used for $\pi/2< \theta\le \pi$. Evidently in the region $0<\theta<\pi$, both the potentials in \eqref{dmudxmu3} can describe the electromagnetic field. In such region, requiring the gauge transformation (of the wave function) between them be single-valued implies the Dirac quantization condition \cite{Wu:1975es}. We would like to mention that no matter which the potential in \eqref{dmudxmu3} (e.g., the upper sign, corresponding to \eqref{dmudxmu2}) is used in the region $0<\theta<\pi$, the observations are the same. Let us prove this statement firstly for the electromagnetic phase difference. Denote the two potentials in \eqref{dmudxmu3} by $A^{+}_\mu$ and $A^{-}_\mu$ respectively. Obviously we have $A^{-}_\mu-A^{+}_\mu=2P$. Then we get
\be
(\delta\phi^e)^{-}-(\delta\phi^e)^{+}
=\frac{q}{\hbar}\Bigl(\int_{l_2} 2P \dd \varphi-\int_{l_1} 2P \dd \varphi\Bigr)
=\frac{q}{\hbar}\oint 2P \dd \varphi.
\label{minus plus}
\ee
The rectangle in FIG.~\ref{experimentKN} does not enclose the $z$ axis, hence the loop integral in \eqref{minus plus} equals zero. Even if the loop encloses the $z$ axis, the observations are not different between $A^{-}_\mu$ and $A^{+}_\mu$, because the Dirac quantization condition \eqref{Dirac quantization} together with \eqref{minus plus} ensures that $(\delta\phi^e)^{-}-(\delta\phi^e)^{+}=2n\pi$. As for the gravitational phase difference, we rewrite it as (see \eqref{phi_g})
\be
\delta\phi^g=\int_{l_2}g_{\nu\mu} P^\mu {B^\nu}_\beta\dd x^\beta
-\int_{l_1}g_{\nu\mu} P^\mu {B^\nu}_\beta\dd x^\beta.
\label{deltatphig}
\ee
Evidently, the metric and the quantity ${B^\nu}_\beta$ are totally determined by the gravitational field. On the other hand, the equation of motion \eqref{EOMM} relates to the metric and the electromagnetic field tensor, while the gauge transformation of the potential does not change the electromagnetic field. Hence $P^\mu$ and the paths in \eqref{deltatphig} are independent of the electromagnetic gauge we choose. Therefore, it makes no difference which electromagnetic gauge we use for the gravitational phase difference.

\subsection{Electromagnetic gauge invariance}
In this subsection we discuss the gauge invariance for the electromagnetic phase shift
\be
\delta\phi^{\rm EM}=\frac{q}{\hbar} \oint A_\mu \dd x^\mu.
\label{delta phiEM}
\ee
A gauge transformation should keep the electromagnetic phase factor $\Phi=\exp(i\delta\phi^{\rm EM})$ invariant. Such requirement is also derived from the single-valued property of the wave function~\cite{Wu:1975es}. For example, consider a transformation
\be
\psi' =e^{i K} \psi,
\qquad A'_\mu =A_\mu +\frac{\hbar}{q}\partial_\mu K.
\label{a transformation}
\ee
Then one gets
\be
\delta\phi'^{\rm EM}-\delta\phi^{\rm EM}
=\frac{q}{\hbar} \oint (A'_\mu-A_\mu) \dd x^\mu
=\oint \partial_\mu K \dd x^\mu
=\Delta K. 
\ee
To keep the wave function $\psi$ single-valued, one should have $\Delta K=2\pi n$, namely
\be
\delta\phi'^{\rm EM}-\delta\phi^{\rm EM}=2\pi n,
\label{gauge requirement}
\ee
where $n$ is an integer. Obviously, if $K$ is a single-valued function, we have $n=0$ and the gauge invariance holds. The multiple-valued case is more complicated.

The question we want to ask is: Which kinds of transformations of the electromagnetic potential in dyonic Kerr-Newman spacetime satisfy the requirement \eqref{gauge requirement}? Obviously, a transformation $A'_\mu =A_\mu +(\hbar/q)\partial_\mu K$ with $K$ a continuous and differentiable function (it is of course single-valued) satisfies this requirement. How about other kinds of transformations? To be honest, we do not know the exact solution. However, we can answer this question in a very simplified case, i.e., a static magnetic monopole, corresponding to a non-rotating magnetic charged black hole. 

As we mentioned previously, there are two ways to describe the electromagnetic field in the presence of magnetic charges. The first way is to modify the definition by a string tensor, as \eqref{field string} shows, while the second way is to use a double patches potential, as \eqref{dmudxmu3} shows. We discuss the string scenario firstly. For simplicity, consider a static magnetic monopole with a magnetic charge $P$ and choose $A_0=0$. The solution of the gauge potential is~\cite{Shnir:2005vvi}
\be
\vec{A}(\vec{r})
=P\int_{L} \frac{(\vec{r}-\vec{r}')\times \dd \vec{r}'}{|\vec{r}-\vec{r}'|^3},
\ee 
where the integral is along the Dirac string $L$ from the monopole to infinity. Now deform the string to be $L'$, so we have
\be
\vec{A}'(\vec{r})-\vec{A}(\vec{r})
=P\int_{L'} \frac{(\vec{r}-\vec{r}')\times \dd \vec{r}'}{|\vec{r}-\vec{r}'|^3}-P\int_{L} \frac{(\vec{r}-\vec{r}')\times \dd \vec{r}'}{|\vec{r}-\vec{r}'|^3}.
\label{A prime minus}
\ee
In Ref.~\cite{Shnir:2005vvi}, the right hand side of \eqref{A prime minus} is proved to proportional to the divergence of a solid angle, i.e.
\be
\vec{A}'(\vec{r})-\vec{A}(\vec{r})
=P \nabla \Omega (\vec{r}),
\label{relation Omega}
\ee
where $\Omega (\vec{r})$ is the solid angle of the surface spanned by the strings $L$ and $L'$ viewed from $\vec{r}$. Obviously, the function $K$ is not single-valued at such surface, considering $\partial_\mu K=(q/\hbar)g (A'_\mu-A_\mu)$ and \eqref{relation Omega}. Indeed, when an object crosses the surface, the solid angle $\Omega (\vec{r})$ seen from it jumps by $4\pi$. Then we get
\be
\delta\phi'^{\rm EM}-\delta\phi^{\rm EM}
=\frac{q P}{\hbar} \oint \nabla \Omega(\vec{r})\, \cdot \dd\vec{r}.
\label{loop integral}
\ee
If the loop of \eqref{loop integral} does not cross the surface, the solid angle variates continuously such that $\oint \nabla \Omega(\vec{r})\, \cdot \dd\vec{r}=0$. In this case the requirement \eqref{gauge requirement} surely holds. If the loop crosses the surface, every time it crosses the latter, the solid angle jumps by $4\pi$, such that $\oint \nabla \Omega(\vec{r})\, \cdot \dd\vec{r}=4\pi k$, where $k$ is an integer. Hence, in this case we have
\be
\delta\phi'^{\rm EM}-\delta\phi^{\rm EM}
=\frac{ 4\pi k\, q P}{\hbar}
=2\pi j k,
\label{cross}
\ee
where we used the Dirac quantization condition \eqref{Dirac quantization} in the last step. In other words, the Dirac quantization condition ensures that the requirement \eqref{gauge requirement} is also satisfied in this case. In conclusion, the deformations of the Dirac string for a magnetic monopole are also gauge transformations.

\begin{figure}[h]
\centering
\includegraphics[width=0.8\textwidth]{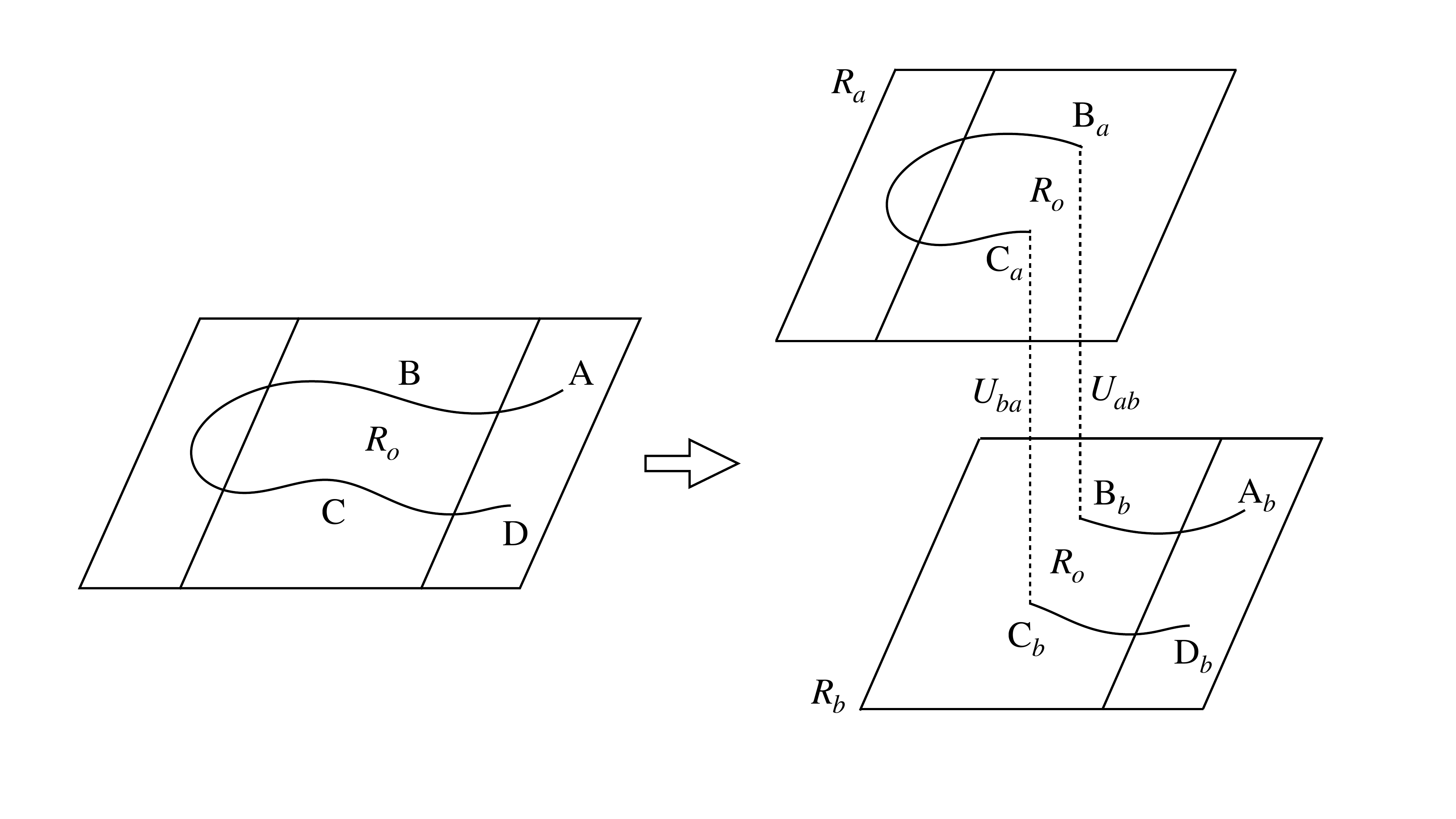}
\caption{The procedure to define a phase factor. This figure is drawn based on the second figure of Ref.~\cite{Wu:1975es}.}
\label{Phase factor figure}
\end{figure}

Now we talk about the double patches potential way. In such formulation, a difficulty is how to define an electromagnetic phase when the path crosses the boundary between two regions described by two different patches of the potential, compared with the previous definition
\be
\phi^{\rm EM}=\frac{q}{\hbar} \int A_\mu \dd x^\mu.
\label{Phase phiEM}
\ee
Wu and Yang found a way to resolve such difficulty~\cite{Wu:1975es}, but slightly differently, what they dealt with was how to define the phase factor $\Phi=\exp(i\phi^{\rm EM})$ for the above situation. In their work, the method is to ``glue'' the endpoints of different segments of the path by a transformation $U$ at the boundary. Let us make a brief introduction of their way. As for a static magnetic monopole, adopt two patches for the gauge potential:
\be
\vec{A}
=\begin{cases}
\vec{A}_a=  \vec{e}_\varphi\frac{P(1-\cos\theta)}{r\sin\theta}, 
&\text{at the region $R_a$:}\quad 0\le \theta< \frac{\pi}{2}+\varepsilon,
\\
\vec{A}_b= -\vec{e}_\varphi\frac{P(1+\cos\theta)}{r\sin\theta},
&\text{at the region $R_b$:}\quad \frac{\pi}{2}-\varepsilon< \theta \le \pi.
\end{cases}
\label{WuYangWay}
\ee 
There is an overlapping region $R_o$: $\,\pi/2-\varepsilon< \theta<\pi/2+\varepsilon$ in \eqref{WuYangWay}. Obviously, the expression \eqref{dmudxmu3} reduces to \eqref{WuYangWay} when $Q=a=0$, except that the former has no overlap. Outside the region $R_o$, the phase factor is simply defined as
\be
\Phi_{CB} =\exp\Bigl(\frac{q}{\hbar} \int_B^C  A_\mu \dd x^\mu \Bigr).
\label{definition Phi Tr}
\ee
Inside such region, both $\vec{A}_a$ and $\vec{A}_b$ are applicable to this definition, and the Dirac quantization condition ensures that these two patches are physically indistinguishable. Especially, in such region one has
\be
\Phi_{C_b B_b} =U_{ba}(C)\, \Phi_{C_a B_a} U_{ab}(B),
\label{relation ab}
\ee
where $U_{ba}=U^{-1}_{ab}=\exp(iK)$ with $\psi_b =U_{ba}\psi_a$. One can check this relation by \eqref{a transformation}. The relation \eqref{relation ab} means that in the overlapping region, one can use $U$ at the two endpoints of $\Phi_a$ to transform $\Phi_a$ to $\Phi_b$. For example, as for three segments $\Phi_{D_b C_b}$, $\Phi_{C_a B_a}$, and  $\Phi_{B_b A_b}$ in the region $R_o$, if we want to ``glue'' them together, we can firstly transform $\Phi_{C_a B_a}$ to $\Phi_{C_b B_b}$ using the relation \eqref{relation ab}, then put them together:
\be
\Phi_{DCBA}= \Phi_{D_b C_b}\, 
 U_{ba}(C)\, \Phi_{C_a B_a}\,
  U_{ab}(B) \,
 \Phi_{B_b A_b}.
 \label{glue segments}
\ee
In other words, we can use the transformation $U$ to ``glue'' the segments belonging to $\vec{A}_a$ and $\vec{A}_b$ respectively. The procedure \eqref{glue segments} was used to define the phase factor $\Phi_{DCBA}$ in Ref.~\cite{Wu:1975es}, even these segments are not only in $R_o$. We show such procedure in an example in FIG.~\ref{Phase factor figure}, where a path crisscrosses in and out of the overlap $R_o$. Cut off the path in $R_o$. It is divided into segments which cannot simultaneously touch the regions $R_a$ and $R_b$. Then apply the definition \eqref{definition Phi Tr} to every segment. For segments which touch $R_a$, use $\vec{A}_a$ as the gauge potential, while for those touching $R_b$, use $\vec{A}_b$. Finally, use $U$ to ``glue'' these segments together. By such way, the phase factor is globally well defined.

Correspondingly, we can use such way to define the electromagnetic phase of the path $ABCD$ in FIG.~\ref{Phase factor figure}:
\be
\phi_{DCBA}= \phi_{D_b C_b}
+E_{ba}(C)
+\phi_{C_a B_a} 
 +E_{ab}(B) 
 +\phi_{B_b A_b},
 \label{glue segments phase}
\ee
where every segment is defined by \eqref{Phase phiEM} (for example,\footnote{Note that $(q/\hbar)\int_{C}^{D} A_\mu \dd x^\mu$ is defined as $\phi_{C D}$ in the other places of this paper, but here we would like to follow the convention in Ref.~\cite{Wu:1975es}.} 
$\phi_{D_b C_b} =(q/\hbar)\int_{C_b}^{D_b} A^b_\mu \dd x^\mu$), and $E_{ba}(x)=-E_{ab}(x)=K(x)$. According to Ref.~\cite{Wu:1975es}, by the definition in \eqref{glue segments}, the phase factor around a loop is $\exp(i q f /\hbar)$, where $f$ is the magnetic flux through the surface of such loop. In other words, the phase shift around a loop is $q f /\hbar$. Therefore, in the definition \eqref{glue segments phase}, any transformation which does not change the electromagnetic field would keep the phase shift invariant. 

By the way, Ref.~\cite{Wu:1975es} pointed out that the gauge field theories (including electromagnetism) could be described by the fiber bundle theory in mathematics. In other words, the terminology of electromagnetism can be translated to that of the fiber bundle theory~\cite{Wu:1975es}, e.g., the gauge potential is called ``connection on a principal fiber bundle'', the phase factor is called ``parallel displacement'', $U$ is called ``transition function'', and the field strength is called ``curvature''.

Finally, we emphasize that the above discussions of the gauge invariance in the string scenario and the double patches potential scenario are only applied to a static magnetic monopole, which is just a very reduced version of the dyonic Kerr-Newman black hole. Whether the gauge invariance also holds for the latter remains unresolved for the author. To deal with the latter, we need to generalize these analyses to a general electromagnetic potential $A_\mu$. The string scenario would be very complicated, because we need to consider the deformations of a string in spacetime and \eqref{relation Omega} needs to be extended to a four-dimensional form. As for the double patches potential recipe, we also need to examine the definitions \eqref{glue segments} and \eqref{glue segments phase} in general cases.

\subsection{Massless particles\label{massless particles}}
In the above contents we always considered charged massive particles. Nevertheless, the calculations of the phase differences in Sec.~\ref{interference} are also applicable for uncharged massless particles. As for the latter, the quantities \eqref{mathcalE} and \eqref{L given} should change to
\bea
&&\mathcal{E}=
\sqrt{g_{00}}\, \hbar \omega,
\label{mathcalEMassless}
\\
&&\mathcal{L}=\frac{1}{g_{00}}\bigl(
-g_{03} +v^\varphi \Gamma_{33}  \sqrt{g_{00}}
\bigr) \mathcal{E},
\label{L givenMassless}
\eea
and we need to let $v=1$ in \eqref{t varphi}. (combining \eqref{vL} with \eqref{v components} and $g_{\mu\nu}\dd x^\mu \dd x^\nu=0$ for massless particles, one can prove $v=1$). By the way, though \eqref{t varphi} is derived for massive particles, it also holds for massless particles ($v=1$), following a derivation same as the one in Ref.~\cite{Mo}. The quantity $\omega$ in \eqref{mathcalEMassless} is the frequency measured by a stationary observer who has a four-velocity parallel to the asymptotically time-translation Killing vector in a stationary spacetime~\cite{Carroll:2004st}. The equation \eqref{mathcalEMassless} follows from the relation $\hbar \omega =\mathcal{E}(K^\mu K_\mu)^{-1/2}$ (see Ref.~\cite{Carroll:2004st}), where $K^\mu$ is the asymptotically time-translation Killing vector. Using $K^\mu=(1,0,0,0)$, we get $K^\mu K_\mu=g_{00}$ such that \eqref{mathcalEMassless} holds. Correspondingly, the gravitational phase difference for uncharged massless particles in dyonic Kerr-Newman spacetime is
\be
\delta\phi^{\rm g}
\approx
\delta\phi^{\rm g}_0
+\delta\phi^{\rm g}_{QP},
\label{g phase difference Massless}
\ee
where
\bea 
\delta\phi^{\rm g}_0
&=&\frac{ l s \omega}{ r_1 } \Bigl\{
\cos\gamma\,\Bigl(\frac{r_g}{2r_1}+\frac{r_g^2}{2r_1^2}\Bigr)
-\frac{a r_g}{r_1^2}( 2\sin\gamma\,\cos\theta_1 +\cos\gamma\,\sin\theta_1 )
\nonumber\\
&&
+\cos\gamma\,\frac{r_g^3}{2r_1^3}
-\frac{a r_g^2}{r_1^3}
\Bigl(
  \cos\theta_1\,\sin\gamma
 +\frac{3}{2}\cos\gamma\,\sin\theta_1
 \Bigr)
\nonumber\\
&&
+\frac{a^2 r_g}{r_1^3}
\Bigl[ -\frac{1}{16}\Bigl(
2 \cos\gamma+ \cos(\gamma- 2 \theta_1) + 
 21 \cos(\gamma + 2 \theta_1)
\Bigr)
+\frac{1}{2}\sin\theta_1 \sin(\theta_1-\gamma)
\Bigr]
\Bigr\},
\label{delta 0 dyonMassless}
\\
\delta\phi^{\rm g}_{QP}
&=&\frac{ l s \omega }{ r_1 } \Bigl[
-\cos\gamma\,\Bigl(
\frac{Q^2 \! + \! P^2}{r_1^2}  \! + \! \frac{3r_g (Q^2 \! + \! P^2)}{2 r_1^3}
\Bigr)
+2\sin(\theta_1 \! + \! \gamma)\frac{ a (Q^2 \! + \! P^2)}{r_1^3}
\Bigr],
\label{delta Q dyonMassless}
\eea
and the free phase difference is
\bea
\delta\phi^{\rm f}
&=&\frac{l s\omega }{ r_1}
\Bigl\{
 -\frac{\cos\gamma}{4}\frac{r_g^2}{r_1^2}
+\frac{1}{16}
\Bigl[12\cos\gamma \frac{r_g (Q^2+P^2)}{ r_1^3}
- 5\cos\gamma \frac{r_g^3}{ r_1^3}
+ 8\cos\gamma \sin\theta_1  \frac{a r_g^2}{r_1^3}
\nonumber\\
&&
- 8 \sin\theta_1\,(2\cos\theta_1\sin\gamma +3\sin\theta_1\cos\gamma)
\frac{a^2 r_g}{r_1^3}
\Bigr]
\Bigr\}.
\label{deltaphiFFML2}
\eea
Flipping the rectangle (around its height) yields new gravitational phase difference $\delta{\phi}'^{\rm g}$ and new free phase difference $\delta{\phi}'^{\rm f}$ (which are obtained by applying the replacements \eqref{replacements flip} to \eqref{g phase difference Massless} and \eqref{deltaphiFFML2}). Such operation causes a fringe shift
\be
n=\frac{\delta{\phi}'^{\rm g}+\delta{\phi}'^{\rm f}-(\delta\phi^{\rm g}+\delta{\phi}^{\rm f})}{2\pi} 
=n^{\rm g}_0 +n^{\rm g}_{QP}+n^{\rm f},
\label{fringe shift nDyon Massless}
\ee
where
\bea
n^{\rm g}_0 &=&
\frac{ l s \omega }{\pi r_1 }\Bigl[
\frac{a r_g}{r_1^2}\bigl(
2\sin\gamma\,\cos\theta_1 +\cos\gamma\,\sin\theta_1
 \bigr)
 \nonumber\\
 &&
+\frac{a r_g^2}{r_1^3}\Bigl(\sin(\gamma+\theta_1)
+\frac{1}{2}\cos\gamma\,\sin\theta_1
\Bigr)
\Bigr],
\label{n00g}
\\
n^{\rm g}_{QP} &=& 
-\frac{2 l s \omega }{\pi r_1 } \sin(\gamma+\theta_1)
\frac{a (Q^2+P^2)}{r_1^3},
\label{n00QP}
\\
n^{\rm f}&=&
-\frac{l s \omega}{2\pi  r_1  }
\cos\gamma\sin\theta_1\, 
\frac{a r_g^2}{r_1^3}.
\eea
Besides, if we rotate the rectangle around AB from $\gamma=0$ to $\gamma=\pi$, we get a fringe shift $N=-\pi^{-1}[\delta\phi^{\rm g}(\gamma=0)+\delta\phi^{\rm f}(\gamma=0)]$. Comparing \eqref{g phase difference} and \eqref{fringe shift nDyon} with \eqref{g phase difference Massless} and \eqref{fringe shift nDyon Massless}, we notice that the latter expressions can be obtained by applying $q=0$, $v=1$, and the replacement $m(1-v^2)^{-1/2}\rightarrow \hbar \omega$ to the former expressions. It meets our expectation because we replace \eqref{mathcalE} with \eqref{mathcalEMassless} for uncharged massless particles. As for Kerr-Newman spacetime, we just need to let $P=0$ in \eqref{g phase difference Massless}, \eqref{deltaphiFFML2}, and \eqref{fringe shift nDyon Massless} to get the gravitational phase difference, free phase difference, and fringe shift.

\subsection{Extension to dyonic particles\label{dyonic particles}}
Now we extend the results in Sec.~\ref{massive particles} to the case of dyonic particles in dyonic Kerr-Newman spacetime. Firstly, we need to find the action of a dyonic particle which has an electric charge $q$ and a magnetic charge $g$. For this purpose, let us recall the equation of motion in flat spacetime~\cite{Bruce}\footnote{One can check that in non-relativistic case, \eqref{relativistic particle} reduces to the equation of motion in Ref.~\cite{Schwinger:1969ib}.}
\be
m
\frac{\dd u^\alpha}{\dd s} 
=(q F^{\alpha\mu}
+g \tilde{F}^{\alpha\mu}) u_\mu,
\label{relativistic particle}
\ee
where $\tilde{F}_{\mu\nu} := \frac{1}{2}\epsilon_{\mu\nu \alpha\beta} F^{\alpha\beta}$ is the dual of the electromagnetic field. Generalize \eqref{relativistic particle} into generic spacetime, i.e. 
\be
m\Bigl(
\frac{\dd u^\alpha}{\dd s} 
+{\Gamma^\alpha}_{\mu\nu} u^\mu u^\nu
\Bigr)
=q {F^\alpha}_{\mu} u^\mu
+g \tilde{F}{{^\alpha}_\mu} u^\mu.
\label{generic}
\ee
One can find that the action
\be
S=\int (-m \dd s -q A_\mu \dd x^\mu -g \tilde{A}_\mu\dd x^\mu),
\label{actionS}
\ee
fits with the equation \eqref{generic}, where $\tilde{A}_\mu$ is the potential of $\tilde{F}_{\mu\nu}$. In dyonic Kerr-Newman spacetime, we can obtain $\tilde{A}_\mu$ by directly making the transformations $Q\rightarrow P$ and $P\rightarrow -Q$ in the electromagnetic potential~\cite{Hackmann:2013pva}.

For the interference experiment, we do not need to repeat the calculations in Sec.~\ref{interference}. Instead, we take a simpler way. Following Ref.~\cite{Hackmann:2013pva}, we define new charges
\be
\hat{Q}=\frac{q Q +g P}{\sqrt{q^2+g^2}},
\qquad
\hat{P}=\frac{q P -g Q}{\sqrt{q^2+g^2}},
\qquad
\hat{q}=\sqrt{q^2+g^2},
\ee
new metric $\hat{g}_{\mu\nu}$, and new electromagnetic potential $\hat{A}_\mu$, where $\hat{g}_{\mu\nu}$ and $\hat{A}_\mu$ are obtained by using both the replacements $Q\rightarrow \hat{Q}$ and $P\rightarrow \hat{P}$ in \eqref{R0} and \eqref{dmudxmu2}. One can check
\be
g_{\mu\nu}=\hat{g}_{\mu\nu},
\qquad
q A_\mu +g \tilde{A}_\mu
=\hat{q} \hat{A}_\mu.
\ee
With such way, the action \eqref{actionS} is written as
\be
S=\int (-m \dd s -\hat{q} \hat{A}_\mu \dd x^\mu).
\label{actionSS}
\ee
Evidently the action \eqref{actionSS} is not different from the case of a particle without magnetic charge, except that $Q$, $P$, $q$, $g_{\mu\nu}$, and $A_\mu$ should be replaced by $\hat{Q}$, $\hat{P}$, $\hat{q}$, $\hat{g}_{\mu\nu}$, and $\hat{A}_\mu$ respectively. Given that the gravitational phase, the electromagnetic phase, and the free phase come from the action (recall \eqref{phi B}, \eqref{EMphase}, and \eqref{free phase0}), they are obtained also by using the above substitutions to the phases of electrically charged particles. Accordingly, for the gravitational phase difference, the electromagnetic phase difference, and the free phase difference in the interference experiment, we only need to replace $Q$ with $\hat{Q}$, $P$ with $\hat{P}$, and $q$ with $\hat{q}$ in the expressions \eqref{delta 0 dyon}, \eqref{delta Q dyon}, \eqref{delta q0 dyon}, \eqref{delta e dyon}, and \eqref{deltaphiFF2}. Therefore, the total quantum phase difference and the fringe shifts are
\bea
\delta\phi^{\rm dyon}_{\rm total} &=& \delta\phi^{\rm g} (Q\!\rightarrow\hat{Q}, P\!\rightarrow\hat{P}, q\!\rightarrow\hat{q})
+\delta\phi^{\rm e} (Q\!\rightarrow\hat{Q}, P\!\rightarrow\hat{P}, q\!\rightarrow\hat{q})
+\delta\phi^{\rm f} (Q\!\rightarrow\hat{Q}, P\!\rightarrow\hat{P}, q\!\rightarrow\hat{q}),
\\
n^{\rm dyon} &=& n(Q\rightarrow\hat{Q}, P\rightarrow\hat{P}, q\rightarrow\hat{q}),
\\
N^{\rm dyon} &=& N(Q\rightarrow\hat{Q}, P\rightarrow\hat{P}, q\rightarrow\hat{q}),
\eea
where $\delta\phi^{\rm g}$, $\delta\phi^{\rm e}$, $\delta\phi^{\rm f}$, $n$, and $N$ are given by \eqref{g phase difference}, \eqref{delta e dyon}, \eqref{deltaphiFF2}, \eqref{fringe shift nDyon}, and \eqref{N dyon}.

\subsection{Discussion\label{DiscussionDyon}}
We will discuss two things: (I) how the black hole's magnetic charge affects the quantum interference of electrically charged particles; (II) how to test the Dirac quantization condition we mentioned in Sec.~\ref{massive particles}. 

For simplicity, consider a non-rotating dyonic black hole. Letting $a=0$ in \eqref{fringe shift nDyon}, only the first term of $n^{\rm e}_0$ is left, such that
\be
n=
-\frac{q l s}{\pi\hbar r_1^2} P\sin\gamma.
\label{nP}
\ee
It looks a little strange that $n$ is non-zero, considering that the spacetime is spherically symmetric. The expression \eqref{nP} is independent of the gravitational constant in SI units, hence it can only originate from electromagnetic interaction rather than gravitation. To show it explicitly, we write the electromagnetic phase difference for the rectangle in FIG.~\ref{experimentKN} as
\bea
\delta\phi^e&=&\phi_{ADC}-\phi_{ABC}
\nonumber\\
&=&\frac{q}{\hbar}\Bigl[
\int_{l_2} A_0(\vec{r}) \dd t -\int_{l_1}A_0(\vec{r}) \dd t +\oint A_i(\vec{r}) \dd x^i
\Bigr],
\label{delta phiea}
\eea
where $l_1$ and $l_2$ stand for the path ABC and ADC respectively. Make a rough estimate for the last term
\bea
(\delta\phi^e)_{\rm last}
&=&
\frac{q}{\hbar}\oint A_i(\vec{r}) \dd x^i
\nonumber\\
&\approx& -\frac{q}{\hbar}\oint \vec{A}(\vec{r})\cdot \dd \vec{r}
\nonumber\\
&=&
-\frac{q}{\hbar}\iint_{\sigma} \vec{B}(\vec{r})\cdot \dd \vec{\sigma}
\nonumber\\
&=&
-\frac{q}{\hbar} P \iint_{\sigma}\frac{\vec{r}\cdot\dd\vec{\sigma}}{r^3}
\nonumber\\
&\approx&\frac{q}{\hbar} P \frac{ ls \sin\gamma}{r_1^2},
\label{delta last}
\eea
where $\dd\vec{\sigma}$ is the infinitesimal area vector of the surface enclosed by the loop. In deriving \eqref{delta last}, we used Stokes' theorem on the third line, and used $\vec{B}=P\vec{r}/r^3$ for the magnetic field (see Appendix~\ref{EMpotential}) on the last second line. After flipping the rectangle, the direction of the area vector is reversed, such that $(\delta\phi^e)_{\rm last}$ changes to $(\delta\phi'^e)_{\rm last}=-(\delta\phi^e)_{\rm last}$. This process causes a fringe shift
\be
n_{\rm last}=\frac{(\delta\phi'^e)_{\rm last}-(\delta\phi^e)_{\rm last}}{2\pi}
=-\frac{qls}{\pi \hbar r_1^2} P\sin\gamma,
\ee
consistent with \eqref{nP}. In summary, the non-zero fringe shift $n$ for a non-rotating dyonic black hole originates from the magnetic field of the black hole's magnetic charge. Such conclusion also holds for a non-rotating magnetically charged black hole. 

As mentioned in the last second paragraph of Sec.~\ref{massive particles}, in the presence of the Dirac string, the definition of the electromagnetic field tensor should be modified as \eqref{field string}. But on the third line of \eqref{delta last} we still used the definition $\vec{B}=\nabla\times \vec{A}$ for the magnetic field. This is because the rectangle in FIG.~\ref{experimentKN} does not enclose the Dirac string (the negative semi-axis $z$) such that we can use the above definition directly. However, as discussed in Sec.~\ref{massive particles}, even though the rectangle encloses the Dirac string, the latter cannot be detected through the gravitational phase difference, and cannot be observed from the electromagnetic phase difference if the Dirac quantization condition holds. Therefore, it is necessary to find a way to test such condition experimentally. 

\begin{figure}[h]
\centering
\includegraphics[width=0.8\textwidth]{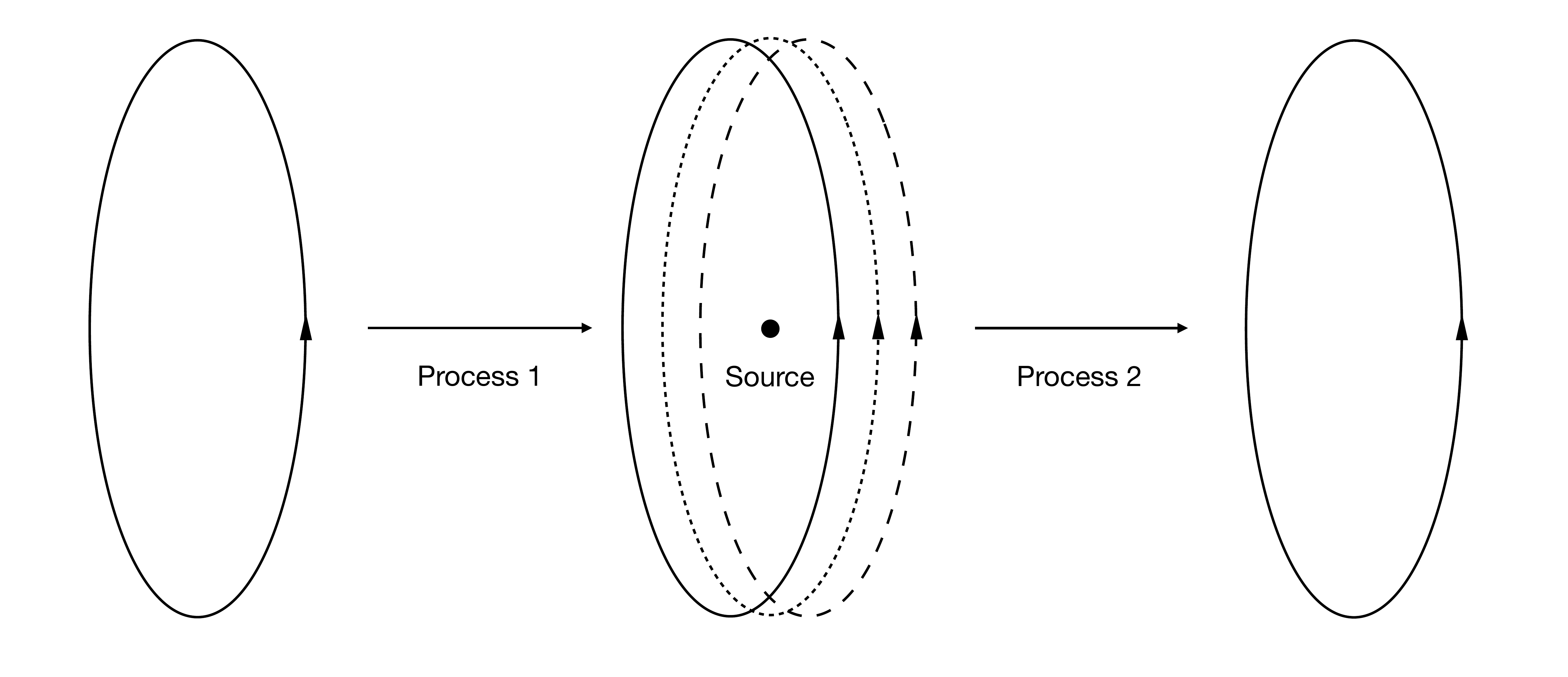}
\caption{A gedanken experiment to test the Dirac quantization condition. A loop is initially put at infinity, then it is placed at the positions closer and closer to a magnetic monopole source, afterwards it is placed at the positions farther and farther away from the source. By measuring the phase shifts of the loop we can test the Dirac quantization condition.}
\label{Monopole1}
\end{figure}

Here we show a simple idea to test the Dirac quantization condition if magnetic monopoles do exist. Consider a closed loop in the field of a static non-rotating magnetic monopole which locates at the origin. For simplicity, we ignore gravitation and assume that both the Dirac string and the axis of the loop are along negative semi-axis $z$. The phase shift of the loop is
\bea
\delta\phi^{\rm em}
&=&\frac{q}{\hbar}\oint A_\mu \dd x^\mu
\nonumber\\
&=&-\frac{q}{\hbar}\oint \vec{A}\cdot \dd \vec{x}
\nonumber\\
&=&-\frac{q}{\hbar}\iint (\nabla\times\vec{A})\cdot \dd \vec{\sigma}.
\label{deltaem0}
\eea
The curl is~\cite{Shnir:2005vvi,Jackson:1998nia,Heras:2018uub} 
\be
\nabla\times\vec{A}=\vec{B}_{\rm mono}+\vec{B}_{\rm st},
\label{nablaA}
\ee
where $\vec{B}_{\rm mono}=P\vec{r}/r^3$ is the magnetic field of the monopole, and $\vec{B}_{\rm st}=4\pi P\delta(x)\delta(y)\Theta(-z)\vec{e}_z$ is the magnetic field on the Dirac string. Here $\Theta(z)$ is the step function, and $\vec{e}_z$ is the unit vector along the positive semi-axis $z$. Plugging \eqref{nablaA} into \eqref{deltaem0} yields
\be
\delta\phi^{\rm em}=\delta\phi^{\rm mono}+\delta\phi^{\rm st},
\label{Phiem0}
\ee
where
\be
\delta\phi^{\rm mono}=
\begin{cases}
-\frac{q P}{\hbar} \Omega, &\text{if the loop is right-handed,}\\
\frac{q P}{\hbar} \Omega, &\text{if the loop is left-handed,}
\end{cases}
\label{deltamono}
\ee
and~\cite{Heras:2018uub}
\be
\delta\phi^{\rm st}=
\begin{cases}
\frac{4\pi q P}{\hbar}, &\text{if the loop is right-handed and encloses the string,}
\\
-\frac{4\pi q P}{\hbar}, &\text{if the loop is left-handed and encloses the string,}
\\
0, &\text{if the loop does not enclose the string,}
\end{cases}
\label{deltaphist}
\ee
where $\Omega$ is  the solid angle, and here ``right-handed'' means that the loop is a right-handed helix with respect to the displacement vector from the source to the center of the loop, while ``left-handed'' corresponds to left-handed helix. Evidently the second expression in \eqref{deltaphist} is just the result \eqref{electromagnetic phase shift2}. Applying the Dirac quantization condition \eqref{Dirac quantization} to \eqref{deltamono} and \eqref{deltaphist}, we get
\bea
&&\delta\phi^{\rm mono}=
\begin{cases}
-\frac{j}{2} \Omega, &\text{for right-handed loop,}\\
\frac{j}{2} \Omega, &\text{for left-handed loop,}
\end{cases}
\label{deltaphiEM2}
\\
&&\delta\phi^{\rm st}=
\begin{cases}
2 j\pi, &\text{for right-handed loop which encloses the string,}
\\
-2 j\pi, &\text{for left-handed loop which encloses the string,}
\\
0, &\text{if the loop does not enclose the string.}
\end{cases}
\label{deltaphiEMst}
\eea
In what follows, we apply \eqref{deltaphiEM2} and \eqref{deltaphiEMst} to a gedanken experiment shown in FIG.~\ref{Monopole1}. 
In FIG.~\ref{Monopole1}, when $z<0$, the loop is right-handed, while when $z>0$, the loop is left-handed. Accordingly we get
\be
\delta\phi^{\rm em}=
\begin{cases}
-\frac{j}{2} \Omega+2j\pi, &\text{when $z<0$,}
\\
\frac{j}{2} \Omega, &\text{when $z>0$.}
\end{cases}
\label{phasshiftem}
\ee
Therefore, to test the Dirac quantization condition, experimentally we just need to test \eqref{phasshiftem}. For example, for a circular loop with a radius $R$, one can find
\be
\Omega=2\pi \Bigl(1-\frac{D}{\sqrt{R^2+D^2}}\Bigr),
\ee
where $D=|z|$ is the distance between the loop and the source. In practice, it is more convenient to measure the fringe shift. According to \eqref{phasshiftem}, if we change the position of the loop from $z_a$ to $z_b$, the fringe shift is
\be
n=
\begin{cases}
-\frac{j (\Omega_b-\Omega_a)}{4\pi}, &\text{when $z_a<0$ and $z_b<0$,}
\\
\frac{j (\Omega_b-\Omega_a)}{4\pi}, &\text{when $z_a>0$ and $z_b>0$.}
\end{cases}
\label{fringeshiftem}
\ee
In the process 1 of FIG.~\ref{Monopole1}, the position of the loop is changed gradually from $z=-\infty$ to $z=0^-$. While in the process 2, the position is changed gradually from $z=0^+$ to $z=\infty$. The total fringe shifts of these two processes are respectively 
\be
n_1=-\frac{j}{2}, \quad
n_2=-\frac{j}{2}.
\label{n1n2}
\ee
Thus the total fringe shift from $z=-\infty$ to $z=\infty$ is $n_{\rm total}=-j$. Therefore, to test the Dirac quantization condition, one can test $n_1$, $n_2$ or $n_{\rm total}$. 

\begin{figure}[h]
\centering
\includegraphics[width=0.8\textwidth]{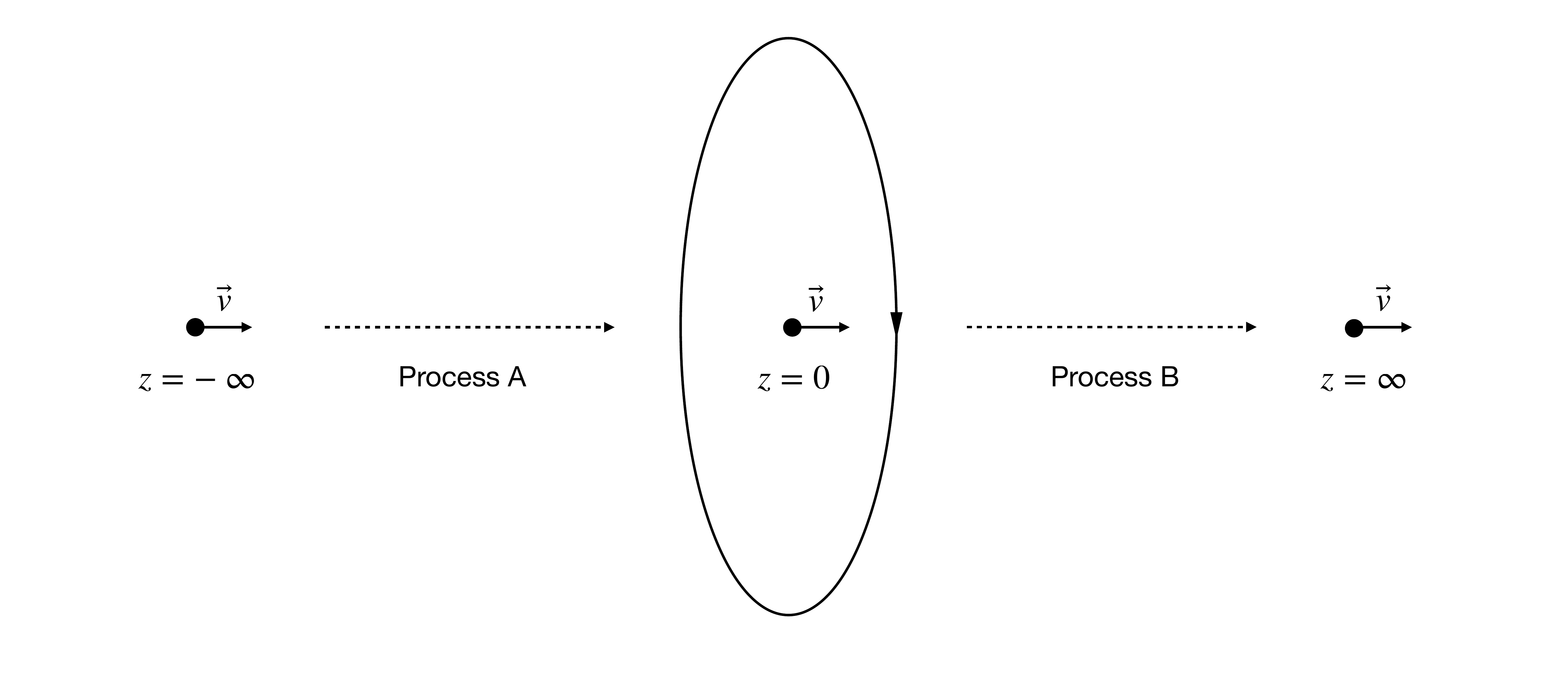}
\caption{A gedanken experiment to test the Dirac quantization condition by measuring the phase shifts of a loop when a moving magnetic monopole approaches it, crosses it, and goes away from it. The loop is placed at $z=0$, and it is counterclockwise with respect to the positive semi-axis $z$. The velocity $\vec{v}$ of the magnetic monopole is along the positive semi-axis $z$. In the process A, the magnetic monopole goes from $z=-\infty$ to $z=0$. In the process B, it goes from $z=0$ to $z=\infty$.}
\label{Monopole2}
\end{figure}

Obviously the experiment in FIG.~\ref{Monopole1} is equivalent to the following scenario: Fix the position of the loop but change the position of the magnetic monopole source such that the latter gradually crosses the former. Notice that the source is static at every point. However, if the source is moving, e.g., it has a constant velocity along the $z$ axis, the formula \eqref{deltamono} no longer holds. Nonetheless, considering that $\oint A_\mu \dd x^\mu$ is Lorentz invariant, we can choose a reference frame $(t',x',y',z')$ in which the source is static such that the above results are still valid. Concretely, consider a scenario shown in FIG~\ref{Monopole2}: A magnetic monopole source (with a constant velocity $\vec{v}=v \vec{e}_z$) goes from $z=-\infty$, crosses a circular loop (whose axis is along the positive semi-axis $z$) at $z=0$, and goes to $z=\infty$. For simplicity we assume that the Dirac string is along the direction of $\vec{v}$. 
In the frame $(t',x',y',z')$, the formulas \eqref{Phiem0}, \eqref{deltamono}, and \eqref{deltaphist} still hold, therefore we have
\bea
&&\delta\phi'^{\rm mono}=
\begin{cases}
-\frac{q P}{\hbar} 2\pi \Bigl(1-\frac{D'}{\sqrt{R^2+D'^2}}\Bigr), &\text{when $z^{\rm mono}<0$,}\\
\frac{q P}{\hbar} 2\pi \Bigl(1-\frac{D'}{\sqrt{R^2+D'^2}}\Bigr), &\text{when $z^{\rm mono}>0$,}
\end{cases}
\label{deltamonoPrime}
\\
&&\delta\phi'^{\rm st}=
\begin{cases}
\frac{4\pi q P}{\hbar}, &\text{when $z^{\rm mono}<0$,}
\\
0, &\text{when $z^{\rm mono}>0$,}
\end{cases}
\label{deltaphistPrime}
\eea
where $D'$ is the distance between the source and the loop in the frame $(t',x',y',z')$, and $z^{\rm mono}$ is the $z$ coordinate of the magnetic monopole in the frame $(t,x,y,z)$. Evidently we have $D'=\gamma_v D$, where $\gamma_v=(1-v^2/c^2)^{-1/2}$. The result \eqref{deltamonoPrime} agrees with the magnetic flux formula in Ref.~\cite{Cabrera:1982gz}. Following Ref.~\cite{Cabrera:1982gz}, we let $z=-\infty$ corresponds to $t=-\infty$, $z=0$ corresponds to $t=0$, and $z=\infty$ corresponds to $t=\infty$, such that $D=|vt |$. In such way, the phase shifts can be rewritten as functions of time. Plugging the Dirac quantization condition \eqref{Dirac quantization} into \eqref{deltamonoPrime} and \eqref{deltaphistPrime}, and recalling that $\delta\phi^{\rm em}=\delta\phi'^{\rm em}$, we get
\bea
&&\delta\phi'^{\rm mono}=-j\pi \Bigl(1-2\Theta(t)+\frac{\gamma_v vt}{\sqrt{R^2+(\gamma_v vt)^2}}\Bigr),
\\
&&\delta\phi'^{\rm st}=2j\pi\bigl(1-\Theta(t) \bigr),
\\
&&\delta\phi^{\rm em}
=j\pi \Bigl(1-\frac{\gamma_v vt}{\sqrt{R^2+(\gamma_v vt)^2}}\Bigr).
\label{deltaphiem3}
\eea
Therefore, to test the Dirac quantization condition, we just need to test the formula \eqref{deltaphiem3}. Differently from Ref.~\cite{Cabrera:1982gz} in which the author searched for magnetic monopoles by monitoring the current of a superconducting loop, here we can monitor the phase shift of a loop through the quantum interference, e.g, recording the fringe shifts. For example, using \eqref{deltaphiem3} one can find $\delta\phi^{\rm em}|_{t=-\infty}=2j\pi$, $\delta\phi^{\rm em}|_{t=0}=j\pi$, and $\delta\phi^{\rm em}|_{t=\infty}=0$. Accordingly the total fringe shifts of the process A and process B are respectively
\be
n_{A}=-\frac{j}{2}, \quad
n_{B}=-\frac{j}{2},
\ee
thus $n_{\rm total}=-j$, same as the results for FIG~\ref{Monopole1}. Generally, according to \eqref{deltaphiem3}, the fringe shift between $t_1$ and $t_2$ is
\be
n_{12}=-\frac{j}{2}\Bigl(\frac{\gamma_v vt_2}{\sqrt{R^2+(\gamma_v vt_2)^2}}
-\frac{\gamma_v vt_1}{\sqrt{R^2+(\gamma_v vt_1)^2}}
\Bigr).
\ee

%%%%%%%%%
\section{Gravitation without universality\label{Non-Universal}}

\subsection{A brief introduction\label{brief introduction}}
Our calculations in all the previous sections are based on the TG theory which is equivalent to general relativity. However, there are other versions of TG different from general relativity, e.g., a version in which the weak equivalence principle breaks down~\cite{Aldrovandi:2003pa,Aldrovandi:2004fy,Aldrovandi:2005mz}. It is not surprising, because TG is not a geometric description of gravity and this principle is not a necessary requirement in TG~\cite{Aldrovandi:2013wha}. In this subsection, I make a brief introduction of such theory which violates the weak equivalence principle, based on Chap.~11 of Ref.~\cite{Aldrovandi:2013wha}. For simplicity, we only consider uncharged particles. 

TG is a gauge theory in which gravitation is generated from the translation group in tangent space. Correspondingly, the covariant derivative of a field $\Psi$ which represents a particle is
\be
\tilde{h}_\mu \Psi \, = (\partial_\mu x^a+\dot{A}{^a}_{b\mu}x^b + \sigma B{^a}_\mu)\partial_a \Psi
\, =\bigl[ {h^a}_\mu +\xi B{^a}_\mu\bigr] \partial_a \Psi,
\ee
where $\sigma=m_{\rm g}/m_{\rm i}$ and $\xi=\sigma -1$, with $m_{\rm g}$ and $m_{\rm i}$ the gravitational mass and inertial mass respectively, and ${h^a}_\mu$ is the tetrad defined in \eqref{hdotAB}. In this theory, $\dot{A}{^a}_{b\mu}$ stands for the inertial effects of the frame, while $B{^a}_\mu$ represents the gravitational field. When $\xi=0$, such theory reduces to the version we used in the previous sections. The action of the particle is
\bea
\mathcal{S}
&=&-\int u_a \Bigl[ m_{\rm i}  (\partial_\mu x^a+\dot{A}{^a}_{b\mu}x^b)
+m_{\rm g} \sigma B{^a}_\mu
\Bigr] \dd x^\mu
\nonumber\\
&=&-m_{\rm i}\int u_a h{^a}_\mu \dd x^\mu
-m_{\rm i}\xi \int u_a B{^a}_\mu \dd x^\mu
\nonumber\\
&=&-m_{\rm i}\int\dd s
-m_{\rm i}\xi \int u_a B{^a}_\mu \dd x^\mu,
\label{action violation}
\eea
where \eqref{s three} is used in the last step. Generally, the equation of motion from \eqref{action violation} is not equivalent to the geodesic equation in general relativity, because of the presence of $\xi$. Indeed, such equation is given by
\be
\frac{\dd u_a}{\dd s}
-\dot{A}{^b}_{a\rho} u_b u^\rho
=-\dot{K}{^b}_{a\rho}  u_b u^\rho
-\xi\, {h_a}^\mu \Bigl[
(\delta^\rho_\mu -u^\rho u_\mu) {B^b}_\rho \frac{\dd u_b}{\dd s} -(\partial_\mu{B^b}_\rho-\partial_\rho {B^b}_\mu ) u_b u^\rho
\Bigr],
\label{EoM violation}
\ee
where $\dot{K}{^b}_{a\rho}$ is the contortion. Only when $\xi=0$, the equation \eqref{EoM violation} is equivalent to the geodesic equation (read Sec. 6.2.3 in Ref.~\cite{Aldrovandi:2013wha} for the proof).

Though \eqref{EoM violation} depends on $\xi$, the Lorentz connection $\dot{A}{^a}_{b\mu}$ and gauge potential ${B^a}_\mu$ do not, as mentioned in Ref.~\cite{Aldrovandi:2013wha}. Accordingly, the tetrad ${h^a}_\mu$ is also independent of the particle's property. In other words, the gravitational field strength (represented by the torsion \eqref{hT expression}) in TG is independent of $\xi$. Consequently, regardless of $\xi=0$ or $\xi\ne 0$, the gravitational action \eqref{dotL} is the same such that the field equation in TG is the same. It means that whether the weak equivalence principle holds or not, the field equation in TG is equivalent to the Einstein field equation. Besides, even if $\xi\ne 0$, the relations $\dd s=\sqrt{g_{\mu\nu}\dd x^\mu \dd x^\nu}$ and $g_{\mu\nu}=\eta_{ab} {h^a}_\mu{h^a}_\nu$ still hold~\cite{Aldrovandi:2004fy}. Therefore, we can still use the metric solutions of Einstein field equation to find the tetrad in TG. 
Now that the gauge potential and tetrad are independent of the weak equivalence principle, the equation \eqref{B field} still holds even if $\xi\ne 0$. Let us quickly re-derive such equation. According to \eqref{hdotAB}, we have
\be
B{^a}_\mu=
{h^a}_\mu -\partial_\mu x^a-\dot{A}{^a}_{b\mu}
={h^a}_\mu -(\partial_{\mu'} x^a)\, \partial_\mu x^{\mu'}-\dot{A}{^a}_{b\mu}.
\label{Bamu general}
\ee
In the interference experiment we always choose  proper frames such that $\dot{A}{^a}_{b\mu}$ vanishes and we let $x^{\mu'}$ be the cartesian coordinates, by which \eqref{Bamu general} reduces to \eqref{B field}. 

In summary, when the weak equivalence principle is violated, the particle's action and motion in TG are not equivalent to those in general relativity, but the gravitational action and field equation are still equivalent to those in general relativity. Accordingly, the solutions of the gravitational field in TG can be still obtained from the metric solutions of Einstein field equation. Especially, the gravitational gauge potential, the tetrad, and the relation between them are not changed. Therefore, the expression of $B{^a}_\mu$ is not changed compared with the case where the weak equivalence principle holds. This conclusion is crucial when we compute the $S_\beta$ function in Appendix~\ref{without}.

\subsection{Phase differences and fringe shifts\label{PandF}}
Here we study the situation that the universality of gravity is broken down for the interference experiment. Before proceeding, let us present the motivation. We notice that both the gravitational phase difference \eqref{delta phi g} and the electromagnetic phase difference \eqref{delta e} do not contain the charge-to-mass ratio $q/m$. It looks strange because \eqref{phi B} and \eqref{EMphase} can be written as
\bea
\phi_g &=&\frac{m}{\hbar}\int  B{^\nu}_\beta u_\nu  u^\beta \dd s,
\label{phi Bnew}
\\
\phi_e &=&\frac{q}{\hbar} \int A_\beta u^\beta\dd s,
\label{phi enew}
\eea
with the four-velocity satisfying the equation of motion~\cite{Landau:1975pou}
\be
\frac{\dd u^\mu}{\dd s} +\Gamma^\mu_{\alpha\beta} u^\alpha u^\beta
=\frac{q}{m} F^{\mu\sigma} u_\sigma,
\label{EOMM}
\ee
which shows that the solution of $u^\mu$ should depend on $q/m$. The reason why $q/m$ does not appear in these phase differences can be explained by the following qualitative argument. The integrand of \eqref{phi Bnew} contains a quadratic term of the four-velocity, for which we may expect $q/m$ and $(q/m)^2$ appear in $\phi_g$. However, as for the term of $q/m$, its mass is cancelled by the multiplier in front of the integral in \eqref{phi Bnew}. Besides, the terms proportional to $q^2$ are of the order of $O(l^2/r_1^2)$, which is neglected in our approximations. The reason is as follows. According to \eqref{delta phi ag}, \eqref{delta phi bg}, \eqref{delta phi cg}, \eqref{S beta3}, \eqref{S beta3-2}, and \eqref{t varphi}, the terms of $q^2$ can only appear in $\delta\phi_b$. In \eqref{delta phi bg}, both $(S_0^D)_{DC} -(S_0^A)_{AB}$ and $t_{DC}-t_{AB}$ contain terms of $q$, where $t_{DC}$ is given by
\be
t_{DC}\approx s\Bigl(
\frac{1}{v_2\sqrt{g_{00}(r_2,\theta_2) }}-\frac{g_{03}(r_2,\theta_2) }{g_{00}(r_2,\theta_2)\sqrt{\Gamma_{33}(r_2,\theta_2)} }
\Bigr),
\ee
with $v_2$ the velocity on the path DC solved through \eqref{mathcalE}, i.e.
\be
v_2=\sqrt{1-g_{00}(r_2,\theta_2)\Bigl[\frac{\sqrt{g_{00}(r_1,\theta_1)} }{\sqrt{1-v^2} } +\frac{q}{m}\Bigl(A_0(r_1,\theta_1)-A_0(r_2,\theta_2) \Bigr)\Bigr]^{-2} },
\label{v2}
\ee
where $v$ is the velocity on AB. However, after expanding \eqref{delta phi bg} by \eqref{r2 theta2}, the term $\delta\phi_b$ is of the order of $O(l^2/r_1^2)$ which is neglected in our approximations. Therefore the terms of $q^2$ is neglected. Up to now, we have explained why the factor $q/m$ does not appear in \eqref{delta phi g}. As for the electromagnetic phase difference, considering the phase \eqref{phi enew} contains the four-velocity, one may expect $\delta\phi^e$ contains $q^2/m$. However, similar to the case of the gravitational phase difference, the terms of $q^2/m$ only appear in $\delta\phi^e_b$ (see \eqref{delta phi bgEM}) so that they are of the order of $O(l^2/r_1^2)$. Such terms are neglected in our approximation, hence the factor $q/m$ is also absent from the electromagnetic phase difference.

Is it possible to reintroduce the charge-to-mass ratio to the phase differences? The answer is yes. We find that it could happen for a theory which breaks the universality of gravitation (or more precisely, violates the weak equivalence principle). To be explicit, let us write down the equation of motion for a charged particle (see Appendix~\ref{without})
\bea
&&\frac{\dd u^\mu}{\dd s}
+ {\Gamma^\mu}_{\nu\rho} u^\nu u^\rho
= \frac{q}{m_{\rm i}} F{^\mu}_\nu u^\nu
\nonumber\\
&&\quad +\frac{m_{\rm i} - m_{\rm g} }{m_{\rm i} } g^{\mu\sigma}
\Bigl[ ( {\delta^\rho}_\sigma -  u^\rho u_\sigma) {B^b}_\rho \frac{\dd u_b}{\dd s}  - (\partial_\sigma {B^b}_\rho -\partial_\rho {B^b}_\sigma ) u_b u^\rho
\Bigr].
\label{EOMwithoutUn}
\eea
According to \eqref{EOMwithoutUn}, we would expect the factors $q/m_{\rm i}$ and $(m_{\rm i}-m_{\rm g})/m_{\rm i}$ appear in the phase differences, considering that the phase differences relate to the four-velocity. Indeed, for non-relativistic particles in Kerr-Newman spacetime, the gravitational phase difference is (see Appendix~\ref{without} for the derivations)
\be
\delta\phi_{\rm g}
\approx
\delta\phi^{\rm g}_0
+\delta\phi^{\rm g}_Q
+\delta\phi^{\rm g}_q
+\delta\phi^{\rm g}_{\rm ad},
\label{delta phi gW0}
\ee
where
\bea
\delta\phi^{\rm g}_0
&=&\frac{m_{\rm g} l s}{ \hbar r_1 \sqrt{1-v^2}} \Bigl\{
\frac{1}{v}
\cos\gamma\,\Bigl(\frac{r_g}{2r_1}+\frac{r_g^2}{2r_1^2}\Bigr)
-\frac{a r_g}{r_1^2}\Bigl(2\cos\theta_1\, \sin\gamma+\cos\gamma\,\sin\theta_1
\Bigr)
\nonumber\\
&&
+\frac{1}{v}\cos\gamma\,\frac{r_g^3}{2r_1^3}
-\frac{a r_g^2}{r_1^3}
\Bigl(
  \cos\theta_1\,\sin\gamma
 +\frac{3}{2}\cos\gamma\,\sin\theta_1
 \Bigr)
\nonumber\\
&&
+\frac{a^2 r_g}{r_1^3}
\Bigl[ \frac{1}{v}\Bigl(\frac{1}{4}\cos\gamma\,-\frac{7}{4}\cos\gamma\,\cos^2\theta_1
+\frac{1}{2}\sin(2\theta_1)
\sin\gamma
\Bigr)
\nonumber\\
&&
+ v\sin\theta_1\,\Bigl(
\frac{3}{2} \cos\gamma\, \sin\theta_1
+\sin\gamma\, \cos\theta_1
\Bigr)
\Bigr]
\Bigr\},
\label{delta g0W0}
\\
\delta\phi^{\rm g}_Q
&=&\frac{m_{\rm g} l s}{ \hbar r_1 \sqrt{1-v^2}} \Bigl[
-\frac{1}{v}\cos\gamma\, \Bigl(
\frac{Q^2}{r_1^2} +\frac{3r_g Q^2}{2 r_1^3}
\Bigr)
+2\sin(\theta_1+\gamma)\frac{a Q^2}{r_1^3}
\Bigr],
\label{delta gQW0}
\\
\delta\phi^{\rm g}_q
&=&\frac{m_{\rm g} }{m_{\rm i}} \frac{q l s}{ \hbar r_1} \Bigl[
\frac{1}{v}\cos\gamma\, \Bigl(
-\frac{r_g Q}{2r_1^2} 
-\frac{3r_g^2 Q}{8 r_1^3}
+\frac{Q^3}{2r_1^3} 
\Bigr)
+\cos\gamma\,\sin\theta_1\, \frac{a r_g Q}{r_1^3}
\Bigr],
\label{delta gqqW0}
\\
\delta\phi^{\rm g}_{\rm ad}
&=&\frac{m_{\rm g} }{m_{\rm i}} \frac{(m_{\rm g}-m_{\rm i}) l s }{\hbar r_1} 
\Bigl[
\frac{1}{v}\cos\gamma\, \Bigl(
\frac{r_g^2}{4r_1^2} 
+\frac{5 r_g^3}{16 r_1^3}
-\frac{3 r_g Q^2}{4r_1^3} 
\Bigr)
-\cos\gamma\sin\theta_1 \frac{a r_g^2}{2 r_1^3}
\Bigr].
\label{delta g ad0}
\eea
The expression \eqref{delta phi gW0} does not hold for relativistic particles because we have used the non-relativistic approximation to linearize the relation between the generalized momentum and the four-velocity (see \eqref{pimu2}). For relativistic particles, such relation is non-linear (see \eqref{pimu}), which makes it difficult to compute the gravitational phase difference in this case. Comparing \eqref{delta phi gW0} with \eqref{delta phi g}, we can find that the new gravitational phase difference $\delta\phi_{\rm g}$ is obtained by replacing $m$ with $m_{\rm g}$ in \eqref{delta g0} and \eqref{delta gQ}, multiplying \eqref{delta gqq} by $m_{\rm g}/m_{\rm i}$, and adding a term $\delta\phi^{\rm g}_{\rm ad}$. The expression \eqref{delta phi gW0} reduces to \eqref{delta phi g} when $m_{\rm i}=m_{\rm g}$. According to \eqref{delta gqqW0} and \eqref{delta g ad0}, the charge-to-mass ratio $q/m_{\rm i}$ and the factor $(m_{\rm i}-m_{\rm g})/m_{\rm i}$ appear in the gravitational phase difference, as we expect. However, the electromagnetic phase difference is still given by \eqref{delta e}, namely (for the explanation, see the last paragraph in Appendix~\ref{without})
\bea
\delta\phi_{\rm e}
&\approx&
\frac{q l s}{ \hbar r_1} \Bigl\{
-\frac{1}{v}
\frac{Q}{r_1} \cos\gamma
+\frac{a Q}{r_1^2}
\bigl(2\sin\gamma\,\cos\theta_1 +\cos\gamma\,\sin\theta_1\bigr)
\nonumber\\
&&
+\frac{1}{v} \frac{a^2 Q}{4 r_1^3}
\Bigl[
\cos\gamma\, \bigl(5 \! + \! 7\cos(2\theta_1) \bigr)
 \! - \! 4\sin\gamma\,\sin(2\theta_1)
\Bigr]
\! + \! \frac{a r_g Q}{2r_1^3}\cos\gamma\,\sin\theta_1
\Bigr\},
\label{deltaPhi ee}
\eea
such that it does not contain the factors $q/m_{\rm i}$ and $(m_{\rm i}-m_{\rm g})/m_{\rm i}$. This is because in $\delta\phi_{\rm e}$, these factors can only appear in the terms of the order $O(l^2/r_1^2)$ which we neglect in the calculations. Finally, the free phase difference is (see Appendix~\ref{without})
\bea 
\delta\phi_{\rm f} 
&=&\frac{l s}{ \hbar r_1}
\Bigl\{
\frac{1}{v}
\Bigl[
 -\frac{m_{\rm i} \cos\gamma}{4\sqrt{1-v^2}}\frac{r_g^2}{r_1^2}
 +\frac{q\cos\gamma}{2} \frac{r_g Q}{r_1^2}
 +\frac{(m_{\rm i} - m_{\rm g})\cos\gamma}{4}\frac{r_g^2}{r_1^2}
\Bigr]
+\frac{m_{\rm i}}{16\sqrt{1-v^2}}
\Bigl[12\cos\gamma \frac{r_g Q^2}{ v r_1^3}
\nonumber\\
&&
- 5\cos\gamma \frac{r_g^3}{ v r_1^3}
+ 8\cos\gamma \sin\theta_1  \frac{a r_g^2}{r_1^3}
- 8v \sin\theta_1\,(2\cos\theta_1\sin\gamma +3\sin\theta_1\cos\gamma)
\frac{a^2 r_g}{r_1^3}
\Bigr]
+\frac{q}{8 v}\Bigl[
-4 \cos\gamma\frac{Q^3}{r_1^3}
\nonumber\\
&&
-4 \cos\gamma\frac{Q P^2}{r_1^3}
+ 3\cos\gamma\frac{Q r_g^2}{r_1^3}
- 8 v \cos\gamma\sin\theta_1\frac{a r_g Q}{r_1^3}
\Bigr]
\nonumber\\
&&+\frac{(m_{\rm g}-m_{\rm i})\cos\gamma}{16v}
\Bigl(
12 \frac{ P^2 r_g}{r_1^3}
- 5 \frac{r_g^3}{r_1^3}
+ 12 \frac{Q^2 r_g}{r_1^3}
+ 8 v\sin\theta_1 \frac{a r_g^2}{r_1^3}
\Bigr)
\Bigr\}.
\label{deltaphiFF1}
\eea

Flipping the rectangle (such that A and B swap) gives new phase differences $\delta{\phi}'_{\rm g}$ and $\delta{\phi}'_{\rm e}$. The expressions of them can be found by using the replacements \eqref{replacements flip} in \eqref{delta phi gW0}, \eqref{deltaPhi ee}, and \eqref{deltaphiFF1}. Such process produces a fringe shift
\be
n= n^g_0 +n^g_{Q}+n^g_{q} +n^e_0 +n^e_G +n^f,
\label{fringe shift nU}
\ee
where
\bea
n^g_0 &=&
\frac{m_{\rm g} l s}{\pi\hbar r_1\sqrt{1-v^2} }\Bigl[
\frac{a r_g}{r_1^2}\bigl(
2\sin\gamma\,\cos\theta_1 +\cos\gamma\,\sin\theta_1
 \bigr)
+\frac{a r_g^2}{r_1^3}\Bigl( \sin(\gamma+\theta_1)
+\frac{1}{2}\cos\gamma\,\sin\theta_1
\Bigr)
\Bigr],
\\
n^g_{Q} &=& 
-\frac{2 m_{\rm g} l s}{\pi\hbar r_1\sqrt{1-v^2} } \sin(\gamma+\theta_1)
\frac{a Q^2}{r_1^3},
\\
n^g_{q} &=&
-\frac{m_{\rm g} }{m_{\rm i} }\frac{q l s}{\pi\hbar r_1} \cos\gamma\, \sin\theta_1
\frac{a r_g Q}{r_1^3},
\\
n^g_{\rm ad} &=&
\frac{m_{\rm g} }{m_{\rm i} }
\frac{(m_{\rm g}-m_{\rm i}) l s }{2\pi \hbar r_1} \cos\gamma\sin\theta_1 \frac{a r_g^2}{ r_1^3},
\\
n^e_0 &=&
-\frac{q l s}{\pi\hbar r_1}
\frac{a Q}{r_1^2}
(2\sin\gamma\cos\theta_1
+\cos\gamma\sin\theta_1),
\\
n^e_G &=&
-\frac{q l s}{2\pi\hbar r_1}
\cos\gamma\, \sin\theta_1\, 
\frac{a r_g Q}{r_1^3},
\\
n^f&=&
\frac{l s}{\pi \hbar r_1  }\Bigl[ 
\frac{1}{2} \cos\gamma\sin\theta_1\, \Bigl( m_{\rm i}-m_{\rm g} -\frac{m_{\rm i}}{\sqrt{1-v^2}}\Bigr)
\frac{a r_g^2}{r_1^3}
 +q\cos\gamma\sin\theta_1\frac{a r_g Q}{r_1^3}
\Bigr].
\label{nff1}
\eea
And the fringe shift $N$ (rotate the rectangle around AB from $
\gamma=0$ to $\gamma=\pi$) is $N=-\pi^{-1} \bigl[\delta\phi_{\rm g}(\gamma=0)+\delta\phi_{\rm e}(\gamma=0) + \delta\phi_{\rm f}(\gamma=0)  \bigr]$.

By the way, for dyonic Kerr-Newman spacetime, we can repeat the derivations in Appendix \ref{without} to derive the phase differences. In the case which violates the universality of gravity, the gravitational phase difference for non-relativistic particles is
\be
\delta\phi^{\rm g}_{\rm dyon}
\approx
\delta\phi^{\rm dyon}_0
+\delta\phi^{\rm dyon}_{QP}
+\delta\phi^{\rm dyon}_q
+\delta\phi^{\rm dyon}_{\rm ad},
\label{g phase differenceU}
\ee
where
\bea 
\delta\phi^{\rm dyon}_0
&=&\frac{m_{\rm g} l s}{ \hbar r_1 \sqrt{1-v^2}} \Bigl\{
\frac{1}{v}
\cos\gamma\,\Bigl(\frac{r_g}{2r_1}+\frac{r_g^2}{2r_1^2}\Bigr)
-\frac{a r_g}{r_1^2}\Bigl(2\cos\theta_1\, \sin\gamma+\cos\gamma\,\sin\theta_1
\Bigr)
\nonumber\\
&&
+\frac{1}{v}\cos\gamma\,\frac{r_g^3}{2r_1^3}
-\frac{a r_g^2}{r_1^3}
\Bigl(
  \cos\theta_1\,\sin\gamma
 +\frac{3}{2}\cos\gamma\,\sin\theta_1
 \Bigr)
\nonumber\\
&&
+\frac{a^2 r_g}{r_1^3}
\Bigl[ \frac{1}{v}\Bigl(\frac{1}{4}\cos\gamma\,-\frac{7}{4}\cos\gamma\,\cos^2\theta_1
+\frac{1}{2}\sin(2\theta_1)
\sin\gamma
\Bigr)
\nonumber\\
&&
+v\sin\theta_1\,\Bigl(
\frac{3}{2} \cos\gamma\, \sin\theta_1
+\sin\gamma\, \cos\theta_1
\Bigr)
\Bigr]
\Bigr\},
\label{delta 0 dyonU}
\\
\delta\phi^{\rm dyon}_{QP}
&=&\frac{m_{\rm g} l s}{ \hbar r_1 \sqrt{1-v^2}} \Bigl[
-\frac{1}{v}\cos\gamma\, \Bigl(
\frac{Q^2+P^2}{r_1^2} +\frac{3r_g (Q^2+P^2)}{2 r_1^3}
\Bigr)
+2\sin(\theta_1+\gamma)\frac{a (Q^2+P^2)}{r_1^3}
\Bigr],
\label{delta Q dyonU}
\\
\delta\phi^{\rm dyon}_q
&=&\frac{m_{\rm g}}{m_{\rm i}} \frac{q l s}{ \hbar r_1} \Bigl[
\frac{1}{v}\cos\gamma\, \Bigl(
-\frac{r_g Q}{2r_1^2} 
-\frac{3r_g^2 Q}{8 r_1^3}
+\frac{Q^3}{2r_1^3} 
\Bigr)
+\cos\gamma\,\sin\theta_1\, \frac{a r_g Q}{r_1^3}
\nonumber\\
&&+\frac{\cos\gamma}{2v} 
\frac{P^2 Q}{r_1^3}
+
\frac{1}{4v}\Bigl(\cos(\gamma-\theta_1)+3\cos(\gamma+\theta_1)\Bigr)\frac{P a r_g}{r_1^3}
\Bigr],
\label{delta q0 dyonU}
\\
\delta\phi^{\rm dyon}_{\rm ad}
&=&\frac{m_{\rm g} }{m_{\rm i}} \frac{(m_{\rm g}-m_{\rm i}) l s }{\hbar r_1} 
\Bigl[
\frac{1}{v}\cos\gamma\, \Bigl(
\frac{r_g^2}{4r_1^2} 
+\frac{5 r_g^3}{16 r_1^3}
-\frac{3 r_g (Q^2+P^2) }{4r_1^3} 
\Bigr)
-\cos\gamma\sin\theta_1 \frac{a r_g^2}{2 r_1^3}
\Bigr].
\label{delta g adU}
\eea
As for the electromagnetic phase difference $\delta\phi^{\rm e}_{\rm dyon}$, it is still given by \eqref{delta e dyon}, while the free phase difference is
\bea 
\delta\phi^{\rm f}_{\rm dyon}
&=&\frac{l s}{ \hbar r_1}
\Bigl\{
\frac{1}{v}
\Bigl[
 -\frac{m_{\rm i} \cos\gamma}{4\sqrt{1-v^2}}\frac{r_g^2}{r_1^2}
 +\frac{q\cos\gamma}{2} \frac{r_g Q}{r_1^2}
 +\frac{(m_{\rm i} - m_{\rm g})\cos\gamma}{4}\frac{r_g^2}{r_1^2}
\Bigr]
+\frac{m_{\rm i}}{16\sqrt{1-v^2}}
\Bigl[12\cos\gamma \frac{r_g (Q^2+P^2)}{ v r_1^3}
\nonumber\\
&&
- 5\cos\gamma \frac{r_g^3}{ v r_1^3}
+ 8\cos\gamma \sin\theta_1  \frac{a r_g^2}{r_1^3}
- 8v \sin\theta_1\,(2\cos\theta_1\sin\gamma +3\sin\theta_1\cos\gamma)
\frac{a^2 r_g}{r_1^3}
\Bigr]
+\frac{q}{8 v}\Bigl[
-4 \cos\gamma\frac{Q^3}{r_1^3}
\nonumber\\
&&
-4 \cos\gamma\frac{Q P^2}{r_1^3}
+ 3\cos\gamma\frac{Q r_g^2}{r_1^3}
- 8 v \cos\gamma\sin\theta_1\frac{a r_g Q}{r_1^3}
+ 4(\sin\gamma\sin\theta_1-2\cos\gamma\cos\theta_1)\frac{a r_g P}{r_1^3}
\Bigr]
\nonumber\\
&&+\frac{(m_{\rm g}-m_{\rm i})\cos\gamma}{16v}
\Bigl(
12 \frac{ P^2 r_g}{r_1^3}
- 5 \frac{r_g^3}{r_1^3}
+ 12 \frac{Q^2 r_g}{r_1^3}
+ 8 v\sin\theta_1 \frac{a r_g^2}{r_1^3}
\Bigr)
\Bigr\}.
\label{deltaphiFF}
\eea
The total quantum phase difference is $\delta\phi =\delta\phi^{\rm g}_{\rm dyon}+\delta\phi^{\rm e}_{\rm dyon}+\delta\phi^{\rm f}_{\rm dyon}$.

By applying the replacements \eqref{replacements flip} to $\delta\phi^{\rm g}_{\rm dyon}$, $\delta\phi^{\rm e}_{\rm dyon}$, and $\delta\phi^{\rm f}_{\rm dyon}$, we can get the new phase differences $\delta{\phi}'^{\rm g}_{\rm dyon}$, $\delta{\phi}'^{\rm e}_{\rm dyon}$,  and $\delta\phi'^{\rm f}_{\rm dyon}$ of flipping the rectangle around its height. Such process gives a fringe shift
\be
n= n^{\rm g}_0 +n^{\rm g}_{QP}+n^{\rm g}_{q} +n^{\rm g}_{\rm ad} +n^{\rm e}_0 +n^{\rm e}_G 
+n^{\rm f},
\label{fringe shift nDyon2}
\ee
where
\bea
n^{\rm g}_0 &=&
\frac{m_{\rm g} l s}{\pi\hbar r_1\sqrt{1-v^2} }\Bigl[
\frac{a r_g}{r_1^2}\bigl(
2\sin\gamma\,\cos\theta_1 +\cos\gamma\,\sin\theta_1
 \bigr)
+\frac{a r_g^2}{r_1^3}
\Bigl(\sin(\gamma+\theta_1)
+\frac{1}{2}\cos\gamma\,\sin\theta_1
\Bigr)
\Bigr],
\label{nrmg0}%
\\
n^{\rm g}_{QP} &=& 
-\frac{2 m_{\rm g} l s}{\pi\hbar r_1\sqrt{1-v^2} } \sin(\gamma+\theta_1)
\frac{a (Q^2+P^2)}{r_1^3},
\label{nrmgQP}%
\\
n^{\rm g}_{q} &=&
-\frac{m_{\rm g} }{m_{\rm i} }\frac{q l s}{\pi\hbar r_1} \cos\gamma\, \sin\theta_1
\frac{a r_g Q}{r_1^3},
\label{ngq}
\\
n^{\rm g}_{\rm ad}&=&
\frac{m_{\rm g} }{m_{\rm i} }
\frac{(m_{\rm g}-m_{\rm i}) l s }{2\pi \hbar r_1} \cos\gamma\sin\theta_1 \frac{a r_g^2}{r_1^3},
\label{ngad}
\\
n^{\rm e}_0 &=&
\frac{q l s}{\pi\hbar r_1}
\Bigl[
-\frac{P}{r_1}\sin\gamma
\! -\! \frac{a Q}{r_1^2}
(2\sin\gamma\cos\theta_1
\! +\! \cos\gamma\sin\theta_1)
\nonumber\\
&&
+\frac{P a^2}{8 r_1^2}\! \Bigl( 10 \sin\gamma+ 3\sin(\gamma \! -\! 2 \theta_1) 
+11\sin(\gamma 
\! +\! 2\theta_1)\Bigr)
\Bigr],
\\
n^{\rm e}_G &=&
-\frac{q l s}{2\pi\hbar r_1}
\cos\gamma\, \sin\theta_1\, 
\frac{a r_g Q}{r_1^3},
\label{neGG}
\\
n^{\rm f}&=&
\frac{l s}{\pi \hbar r_1  }
\Bigl[ 
\frac{1}{2} \cos\gamma\sin\theta_1\, \Bigl( m_{\rm i}-m_{\rm g} -\frac{m_{\rm i}}{\sqrt{1-v^2}}\Bigr)
\frac{a r_g^2}{r_1^3}
 +q\cos\gamma\sin\theta_1\frac{a r_g Q}{r_1^3}
\Bigr].
\label{nff11}
\eea
We can see that $P$ is absent from \eqref{ngq}, though it appears in \eqref{delta q0 dyonU}. Besides, \eqref{ngad} does not contain $Q$ and $P$, though \eqref{delta g adU} contains them. Actually, the higher-order terms of $n^{\rm g}_q$ and $n^{\rm g}_{\rm ad}$ contain $ Q$ and $P$. Although in this paper we only consider the third- and lower-order terms, to avoid confusion, we write down the fourth-order terms of $n^{\rm g}_q$ and $n^{\rm g}_{\rm ad}$ as follows
\bea
(n^{\rm g}_q)_{\rm 4th}
&=& \frac{m_{\rm g}}{m_{\rm i}} \frac{q l s}{\pi \hbar r_1}
\Bigl[
-\frac{a^2 r_g P}{r_1^4}\sin\theta_1\, (\sin\theta_1\sin\gamma -2 \cos\theta_1\cos\gamma )
-\frac{a r_g^2 Q}{r_1^4}\cos\gamma\sin\theta_1
\nonumber\\
&&
+\frac{a Q P^2}{r_1^4}\cos\gamma\sin\theta_1
+\frac{a Q^3}{r_1^4}\cos\gamma\sin\theta_1
\Bigr],
\\
(n^{\rm g}_{\rm ad})_{\rm 4th}
&=&
\frac{m_{\rm g} }{m_{\rm i} }
\frac{(m_{\rm g}-m_{\rm i} ) l s }{\pi \hbar r_1} 
\cos\gamma\sin\theta_1\Bigl(
\frac{3 a r_g^3 }{4 r_1^4} 
-\frac{3 a r_g Q^2 }{2 r_1^4} 
-\frac{3 a r_g P^2 }{2 r_1^4} 
\Bigr).
\eea

Finally, the fringe shift of rotating the rectangle around AB from $
\gamma=0$ to $\gamma=\pi$ is 
\be
N=-\frac{\delta\phi (\gamma=0)}{\pi}.
\label{NdyonU}
\ee

\subsection{Deviation from the weak equivalence principle}
The above results give theoretical predictions for the phase differences and fringe shifts in the situation $m_{\rm g}\ne m_{\rm i}$. In principle, according to these formulas and the measurements, one can determine the deviation from the weak equivalence principle. 

Let us focus on the interference experiment in dyonic Kerr-Newman spacetime. For convenience, rewrite \eqref{deltaphiFF} as
\bea 
\delta\phi^{\rm f}_{\rm dyon}
&=&
m_{\rm i} \frac{l s}{ \hbar r_1} \Bigl\{
 -\frac{\cos\gamma}{4 v \sqrt{1-v^2}}\frac{r_g^2}{r_1^2}
+\frac{1}{16\sqrt{1-v^2}}
\Bigl[12\cos\gamma \frac{r_g (Q^2+P^2)}{ v r_1^3}
- 5\cos\gamma \frac{r_g^3}{ v r_1^3}
+ 8\cos\gamma \sin\theta_1  \frac{a r_g^2}{r_1^3}
\nonumber\\
&&
- 8v \sin\theta_1\,(2\cos\theta_1\sin\gamma +3\sin\theta_1\cos\gamma)
\frac{a^2 r_g}{r_1^3}
\Bigr]
\Bigr\}
\nonumber\\
&&+m_{\rm i}\Bigl(\frac{m_{\rm g}}{m_{\rm i}}-1\Bigr)
\frac{l s}{ \hbar r_1}
\Bigl\{
\frac{\cos\gamma}{16v}
\Bigl(
12 \frac{ P^2 r_g}{r_1^3}
- 5 \frac{r_g^3}{r_1^3}
+ 12 \frac{Q^2 r_g}{r_1^3}
+ 8 v\sin\theta_1 \frac{a r_g^2}{r_1^3}
\Bigr)
-\frac{1}{v}
\frac{\cos\gamma}{4}\frac{r_g^2}{r_1^2}
\Bigr\}
\nonumber\\
&&+
\frac{l s}{ \hbar r_1} \Bigl\{
\frac{q\cos\gamma}{2v} \frac{r_g Q}{r_1^2}
 +\frac{q}{8 v}\Bigl[
-4 \cos\gamma\frac{Q^3}{r_1^3}
-4 \cos\gamma\frac{Q P^2}{r_1^3}
+ 3\cos\gamma\frac{Q r_g^2}{r_1^3}
- 8 v \cos\gamma\sin\theta_1\frac{a r_g Q}{r_1^3}
\nonumber\\
&&
+ 4(\sin\gamma\sin\theta_1-2\cos\gamma\cos\theta_1)\frac{a r_g P}{r_1^3}
\Bigr]
\Bigr\}
\nonumber\\
&=&
m_{\rm i} K_0 +m_{\rm i}\Bigl(\frac{m_{\rm g}}{m_{\rm i}}-1\Bigr) K_{\rm ad} +K_q,
\label{deltaphiFF3}
\eea
where $K_0$, $K_{\rm ad}$, $K_q$ (which are independent of $m_{\rm g}$ and $m_{\rm i}$) denote the corresponding expressions in \eqref{deltaphiFF3}. Besides, rewrite the terms in the gravitational phase difference as
\be
\delta\phi^{\rm dyon}_0=m_{\rm g} f_0, 
\quad
\delta\phi^{\rm dyon}_{QP}=m_{\rm g} f_{QP}, 
\quad
\delta\phi^{\rm dyon}_q=\frac{m_{\rm g}}{m_{\rm i}} f_q, 
\quad
\delta\phi^{\rm dyon}_{\rm ad}=\frac{m_{\rm g}}{m_{\rm i}}(m_{\rm g}-m_{\rm i}) f_{\rm ad},
\ee
where $f_0$, $f_{QP}$, $f_q$, and $f_{\rm ad}$ (which are independent of $m_{\rm g}$ and $m_{\rm i}$) are given by the corresponding expressions of \eqref{delta 0 dyonU}, \eqref{delta Q dyonU}, \eqref{delta q0 dyonU}, and \eqref{delta g adU} respectively. Accordingly, the total phase difference is rewritten as
\be
\delta\phi
=m_{\rm g} f_0 
+m_{\rm g} f_{QP}
+\frac{m_{\rm g}}{m_{\rm i}} f_q
+\frac{m_{\rm g}}{m_{\rm i}}(m_{\rm g}-m_{\rm i}) f_{\rm ad}
+\delta\phi^{\rm e}_{\rm dyon}
+m_{\rm i} K_0 +m_{\rm i}\Bigl(\frac{m_{\rm g}}{m_{\rm i}}-1\Bigr) K_{\rm ad} +K_q.
\label{deltaphiNon}
\ee
According to \eqref{deltaphiNon}, if $\delta\phi$ is known, we can determine $m_{\rm g}$ by the given parameters (here $m_{\rm i}$ is regarded as a given parameter). What we really care about is how to determine the factor
\be
\xi= \frac{m_{\rm g}}{m_{\rm i}}-1.
\ee
It represents the deviation from the equivalence principle. For this purpose, rewrite \eqref{deltaphiNon} as
\bea
\delta\phi
&=&(\xi+1) [ m_{\rm i}( f_0 
+ f_{QP}) +f_q ]
+(\xi +1)\xi\, m_{\rm i} f_{\rm ad}
+\delta\phi^{\rm e}_{\rm dyon}
+m_{\rm i} K_0 +m_{\rm i}\xi K_{\rm ad} +K_q
\nonumber\\
&\approx& \xi [ m_{\rm i}( f_0 
+ f_{QP} +f_{\rm ad} +K_{\rm ad}  ) +f_q ]
+(\delta\phi)_{\rm universal},
\label{deltaphixi}
\eea
where we have assumed $|\xi| \ll 1$ and neglected its quadratic term, and here $(\delta\phi)_{\rm universal}$ is defined as
\be
(\delta\phi)_{\rm universal}=m_{\rm i}( f_0 + f_{QP}) +f_q 
+\delta\phi^{\rm e}_{\rm dyon}
+m_{\rm i} K_0 +K_q.
\ee
By \eqref{deltaphixi} we get
\be
\xi\approx \frac{\delta\phi-(\delta\phi)_{\rm universal}}{m_{\rm i}( f_0 
+ f_{QP} +f_{\rm ad} +K_{\rm ad}  ) +f_q}.
\label{deltaminusXi}
\ee

In an interference experiment, what we measure are fringe shifts. Therefore we need to find the relations between $\xi$ and the fringe shifts. Equation \eqref{deltaminusXi} gives us a hint. According to \eqref{NdyonU}, we have $N=-\pi^{-1}\delta\phi |_{\gamma=0}$. Hence \eqref{deltaminusXi} implies
\be
\xi\approx
\frac{N-N_{\rm universal}}{-\pi^{-1}\bigl( m_{\rm i} (f_0 + f_{QP}+f_{\rm ad} +K_{\rm ad} )+f_q \bigr)|_{\gamma=0} },
\label{Nxi}
\ee
where $N_{\rm universal}=-\pi^{-1} (\delta\phi)_{\rm universal}|_{\gamma=0}$.
Therefore, by means of \eqref{Nxi}, we can use the measurement of $N$ to determine the deviation from the weak equivalence principle.

We can also use the measurement of $n$ to determine the factor $\xi$. For this purpose, we can rewrite \eqref{nff11} as a form 
\be
n^{\rm f}= m_i U_0 +U_q -m_{\rm i}\Bigl(\frac{m_{\rm g}}{m_{\rm i}}-1
\Bigr) U_{\rm ad},
\ee
and rewrite the terms of $n^g$ in \eqref{nrmg0}, \eqref{nrmgQP}, \eqref{ngq}, and \eqref{ngad} as
\be
n^{\rm g}_0=m_{\rm g} J_0, 
\quad
n^{\rm g}_{QP}=m_{\rm g} J_{QP}, 
\quad
n^{\rm g}_q=\frac{m_{\rm g}}{m_{\rm i}} J_q, 
\quad
n^{\rm g}_{\rm ad}=\frac{m_{\rm g}}{m_{\rm i}}(m_{\rm g}-m_{\rm i}) J_{\rm ad},
\ee
where $U_0$, $U_q$, $U_{\rm ad}$, $J_0$, $J_{QP}$, $J_q$, and $J_{\rm ad}$ are independent of $m_{\rm g}$ and $m_{\rm i}$. Following the same derivations as those of \eqref{deltaminusXi}, we get
\be
\xi\approx \frac{n-n_{\rm universal}}{m_{\rm i}( J_0 
+ J_{QP} +J_{\rm ad} -U_{\rm ad} ) +J_q},
\label{nminusXi}
\ee
where 
\be
n_{\rm universal}
=m_{\rm i}( J_0 
+ J_{QP}) +J_q +n^{\rm e}_0+ n^{\rm e}_G
 + m_{\rm i} U_0 +U_q.
\ee
Therefore $\xi$ can also be determined by the measurement of $n$.

These formulas for $\xi$ also hold for Kerr-Newman spacetime,  Kerr spacetime, Reissner-Nordstr\"om spacetime, and Schwarzschild spacetime,\footnote{Note that \eqref{nminusXi} is not applicable to Reissner-Nordstr\"om spacetime and Schwarzschild spacetime, because its denominator equals to zero for non-rotating black holes.} except that we need to let the corresponding parameters of the black hole equal to zero.

%%%%%%%%%%
\section{Teleparallel gravity versus general relativity\label{comparison}}
Considering that the theory of TG is equivalent to general relativity (GR) in classical level~\cite{Aldrovandi:2013wha}, it is interesting to ask whether these theories are consistent in quantum aspects. A way is to compare the results of these theories in the quantum interference. In Ref.~\cite{Mo}, the authors showed that for Kerr spacetime, the gravitational phase difference evaluated in TG is not identical to the result based on GR (for the latter, see Ref.~\cite{Wajima:1996cu}), though they are consistent in the lowest order term. Inspired by it, in what follows we compare the phase difference \eqref{deltaphige} in Kerr-Newman spacetime with the result in Ref.~\cite{Kagramanova:2008bv}. Furthermore, in Sec.~\ref{comparison2} we explicitly show that in a generic weak gravitational field, the gravitational phase defined in TG does not completely coincide with that in GR.

\subsection{Comparison of phase differences\label{comparison1}}
In Ref.~\cite{Kagramanova:2008bv}, the quantum interference of charged particles is studied in Pleba\'nski–Demia\'nski spacetime, in the framework of GR. The parameters of such spacetime include: a mass-like parameter $M$, a rotation-like parameter $a$, an electric charge $Q$, a magnetic charge $P$, a NUT-like parameter $q_{\rm N}$, an acceleration-like parameter $\alpha$, a twist parameter $w$, and the cosmological constant $\Lambda$. It reduces to Kerr-Newman spacetime when only $M$, $a$, and $Q$ are included. Therefore, it is necessary to compare \eqref{deltaphige} with the result in Ref.~\cite{Kagramanova:2008bv}. 

In Ref.~\cite{Kagramanova:2008bv}, the phase difference between two paths are computed by the formula\footnote{Note that in Ref.~\cite{Kagramanova:2008bv}, they used the word ``phase shift'' instead of ``phase difference'' and set $\hbar=c=1$. Here we restore $\hbar$.}
\be
\delta \Phi=\frac{1}{\hbar}\Bigl(\int_1-\int_2 \Bigr) \bigl(p_\mu +q A_\mu -ig \check{A}_\mu \bigr) \dd x^\mu,
\label{phase shift dyonic}
\ee
where $p_\mu$ is the four-momentum of the particle, $q$ and $g$ are its electric and magnetic charges, with $A_\mu$ and $\check{A}_\mu$ the electromagnetic potential and the dual electromagnetic potential. Now we compare \eqref{deltaphige} with the result in GR. In terms of \eqref{phase shift dyonic}, the authors in Ref.~\cite{Kagramanova:2008bv} found the phase shift for a small interferometer as
\bea
\delta \Phi
&=&\frac{1}{\hbar}
\Bigl\{
\mathscr{E} S \Bigl[
-\frac{\mathscr{E}}{p_0} 
\bigl(\cos\beta \, a_{\hat{r}} 
-\cos\zeta \sin\beta \, a_{\hat{\theta}}
-\sin\zeta \sin\beta \, a_{\hat{\varphi}}
\bigr)
\nonumber\\
&&
-\frac{1}{p_0} \bigl(\cos\beta \, \partial_{\hat{r}} \mathscr{E}_{\rm pot}
-\cos\zeta \sin\beta \, \partial_{\hat{\theta}}\mathscr{E}_{\rm pot}
-\sin\zeta \sin\beta \, \partial_{\hat{\varphi}}\mathscr{E}_{\rm pot}
\bigr)
\nonumber\\
&&+\sin\beta \, \omega_{\hat{\theta}\hat{\varphi}}
+\cos\zeta \cos\beta \, \omega_{\hat{\varphi}\hat{r}}
+\sin\zeta \cos\beta \, \omega_{\hat{r}\hat{\theta}}
\Bigr]
\nonumber\\
&&
+q S \bigl(\sin\beta B_{\hat{r}} +\cos\zeta \cos\beta  B_{\hat{\theta}} +\sin\zeta \cos\beta B_{\hat{\varphi}} \bigr)
\nonumber\\
&&-g S \bigl(\sin\beta E_{\hat{r}} +\cos\zeta \cos\beta  E_{\hat{\theta}} +\sin\zeta \cos\beta E_{\hat{\varphi}} \bigr)
\Bigr\},
\label{PD spacetime}
\eea
where $\mathscr{E}$ is the total conserved energy of the particle, $S$ is the area of the interferometer, $p_0$ is the momentum at the beam splitter, $\beta$ is the tilt angle of the interferometer, $\zeta$ is the angle of the interferometer's baseline with respect to $e_{\hat{\varphi}}$, $a_{\hat{\mu}}$ is the acceleration of the Killing trajectories, $\mathscr{E}_{\rm pot}$ is a non-gravitational potential energy, $\omega_{\hat{\mu}\hat{\nu}}$ is a rotation parameter, $q$ and $g$ are the electric and magnetic charges of the particle, $E_{\hat{i}}$ and $B_{\hat{i}}$ are the components of the electromagnetic field, and $\partial_{\hat{\mu}}$ is defined as $\partial_{\hat{\mu}}=e^\nu_{\hat{\mu}}\partial_\nu$ with $e^\nu_{\hat{\mu}}$ the components of the tetrad.

To compare \eqref{PD spacetime} with our result in Kerr-Newman spacetime, we need to keep the parameters $M$, $a$, and $Q$ but let other parameters of the Pleba\'nski–Demia\'nski black hole equal to zero. We also need to let $g=0$ (we only consider electric charged particles) and $\zeta=0$ (because AB is perpendicular to $\vec{r}_1$ in FIG.~\ref{experimentKN}). Besides, following Ref.~\cite{Kagramanova:2008bv}, the black hole's parameters are assumed to be small and the phase shift is expanded up to the order of $O(M)$, $O(a)$, and $O(Q^2)$, and the particle is assumed to be non-relativistic so that $\mathscr{E}\approx m$ and $p_0 \approx m v_0$, with $v_0$ the matter wave's group velocity at the beam splitter. With these assumptions and the expressions given in Sec.~4 of Ref.~\cite{Kagramanova:2008bv}, we can reduce \eqref{PD spacetime} to
\bea
\delta \Phi^{\rm NR}
&\approx&
\frac{S}{\hbar}\frac{m}{r}\Bigl[ -\frac{\cos\beta}{v_0} \Bigl(\frac{M}{r} -\frac{Q^2}{r^2} \Bigr)
+\frac{Ma}{r^2} \bigl(2\sin\beta\cos\theta +\cos\beta\sin\theta \bigr) \Bigr]
\nonumber\\
&&+\frac{S}{\hbar}\frac{q}{r}\Bigl[ \frac{\cos\beta}{v_0}\Bigl( \frac{Q}{r}-\frac{Q M}{r^2}\Bigr)
- \frac{Q a}{r^2} \bigl(2\sin\beta\cos\theta +\cos\beta\sin\theta \bigr)
\Bigr],
\label{PD NonR}
\eea
for Kerr-Newman spacetime. Additionally, we notice that the phase shift in the Figure 2 of Ref.~\cite{Kagramanova:2008bv} is defined by $\delta\Phi=\Phi_{\rm I}-\Phi_{\rm II}$ (though not written explicitly, it is hinted by (63) of Ref.~\cite{Kagramanova:2008bv}), opposite to our definition. Therefore, for comparison, we reverse the sign of \eqref{PD NonR} to get 
\bea
\delta \Phi'^{\rm NR}
&=&
\frac{S}{\hbar}\frac{m}{r}\Bigl[ \frac{\cos\beta}{v_0} \Bigl(\frac{M}{r} -\frac{Q^2}{r^2} \Bigr)
-\frac{Ma}{r^2} \bigl(2\sin\beta\cos\theta +\cos\beta\sin\theta \bigr) \Bigr]
\nonumber\\
&&+\frac{S}{\hbar}\frac{q}{r}\Bigl[ -\frac{\cos\beta}{v_0}\Bigl( \frac{Q}{r}-\frac{Q M}{r^2}\Bigr)
+\frac{Q a}{r^2} \bigl(2\sin\beta\cos\theta +\cos\beta\sin\theta \bigr)
\Bigr].
\label{PD NonR prime}
\eea
To compare \eqref{deltaphige} with \eqref{PD NonR prime}, we take the non-relativistic approximation $(1-v^2)^{-1/2}\approx 1$, and only keep terms up to the order of $O(M)$, $O(a)$, and $O(Q^2)$, such that \eqref{deltaphige} reduces to
\bea
\delta\phi^{\rm nr}
&=&\frac{S_{\rm pa}}{\hbar}\frac{m}{r_1 } \Bigl[
\frac{\cos\gamma}{v}
\Bigl(\frac{M}{r_1}-\frac{Q^2}{r_1^2}\Bigr)
-\frac{2 M a}{r_1^2}\bigl(2 \sin\gamma\,\cos\theta_1+\cos\gamma\,\sin\theta_1
\bigr)
+\frac{2a Q^2}{r_1^3}\sin(\theta_1+\gamma)
-\frac{3 MQ^2}{2 v r_1^3}\cos\gamma
\Bigr]
\nonumber\\
&&+\frac{S_{\rm pa}}{ \hbar}\frac{q}{r_1} 
\Bigl[  -\frac{\cos\gamma}{v}
\frac{Q}{r_1}
+\frac{Q a}{r_1^2}
\bigl(2\sin\gamma\,\cos\theta_1 +\cos\gamma\,\sin\theta_1\bigr)
+\frac{M a Q}{r_1^3}\cos\gamma\,\sin\theta_1
\Bigr],
\label{deltaPhiNR}
\eea 
where $S_{\rm pa}=l s$ is the area of the rectangle in FIG.~\ref{experimentKN}. Obviously, our result \eqref{deltaPhiNR} is different from \eqref{PD NonR prime}, though they are consistent in the terms of $M/r_1$, $Q/r_1$, $Q^2/ r_1^2$, and $Q a/ r_1^2$. 

By the way, though the term of $Ma/r_1^2$ in \eqref{deltaPhiNR} differs from that in \eqref{PD NonR prime}, it coincides with the term in Ref.~\cite{Wajima:1996cu} (in Ref.~\cite{Mo} we proved such consistency), where the quantum interferometry on earth was studied. This term is caused by the Lense-Thirring effect~\cite{KUROIWA1993330,Mashhoon:1984fj}. Since both Refs.~\cite{Kagramanova:2008bv,Wajima:1996cu} are based on general relativity but predict different coefficients of $Ma/r_1^2$, they could not both be correct. We tried to trace the origin of such difference but failed, because we failed to reproduce the result of Ref.~\cite{Kagramanova:2008bv}. Nonetheless, we briefly introduce the ways they derived this term. For convenience, we denote such term by $\delta\Phi^{Ma}$. As for Ref.~\cite{Kagramanova:2008bv}, the term $\delta\phi^{Ma}$ originates from the part $\mathscr{E} \vec{\omega}\cdot \vec{S}$ of the phase shift (see (68) of Ref.~\cite{Kagramanova:2008bv}). This part generates the third line of \eqref{PD spacetime} and eventually reduces to $\delta\phi^{Ma}$ for non-relativistic case in Kerr-Newman spacetime. However, the components of the rotation parameter (namely $\omega_{\hat{r}\hat{\theta}}$, $\omega_{\hat{r}\hat{\varphi}}$, and $\omega_{\hat{\theta}\hat{\varphi}}$) are given directly in Ref.~\cite{Kagramanova:2008bv} without details of derivations. This is the reason why we failed to reproduce their result of such term. As for Ref.~\cite{Wajima:1996cu}, the phase shift is derived by constructing the Hamiltonian of Schr\"odinger equation in the earth's field up to the first post-Newtonian order. In Ref.~\cite{Wajima:1996cu}, the total phase shift is given by
\be
\delta\Phi=-\frac{1}{\hbar}\int H\dd t,
\ee
and $\delta\phi^{Ma}$ is found as
\be
\delta\phi^{Ma}=
-\frac{1}{\hbar}\int \frac{4MR^2}{5r^3}\vec{\Omega}\cdot \vec{L}
=\frac{2m}{5\hbar}\frac{r_g}{R}\Bigl[
\vec{\Omega}-3\Bigl(\frac{\vec{R}}{R}\cdot \vec{\Omega}\Bigr)\frac{\vec{R}}{R}
\Bigr]\cdot\vec{S},
\label{phiMa2}
\ee
where $M$, $R$, and $\vec{\Omega}$ are the mass, radius, and angular velocity of earth, with $\Omega=5a/(2R^2)$, and $\vec{L}$ is the angular momentum of the particle in flat space. In Ref.~\cite{Mo} we proved that the term of $Ma/r_1^2$ in \eqref{deltaPhiNR} is consistent with \eqref{phiMa2}. Actually the term
\be
\delta\phi^{Ma}=-\frac{S}{\hbar}\frac{m}{r_1 }\frac{2 M a}{r_1^2}\bigl(2 \sin\gamma\,\cos\theta_1+\cos\gamma\,\sin\theta_1
\bigr),
\label{phiMa3}
\ee 
can be derived from a perspective of Gravitoelectromagnetism based on general relativity. Now we show the argument. For simplicity, consider Kerr spacetime. Recalling \eqref{qmphase1}, namely $\phi=(m/\hbar)\int \dd s=-(1/\hbar)\int L \dd t$, the phase shift is given by $\delta\phi=-(1/\hbar)\oint L \dd t$. In Gravitoelectromagnetism, the gravitational field is assumed weak that $g_{\mu\nu}\approx \eta_{\mu\nu}+h_{\mu\nu}$, and the Lagrangian $L$ is given by~\cite{Mashhoon:2003ax}
\bea
L &=&-m \gamma_0^{-1}
-m \gamma_0 (1+v^2) \frac{1}{4}\bar{h}_{00}
-m\gamma_0 \frac{\dd x^i}{\dd t}\bar{h}_{0i}
\nonumber\\
&=&-m \gamma_0^{-1}
+m \gamma_0 (1+v^2) \Gamma
+2m\gamma_0 \frac{\dd x^i}{\dd t}A_i
\nonumber\\
&=&-m \gamma_0^{-1}
+m \gamma_0 (1+v^2) \Gamma
-2m\gamma_0\, \vec{v}\cdot\vec{\mathcal{A}},
\label{L GEM}
\eea
where $\gamma_0=(1-v^2)^{-1/2}$ is the Lorentz factor, and other quantities are defined by $\bar{h}_{\alpha\beta}=h_{\alpha\beta}-{h^\mu}_\mu \eta_{\alpha\beta}/2$, $\Gamma=-\bar{h}_{00}/4=-h_{00}/2$, and $A_i=-\bar{h}_{0i}/2=-h_{0i}/2$. As for Kerr spacetime, we have $h_{01}=h_{02}=0$ and $h_{03}=(a r_g \sin^2\theta) /r$ in spherical coordinates, such that $A_1=0$, $A_2=0$, and $A_3=-(a r_g \sin^2\theta) /(2r)$. In what follows, we show that \eqref{phiMa3} can be derived from the last term of \eqref{L GEM} in the non-relativistic case. For a non-relativistic particle, such term reduces to $L_2=-2m\vec{v}\cdot \vec{\mathcal{A}}$, such that the corresponding phase shift is
\be
\delta\phi_2
=\frac{1}{\hbar}\oint 2m\vec{\mathcal{A}}\cdot \dd \vec{r}
=\frac{2m}{\hbar}\oint (\nabla\times\vec{\mathcal{A}})\cdot \dd \vec{S}.
\label{phi2}
\ee
On the other hand, we have $\vec{\mathcal{A}}=\mathcal{A}_r \vec{e}_r+\mathcal{A}_\theta \vec{e}_\theta+\mathcal{A}_\varphi \vec{e}_\varphi$ such that  
\be
\vec{\mathcal{A}}\cdot \dd \vec{r}
=\mathcal{A}_r \dd r +\mathcal{A}_\theta r\dd\theta +\mathcal{A}_\varphi r\sin\theta \dd\varphi
=-A_i \dd x^i
=\frac{a r_g \sin^2\theta}{2r}  \dd\varphi,
\ee
which implies
\be
\mathcal{A}_r=0,
\qquad
\mathcal{A}_\theta=0,
\qquad
\mathcal{A}_\varphi=\frac{a r_g \sin\theta}{2r^2}.
\label{mathcalA}
\ee
Plugging \eqref{mathcalA} into \eqref{phi2}, we finally get
\bea
\delta\phi_2
&\approx&
\frac{2m}{\hbar}(\nabla\times\vec{\mathcal{A}})\cdot \vec{S}
\nonumber\\
&=&\frac{2m}{\hbar}
\Bigl(\frac{a r_g \cos\theta}{r^3} \vec{e}_r
+\frac{a r_g \sin\theta}{2r^3} \vec{e}_\theta
\Bigr)\cdot \Bigl(S\cos\bigl(\gamma+\frac{\pi}{2}\bigr)\vec{e}_r
+S\cos(\pi-\gamma)\vec{e}_\theta
\Bigr)
\nonumber\\
&=&-\frac{S}{\hbar}
\frac{m}{r}
\frac{2Ma}{r^2}
\bigl(2 \sin\gamma\,\cos\theta+\cos\gamma\,\sin\theta
\bigr).
\label{phi2result}
\eea
This result is consistent with \eqref{phiMa3}. In short, the term of $Ma/r_1^2$ in our result agrees with Ref.~\cite{Wajima:1996cu} and can  be independently reproduced using the gravitoelectromagnetic approximation. However, the origin of its disagreement with Ref.~\cite{Kagramanova:2008bv} remains unresolved. One should not attribute it to a difference between teleparallel gravity and general relativity.

\subsection{Comparison of gravitational phases\label{comparison2}}
In the above discussion we compared our work with Ref.~\cite{Kagramanova:2008bv}. However, as mentioned previously, the formula \eqref{phase shift dyonic} describes the total phase shift of the interference, not the gravitational phase. Now we focus on the latter. In what follows we compare the gravitational phase defined in TG with that in GR for a generic weak gravitational field, and show that they are not completely in accord with each other.

As for the gravitational phase in GR for a weak gravitational field, Stodolsky showed that it depends on the perturbation of the metric~\cite{Stodolsky:1978ks}, by expanding the proper time up to the first order of the perturbation. Hence, to compare it with the gravitational phase in TG, we need to find the relation between such perturbation and the gauge potential ${B^a}_\mu$. For this purpose, we can use the relations~\cite{Aldrovandi:2013wha}
\be
g_{\mu\nu}={h^a}_\mu {h^b}_\nu \eta_{ab},
\quad
{h^a}_\mu={e^a}_\mu+{B^a}_\mu, \label{ghheta}
\ee
where ${e^a}_\mu$ is the tetrad in the absence of gravity. In a weak gravitational field, the potential ${B^a}_\mu$ can be regarded as a perturbation of the tetrad. Besides, we denote the perturbation of the metric by $H_{\mu\nu}$, i.e.
\be
g_{\mu\nu}=\eta_{\mu\nu}+H_{\mu\nu},
\label{getaH}
\ee
where $\eta_{\mu\nu}$ is the Minkowski metric. Combining \eqref{getaH} with \eqref{ghheta}, we get
\be
H_{\mu\nu}
=\eta_{ab} ( {e^a}_\mu {B^b}_\nu + {B^a}_\mu {e^b}_\nu )+\eta_{ab}{B^a}_\mu {B^b}_\nu
=H^{(1)}_{\mu\nu}+H^{(2)}_{\mu\nu},
\label{H1H2}
\ee
where $H^{(1)}_{\mu\nu}$ and $H^{(2)}_{\mu\nu}$ denote respectively the first- and the second-order terms of the gauge potential. Now that $H_{\mu\nu}$ contains $O(B)$ and $O(B^2)$, in what follows we expand the gravitational phases up to the second order of ${B^a}_\mu$.

Therefore, differently from Ref.~\cite{Stodolsky:1978ks}, we need to expand the proper time at least at the second order of $H_{\mu\nu}$, i.e.
\be
\dd s 
\approx
\dd s_0 +\frac{1}{2} H_{\mu\nu} U^\mu \dd x^\nu -\frac{1}{8} H_{\alpha\beta} H_{\mu\nu} U^\alpha U^\beta U^\mu\dd x^\nu,
\label{ds approx}
\ee
with $\dd s_0^2=\eta_{\mu\nu}\dd x^\mu \dd x^\nu$ and $U^\mu=\dd x^\mu/\dd s_0$. Accordingly, the gravitational phase in GR is
\bea
\phi^{\rm GR}
&=& 
\frac{m}{\hbar}\int \Bigl(\frac{1}{2} H_{\mu\nu} U^\mu \dd x^\nu -\frac{1}{8} H_{\alpha\beta} H_{\mu\nu} U^\alpha U^\beta U^\mu\dd x^\nu \Bigr)
\nonumber\\
&\approx& 
\frac{m}{\hbar}\int \Bigl(
\frac{1}{2} \bigl(H^{(1)}_{\mu\nu}+H^{(2)}_{\mu\nu}\bigr) U^\mu \dd x^\nu -\frac{1}{8} H^{(1)}_{\alpha\beta} H^{(1)}_{\mu\nu} U^\alpha U^\beta U^\mu\dd x^\nu
 \Bigr)
 \nonumber\\
&\approx& 
\frac{m}{\hbar}\int \Bigl(
{e^d}_\sigma +\frac{1}{2}{B^d}_\sigma 
-\frac{1}{4} H^{(1)}_{\alpha\beta} U^\alpha U^\beta {e^d}_\sigma
 \Bigr){B^a}_\nu\eta_{ad} U^\sigma \dd x^\nu,
\label{phiGR approx}
\eea
where we used $H^{(1)}_{\mu\nu}\dd x^\mu \dd x^\nu= 2{e^a}_\mu {B^b}_\nu \eta_{ab} \dd x^\mu \dd x^\nu$ in the last step. As for the gravitational phase in TG, recalling \eqref{phi B} we have 
\bea
\phi^{\rm TG}
&=&\frac{m}{\hbar}\int u_a B{^a}_\nu\dd x^\nu
 \nonumber\\
&=&\frac{m}{\hbar}\int 
\eta_{ad} {h^d}_\sigma U^\sigma \frac{\dd s_0}{\dd s} B{^a}_\nu \dd x^\nu.
\label{phiTG approx}
\eea
On the other hand, one can find
\be
\frac{\dd s_0}{\dd s} \approx 1-\frac{1}{2} H_{\mu\nu} U^\mu U^\nu 
+\frac{3}{8} H_{\alpha\beta} H_{\mu\nu} U^\alpha U^\beta U^\mu U^\nu.
\label{dds0dds}
\ee
Then inserting \eqref{dds0dds} and the second equation of \eqref{ghheta} into \eqref{phiTG approx}, we obtain
\be
\phi^{\rm TG}
\approx 
\frac{m}{\hbar}\int 
\Bigl(
{e^d}_\sigma +{B^d}_\sigma 
-\frac{1}{2} H^{(1)}_{\alpha\beta} U^\alpha U^\beta {e^d}_\sigma
 \Bigr){B^a}_\nu\eta_{ad} U^\sigma \dd x^\nu.
\label{phiTG approx2}
\ee
Comparing \eqref{phiTG approx2} with the last line of \eqref{phiGR approx}, we find that the first terms of them are identical but other terms are not the same. In summary, the gravitational phase \eqref{phiGR approx} in GR and the gravitational phase \eqref{phiTG approx} in TG coincide with each other at the first order of ${B^a}_\mu$, but they are different at higher orders.

%%%%%%%%%%
\section{Conclusions\label{summary}}
We found a proper tetrad for which the Lorentz connection vanishes, in Kerr-Newman spacetime, and used it to derive the gravitational gauge potential. Based on these results, we derived the gravitational phase. As an application, we studied a quantum interference experiment for charged particles in Kerr-Newman spacetime, in the region far away from the black hole. We calculated the gravitational phase difference, the electromagnetic phase difference, the free phase difference, and the fringe shifts. In the results, we found that the gravitational phase difference contains three parts $\delta\phi_0^g$, $\delta\phi_Q^g$, and $\delta\phi_q^g$ (see \eqref{delta g0}, \eqref{delta gQ}, and \eqref{delta gqq}). The term $\delta\phi_0^g$ corresponds to the gravitational phase difference of an uncharged black hole, while $\delta\phi_Q^g$ is contributed from the charge of Kerr-Newman black hole. Both $\delta\phi_0^g$ and $\delta\phi_Q^g$ are proportional to the particle's mass but independent of its charge. The term $\delta\phi_q^g$, which is proportional to the particle's charge but independent of its mass, stands for a mix of gravitation and electromagnetic interaction. It comes from the conservation relations \eqref{mathcalE} and \eqref{L given}. The electromagnetic phase difference contains two terms $\delta\phi_0^e$ and $\delta\phi_G^e$ (see \eqref{deltaPhie0} and \eqref{deltaPhieG}). They represent respectively a pure electromagnetic interaction and a mix of electromagnetism and gravitation. The term $\delta\phi_G^e$ originates from the gravitational modifications to time and length (recall the paragraph before \eqref{ADCminusABCf}). Afterwards, we discussed the angle dependences of the gravitational phase difference and the electromagnetic phase difference, and found that both their leading order terms are proportional to $\cos\gamma$ but independent of $\theta$. 
Furthermore, we compared the magnitudes of these phase differences.

We then extended the interference experiment to the case of dyonic black hole, in which we studied the phase differences and fringe shifts for electrically charged particles, dyonic particles, and uncharged massless particles respectively. We discussed the effect of the black hole's magnetic charge in the interference experiment. Besides, we showed that the Dirac string cannot be detected through the gravitational phase difference, and that if the Dirac quantization condition holds, such string also cannot be observed from the electromagnetic phase difference. Additionally, we proposed two gedanken experiments to test the Dirac quantization condition. We also discussed the situation when the universality of gravity breaks down, and found a method to determine the deviation from the weak equivalence principle. Then we compared the phase difference in teleparallel gravity with that in general relativity, for Kerr-Newman spacetime. Besides, for a generic weak gravitational field, we showed that the gravitational phase in teleparallel gravity does not completely coincide with the gravitational phase in general relativity, though they are indeed consistent in the lowest order term of the gauge potential.
 
In what follows, we will discuss: (I) the effect of the black hole's rotation in the interference experiment; (II) the observational aspects about the quantum interference; (III) the gravitoelectromagnetic analogies in terms of the phases.

Let us talk about the effect of the black hole's rotation in terms of the fringe shift~$n$. Firstly, for Reissner-Nordstr\"om spacetime, we find $n=0$ (let $a=0$ in \eqref{fringe shift n}). It is not surprising $n$ vanishes, considering that such spacetime is spherically symmetric and that the rectangle after flipping is the mirror image of the one before flipping (with the plane of $r_1$ and $r_2$ in FIG.~\ref{experimentKN} as the mirror). However, in Kerr-Newman spacetime ($a\ne 0$), as \eqref{fringe shift n} shows, the fringe shift $n$ is non-vanishing. The reason is that in such spacetime, the phases on the rectangle do not respect the above mirror symmetry. For example, because of the black hole's rotation, $A_\varphi$ and $S_\varphi$ are non-zero (see \eqref{dmudxmu} and \eqref{S beta2}), and $t_{AB}\ne t'_{AB}$ (see \eqref{t varphiNew}), such that the phase of AB is different from the phase after flipping. Therefore, the non-vanishing fringe shift $n$ in Kerr-Newman spacetime originates from the black hole's rotation. In other words, such fringe shift is an effect of the gravitational source's rotation.

Obviously such claim also holds for Kerr spacetime in which the fringe shift $n$ is just $n_0^g+n^f$ (recall \eqref{fringe shift n}). For non-relativistic particles in such spacetime, the fringe shift $n$ reduces to (in SI units)
\be
n_{\rm massive} \approx \frac{m c S}{\pi\hbar r_1 }\Bigl[
\frac{a r_g}{r_1^2}\bigl(
2\sin\gamma\,\cos\theta_1 +\cos\gamma\,\sin\theta_1
 \bigr)
 +\frac{a r_g^2}{r_1^3} \sin(\gamma+\theta_1) 
\Bigr],
\label{nonrelativistic}
\ee
with $a=J/(M c)$ and $r_g=2GM/c^2$, where $J$ and $M$ are the angular momentum and the mass of the gravitational source respectively, and $S$ is the area of the rectangle. As an example, let earth be the gravitational source and study the interference of neutrons. For simplicity, we let $\gamma=0$ and $\theta=\pi/2$, and let $r_1$ equals to the equatorial radius of earth. As for the mass of neutron, the parameters of earth (namely $M$, $J$, and $r_1$), and the physical constants in \eqref{nonrelativistic}, we take the values in Ref.~\cite{allen2000allen}. With these values, if we take $S=1\, {\rm m^2}$ for \eqref{nonrelativistic}, the fringe shift is
\be
n_{\rm neutron}\approx 1.7\times 10^{-7}.
\label{nmassive}
\ee
Such value is much smaller than one. According to \eqref{nonrelativistic}, to obtain a larger fringe shift, one can increase the area of the rectangle, e.g., we can take $S=10^7\, {\rm m^2}$ such that $n_{\rm neutron}\approx 1.7$. Besides, we can use more massive particles, e.g., use $^{87}{\rm Rb}$ atom~\cite{Overstreet:2021hea}. Let us also consider massless particles for the above example. Now the fringe shift $n$ is (let $Q=P=0$ in \eqref{fringe shift nDyon Massless})
\bea
n_{\rm massless}&=&
\frac{2 S}{\lambda r_1}\Bigl[
\frac{a r_g}{r_1^2}\bigl(
2\sin\gamma\,\cos\theta_1 +\cos\gamma\,\sin\theta_1
 \bigr)
+\frac{a r_g^2}{r_1^3} \sin(\gamma+\theta_1) 
\Bigr],
\label{relativistic}
\eea
with $\lambda$ the wavelength of the particle. Likewise, we let $\gamma=0$ and $\theta=\pi/2$. If we take $\lambda=1064\, {\rm nm}$ (e.g.,  Nd:YAG laser used for gravitational waves~\cite{LIGOScientific:2016aoc}) and $S=1\, {\rm m^2}$ for massless particles in the above example, the fringe shift is
\be
n_\lambda\approx 2.1\times 10^{-16}.
\label{nmassiveless}
\ee
This result is far less than \eqref{nmassive}. It is not surprising, because according to \eqref{nonrelativistic} and \eqref{relativistic}, we have $n_{\rm massless}/n_{\rm massive}=\hbar\omega/(mc^2)$, and we know that the energy of the photon in \eqref{nmassiveless} is much less than the energy of the neutron in \eqref{nmassive}. 

As shown above, the fringe shift $n$, which originates from the rotation of the black hole, is very small in the examples \eqref{nmassive} and \eqref{nmassiveless}. Considering these numerical values, one may worry whether the quantum interference effects are astronomically observable. Actually, if we measure the fringe shift $N$ (rotate the rectangle from $\gamma=0$ to $\gamma=\pi$) rather than $n$, the observation would be easier. As an example, considering massive particles in Kerr-Newman spacetime, the fringe shift up to the second order is (in SI units)
\bea
N&=&N^{\rm g} +N^{\rm e} +N^{\rm f}, 
\\
N^{\rm g}
&=&-\frac{1}{\pi}
\biggl\{
\frac{m c S}{ \hbar r_1 \sqrt{1-(v/c)^2}}
\Bigl[
\frac{c}{v}
\Bigl(\frac{r_g}{2r_1}+\frac{r_g^2}{2r_1^2}\Bigr)
-\frac{a r_g}{r_1^2}\sin\theta_1
-\frac{G Q^2}{4\pi \varepsilon_0 c^3 v r_1^2}
\Bigr]
-\frac{q Q S}{4\pi \varepsilon_0 \hbar r_1^2 v}
\frac{r_g}{2r_1}
\biggr\},
\label{Ng}
\\
N^{\rm e}
&=&-\frac{1}{\pi}
\frac{q S}{4\pi \varepsilon_0 \hbar r_1} \Bigl(
- \frac{1}{v}
\frac{Q}{r_1}
+\frac{1}{c} \frac{a Q}{r_1^2}\sin\theta_1
\Bigr),
\label{Ne}
\\
N^{\rm f}
&=&
-\frac{1}{\pi}
\biggl(
-\frac{m c S}{ 4\hbar r_1 \sqrt{1-(v/c)^2}}
\frac{c}{v}
\frac{r_g^2}{r_1^2}
+\frac{q Q S}{4\pi \varepsilon_0 \hbar r_1^2 v}
\frac{r_g}{2r_1}
\biggr),
\label{Nf}
\eea
where $N^{\rm g}$, $N^{\rm e}$, and $N^{\rm f}$ are contributed from $\phi^{\rm g}$, $\phi^{\rm e}$, and $\phi^{\rm f}$ respectively. Obviously, both $N^{\rm g}$ and $N^{\rm e}$ have terms of the first order. Hence, $N$ is easier to detect compared with $n$. Actually, in the non-relativistic limit, the first term of $N^{\rm g}$, namely $N^{\rm g}_1=-GMmS/(\pi\hbar r_1^2 v)$, coincides with the earlier result~\cite{Overhauser:1974} which was confirmed by the COW experiment~\cite{Colella:1975dq}. As for other terms, let us make comparisons of them with $N^{\rm g}_1$ in the non-relativistic limit:
\bea
&&
\frac{N^{\rm g}_2}{N^{\rm g}_1}=\frac{r_g}{r_1},
\qquad
\frac{N^{\rm g}_3}{N^{\rm g}_1}=-\frac{2v a}{c r_1}\sin\theta_1,
\qquad
\frac{N^{\rm g}_4}{N^{\rm g}_1}=-\frac{Q^2(4\pi\varepsilon_0 r_1)^{-1}}{Mc^2},
\qquad
\frac{N^{\rm g}_5}{N^{\rm g}_1}=-\frac{qQ(4\pi\varepsilon_0 r_1)^{-1}}{mc^2},
\\
&&
\frac{N^{\rm e}_1}{N^{\rm g}_1}=-\frac{qQ (4\pi\varepsilon_0)^{-1}}{GMm},
\quad
\frac{N^{\rm e}_2}{N^{\rm g}_1}=\frac{qQ (4\pi\varepsilon_0)^{-1}}{GMm} \frac{v a}{c r_1}\sin\theta_1,
\quad
\frac{N^{\rm f}_1}{N^{\rm g}_1}=-\frac{r_g}{2r_1},
\quad
\frac{N^{\rm f}_2}{N^{\rm g}_1}=\frac{qQ(4\pi\varepsilon_0 r_1)^{-1}}{mc^2}.
\eea
The ratios $N^{\rm g}_2/N^{\rm g}_1$, $N^{\rm g}_3/N^{\rm g}_1$, and $N^{\rm f}_1/N^{\rm g}_1$ are obviously far less than one. Both $N^{\rm g}_4/N^{\rm g}_1$ and $N^{\rm g}_5/N^{\rm g}_1$ look like the ratio between a Coulomb potential energy and a rest energy (of the black hole or the particle). They are often very small, considering that $r_1$ is very large. Especially, we have $N^{\rm g}_4/N^{\rm g}_5=(Q/M)/(q/m)$. The terms $N^{\rm g}_4$ and $N^{\rm g}_5$ cancel out when $Q/M=-q/m$. The quantity $N^{\rm e}_1/N^{\rm g}_1$ looks like the ratio between the Coulomb potential energy and Newtonian potential energy. This ratio is the same as the non-relativistic limit of \eqref{ratioge} in Sec.~\ref{brief discussion}, where we discussed a numerical example for  non-relativistic electrons. Finally, the ratio $N^{\rm e}_2/N^{\rm g}_1$ is obviously very small.

Though the predicted phase differences and fringe shifts in this paper remain to be tested, there are many progresses in the observation of quantum interferences. We mention some of them concerning the gravitationally induced interferences. Differently from the COW experiment with earth as the gravitational source~\cite{Overhauser:1974}, the phase shift induced by the gravity of a kilogram-scale source mass was observed in Ref.~\cite{Overstreet:2021hea}. As for massless particles, in 2017, Hilweg  {\em et al.} proposed that the gravitationally induced phase shift of two different paths for a photon, namely $\Delta\phi_{\rm g}=2\pi A g N_c / (\lambda c^2)$ predicted by Ref.~\cite{Zych:2012ut}, can be measured using an optical fiber interferometer~\cite{Hilweg:2016oxz}. Here $A$ is the area enclosed by the paths, $g$ is the gravitational acceleration, $\lambda$ is the wavelength, and $N_c$ is the group effective index of the medium for light. In 2024, Polini {\em et al.} established a large-scale fiber interferometry aiming to detect such phase shift~\cite{Polini:2024krm}.  When $N_c=1$, the phase shift $\Delta\phi_{\rm g}$ is exactly the first term of our result \eqref{delta 0 dyonMassless} at $\gamma=0$ (in SI units) taking earth as the gravitational source, namely $(\delta\phi^{\rm g}_0)_1=( l s \omega r_g)/(2 r_1^2 c)$ (also predicted in Ref.~\cite{Mo}). We hope that the experiment~\cite{Polini:2024krm} can also test the higher order terms of \eqref{delta 0 dyonMassless} in the future. About the weak equivalent principle, there is also an interesting progress, i.e., a proposal to test such principle using the atom interferometer of the Space-Time Explorer and QUantum Equivalence Space Test (STE-QUEST)~\cite{Schubert:2013gux,Aguilera:2013uua}. In terms of the description in Ref.~\cite{Altschul:2014lua} for such proposal, two cooled isotopes of Rb are released into free fall and simultaneously subject to an interferometer, and three laser pulses coherently split, reverse, and recombine the wave packets of them. The interferences reflect the difference between the accelerations of the two isotopes, if the weak equivalent principle is violated. We hope that such experimental scheme is also applicable to testing our results in Sec.~\ref{Non-Universal}. Also read the review~\cite{Yuan:2023evh} to learn more progresses of using quantum interferences to test the weak equivalence principle. Finally, we mention that there are some progresses in searching for dark matter and dark energy by quantum interferences~\cite{Geraci:2016fva,Arvanitaki:2016fyj,Hamilton:2015zga,Elder:2016yxm,Chiow:2019fti}.

%%%%%
Now we talk about the gravitoelectromagnetic analogies. Firstly we discuss the analogy between a rotating mass and a rotating electric charge, in terms of the phases. For this purpose, we directly compare $S_\beta$ of a Kerr black hole with $A_\beta$ of a Kerr-Newman black hole, considering that $S_\beta$ and $A_\beta$ directly relate to the gravitational phase and the electromagnetic phase respectively. Letting $Q=0$ in \eqref{S beta2}, one can find that the non-vanishing $S_\varphi$ of Kerr black hole is caused by the rotation of its mass. On the other hand, \eqref{dmudxmu} shows that the rotation of a Kerr-Newman black hole's electric charge leads to a non-zero $A_\varphi$. In this sense, a rotating mass resembles a rotating electric charge. Likewise, it is appealing to image a rotating ``gravitomagnetic charge'' which resembles a rotating magnetic charge. For a rotating magnetically charged black hole, the electromagnetic potential is given by (let $Q=0$ in \eqref{dmudxmu2})
\be
A_\mu\dd x^\mu
=-P\Bigl[ \frac{a}{\rho^2}\cos\theta\, \dd t +\Bigl( 1-\frac{r^2+a^2}{\rho^2}\cos\theta\Bigr)\dd\varphi
\Bigr].
\label{AmuP}
\ee
Both $A_t$ and $A_\varphi$ are non-zero, but the former is induced by the rotation of the magnetic charge. Analogously, as for a rotating black hole which has a gravitomagnetic charge, we guess that its $S_\mu$ function has similar properties, i.e., both $S_t$ and $S_\varphi$ are non-zero, and the former is caused by the rotation of the gravitomagnetic charge. Such conjecture needs to be examined in future works.

Finally, we would like to mention that ``gravitomagnetic charge'' is not a new concept. As a parameter, it appears in the well known Kerr-Taub-NUT metric \cite{Miller:1973hqu}, which describes the gravitational field of a rotating black hole which has a gravitomagnetic charge \cite{Narzilloev:2023btg}. One extension of our work is to study the interference experiment in FIG.~\ref{experimentKN} for Kerr-Taub-NUT spacetime and compare the role of the black hole's gravitomagnetic charge with the role of a dyonic Kerr-Newman black hole's magnetic charge. Though it is a new project to be done, we have some conjectures for the results. Firstly, we expect that the gravitomagnetic charge contributes to the gravitational phase difference and manifests non-zero effect in the fringe shifts. Besides, we expect that the fringe shift $n$ is non-vanishing even when the gravitomagnetic black hole is non-rotating. If these conjectures are verified, such experiment may be a way to test whether a black hole has a gravitomagnetic charge. Another extension of this work is to study the quantum interference in the spacetime of a rotating black hole with an anisotropic matter field~\cite{Kim:2019hfp,Kim:2021vlk}. Such family of black hole solutions were found by Kim {\em et al.} who generalized the Kerr-Newman solution to include an anisotropic matter. Besides, Lee {\em et al.} found that in such spacetime the inclusion of anisotropic matter has an effect on the shadow cast observables~\cite{Lee:2021sws}. Similarly, if we consider a quantum interference experiment in this spacetime, the anisotropic matter may also influences the results of such experiment. It is an interesting subject for future works.

%%%%%%%%%%%
\begin{acknowledgments}
I would like to thank Ming-Lei Xiao for encouraging me to think more about magnetic monopoles and the electromagnetic potential of a dyonic black hole.
\end{acknowledgments}

%%%%%%%%
\appendix

\section{The expression of the tetrad\label{derive tetrad}}
In this appendix we derive the expression \eqref{b matrix} of Sec.~\ref{KN}. Firstly, a tetrad of Kerr-Newman spacetime is given by~\cite{Arcos:2004tzt}
\be
(H{^a}_\mu)=
\begin{pmatrix}
\gamma_{00} & 0 & 0 &\eta\\
0& \gamma_{11}\sin\theta \cos\varphi & \gamma_{22}\cos\theta\cos\varphi &-\beta\sin\varphi\\
0&  \gamma_{11}\sin\theta\sin\varphi & \gamma_{22}\cos\theta\sin\varphi &\beta\cos\varphi\\
0& \gamma_{11}\cos\theta & -\gamma_{22}\sin\theta &0
\end{pmatrix},
\label{tetrad Kerr}
\ee
with $\eta$, $\beta$, $\gamma_{00}$, and $\gamma_{jj}$ defined as
\be
\eta=g_{03} \gamma_{00}^{-1},
\qquad
\beta=\sqrt{\eta^2-g_{33}},
\qquad
\gamma_{00}=\sqrt{g_{00}},
\qquad
\gamma_{jj}=\sqrt{-g_{jj}},
\label{some quantities}
\ee 
and the inverse tetrad is
\be
(H_a{^\mu})=
\begin{pmatrix}
\gamma_{00}^{-1} & 0 & 0 &0\\
\beta g^{03}\sin\varphi & \gamma_{11}^{-1}\sin\theta \cos\varphi & \gamma_{22}^{-1}\cos\theta\cos\varphi &-\beta^{-1}\sin\varphi\\
-\beta g^{03}\cos\varphi &  \gamma_{11}^{-1}\sin\theta\sin\varphi & \gamma_{22}^{-1}\cos\theta\sin\varphi &\beta^{-1}\cos\varphi\\
0& \gamma_{11}^{-1}\cos\theta & -\gamma_{22}^{-1}\sin\theta &0
\end{pmatrix}.
\label{inverse tetrad0}
\ee
Note that in Ref.~\cite{Arcos:2004tzt}, the equation $\beta^2=\eta^2-g_{33}$, rather than $\beta=\sqrt{\eta^2-g_{33}}$, is given. But we think the latter also holds, because in the case $M=a=Q=0$, the expression \eqref{tetrad Kerr} should reduce to the transformation between the spherical coordinates and the cartesian coordinates. As for $H_a{^\mu}$, checking the relation $\eta_{ab}=g_{\mu\nu} H_a{^\mu}H_b{^\nu}$ (see Ref.~\cite{Aldrovandi:2013wha}), we find that there are some typos in (177) of Ref.~\cite{Arcos:2004tzt}, hence we modify it as \eqref{inverse tetrad0}. 

Unfortunately,  the Lorentz connection of \eqref{tetrad Kerr} does not vanish. The argument is as follows. According to Ref.~\cite{Krssak:2018ywd}, the Lorentz connection is computed by
\be
\dot{A}{^a}_{b\mu}=\frac{1}{2} H^c_{(\text{r})\mu}\Bigl[
{{f_b}^a}_c(H_{(\text{r})})
+{{f_c}^a}_b(H_{(\text{r})})
-{f^a}_{bc}(H_{(\text{r})})
\Bigr],
\label{A formula}
\ee
with $H^c_{(\text{r})\mu}$ the ``reference tetrad'', where the subscript ``$(\rm r)$'' means letting the gravitational constant equal to zero, and the function ${f^a}_{bc}$ is defined as
\be
{f^a}_{bc}
=H_b{^\mu}H_c{^\nu} (\partial_\nu H^a{_\mu}-\partial_\mu H^a{_\nu}).
\label{fabc}
\ee
To use the formula \eqref{A formula}, we need to find the reference tetrad firstly. Considering that in SI units, the mass and the charge in \eqref{rgrho} should be replaced by $GM/c^2$ and $\sqrt{G}Q/(c^2\sqrt{4\pi\varepsilon_0})$ respectively, we can directly let $M=Q=0$ in \eqref{tetrad Kerr} to get the reference tetrad
\be
(H^a_{(\rm r)\mu})=
\begin{pmatrix}
1 & 0 & 0 &0\\
0& \frac{\rho}{\rho_0}\sin\theta \cos\varphi & \rho \cos\theta\cos\varphi &-\rho_0\sin\theta\sin\varphi\\
0&  \frac{\rho}{\rho_0}\sin\theta\sin\varphi & \rho\cos\theta\sin\varphi &\rho_0\sin\theta\cos\varphi\\
0& \frac{\rho}{\rho_0}\cos\theta & -\rho\sin\theta &0
\end{pmatrix},
\label{tetrad KerrR}
\ee
where $\rho$ and $\rho_0$ are defined in \eqref{rgrho} and \eqref{rho0} respectively. Then replacing \eqref{tetrad KerrR} into \eqref{A formula}, we can find that the Lorentz connection has non-vanishing components, e.g., $\dot{A}{^1}_{31}=a^2\sin\theta\cos\theta\cos\varphi \rho_0^{-1}\rho^{-2}$. 

The reason why ${H^a}_\mu$ has a non-vanishing Lorentz connection is that it is a tetrad of non-proper frame. Let us prove such statement. Firstly, as Ref.~\cite{Aldrovandi:2013wha} shows, the Lorentz connection of a tetrad in a generic frame is $\dot{A}{^a}_{b\mu}={\Lambda^a}_c(x) \partial_\mu {\Lambda_b}^c(x)$, where $\Lambda$ is the Lorentz matrix which transforms a proper frame to this generic frame. As for a tetrad of non-proper frame, $\Lambda$ cannot be global, hence the Lorentz connection is non-vanishing. Therefore, we just need to prove that ${H^a}_\mu$ is a tetrad of non-proper frame. For this purpose, let us consider a cartesian coordinates $K'$ firstly. The metric in $K'$ can be derived from
\be
g'_{\mu'\nu'}=[(J^T)^{-1} g J^{-1}]_{\mu'\nu'},
\label{g prime}
\ee
where $J$ is given in \eqref{Jacobi}. And according to the transformation~\cite{Pereira:2001xf}
\be
{H^a}_\mu={H'^a}_{\mu'}\frac{\partial x^{\mu'} }{\partial x^\mu},
\label{h transformation}
\ee
the tetrad in $K'$ can be found by
\be
{H'^a}_{\mu'}={(H J^{-1})^a}_{\mu'},
\label{h prime}
\ee
where $g'$, $g$, $H$, and $H'$ are corresponding matrices. The results of \eqref{g prime} and \eqref{h prime} are very long, so we would not show them here. Nonetheless, it is useful to show the results when the gravitational constant is set to be zero, as follows
\bea
g'_{({\rm r})}&=&{\rm diag}(1,-1,-1,-1),\\
H'_{({\rm r})}&=&
\begin{pmatrix}
1 & 0 & 0 &0\\
0& \zeta\cos^2 \varphi+ 
 \sin^2 \varphi\,
& (\zeta-1)\sin\varphi\cos\varphi &
\kappa\cos\varphi\\
0&  (\zeta-1)\sin\varphi\cos\varphi 
& \cos^2 \varphi+ \zeta \sin^2 \varphi
 &\kappa\sin\varphi\\
0& -\kappa\cos\varphi & -\kappa\sin\varphi &\zeta
\end{pmatrix},
\label{h prime matrix}
\eea
where
\be
\zeta=\frac{r \sin^2 \theta  + \rho_0 \cos^2 \theta}{\rho},
\quad
\kappa=\frac{(\rho_0 - r) \sin 2\theta}{2\rho}.
\ee
We can see that the spacetime reduces to Minkowski spacetime. If we can prove that $H'_{({\rm r})}$ is a tetrad of non-inertial frame, the tetrad $H'$ of course belongs to a non-proper frame, then we can claim that $H$ is a tetrad of non-proper frame (because the coordinates transformation between $H'$ and $H$ does not change the reference frame). As shown in Ref.~\cite{Krssak:2015rqa}, two tetrads of different frames are related by a (local) Lorentz transformation, e.g.,\footnote{Generally, the matrix $\Lambda$ in \eqref{HhTr} is local (i.e., it is point-dependent), except that both ${H^a}_\mu$ and ${h^b}_\mu$ are tetrads of proper frame.}
\be
{H^a}_\mu={\Lambda^a}_b\, {h^b}_\mu.
\label{HhTr}
\ee
Accordingly, if we can find a Lorentz matrix $\Lambda'$ which relates $H'_{({\rm r})}$ with a tetrad $e'$ of inertial frame, and show that $\Lambda'$ is a local transformation, we can say that $H'_{({\rm r})}$ is a tetrad of non-inertial frame. On the other hand, in Minkowski spacetime, for a tetrad of inertial frame, we can use cartesian coordinates such that it takes a simple form~\cite{Aldrovandi:2013wha}
\be
{e'^a}_{\mu'}={\delta^a}_{\mu'}.
\label{e prime mu prime}
\ee
Hence the matrix $\Lambda'$ of the equation $H'^a_{({\rm r})\mu'}={\Lambda'^a}_b \, {e'^b}_{\mu'}$ just takes a form same as the matrix \eqref{h prime matrix}. Obviously, the matrix \eqref{h prime matrix} depends on the coordinates. Thus, $\Lambda'$ also depends on the coordinates, i.e., it is a local transformation. Therefore, the tetrad $H'^a_{({\rm r})\mu'}$ corresponds to a non-inertial frame, which implies that ${H'^a}_{\mu'}$ corresponds to a non-proper frame. Therefore, ${H^a}_{\mu}$ is a tetrad of non-proper frame. We have proved the above statement. 

In calculating the gravitational phase (recall Sec.~\ref{GEphase}), what we need is a tetrad ${h^a}_\mu$ of proper frame, instead of ${H^a}_\mu$ which belongs to a non-proper frame. Nonetheless, according to \eqref{HhTr}, we can derived ${h^a}_\mu$ from the tetrad ${H^a}_\mu$, provided we have found the matrix $\Lambda$. To find $\Lambda$, we can consider the case when the gravitational constant is set to zero, then we have $H^a_{({\rm r})\mu}={\Lambda^a}_b\, {e^b}_\mu$, where ${e^b}_\mu=h^b_{({\rm r})\mu}$ is a tetrad of inertial frame. In such case the task becomes how to find ${e^b}_\mu$, so that we can find ${\Lambda^a}_b$. For this purpose, we just need to transform ${e'^a}_{\mu'}$ (see \eqref{e prime mu prime}) to a tetrad of Boyer-Lindquist coordinates, according to \eqref{h transformation}. The result is
\be
({e^a}_{\mu})=({e'^a}_{\mu'} {J^{\mu'}}_{\mu})
=\begin{pmatrix}
1 & 0 & 0 &0\\
0& \frac{r}{\rho_0}\sin\theta \cos\varphi &\rho_0 \cos\theta\cos \varphi &-\rho_0\sin\theta\sin\varphi\\
0& \frac{r}{\rho_0}\sin\theta \sin\varphi &\rho_0 \cos\theta\sin \varphi &\rho_0\sin\theta\cos\varphi\\
0&\cos\theta &-r\sin\theta &0
\end{pmatrix}.
\label{e inertial}
\ee
Hence, the Lorentz matrix in the equation $H^a_{(\rm r)\mu}={\Lambda^a}_b\, {e^b}_\mu$ is (as for $H^a_{(\rm r)\mu}$, see \eqref{tetrad KerrR})
\be
({\Lambda^a}_b )
=
\begin{pmatrix}
1 & 0 & 0 &0\\
0& \zeta\cos^2 \varphi+ 
 \sin^2 \varphi\,
& (\zeta-1)\sin\varphi\cos\varphi &
\kappa\cos\varphi\\
0&  (\zeta-1)\sin\varphi\cos\varphi 
& \cos^2 \varphi+ \zeta \sin^2 \varphi
 &\kappa\sin\varphi\\
0& -\kappa\cos\varphi & -\kappa\sin\varphi &\zeta
\end{pmatrix}.
\label{Lambda matrix}
\ee
One can find that the expression \eqref{Lambda matrix} is the same as \eqref{h prime matrix}. This is because according to \eqref{h prime}, $H_{(\rm r)}=\Lambda e$, \eqref{e inertial}, and \eqref{e prime mu prime}, we have
\be
H'_{(\rm r)}=H_{(\rm r)} J^{-1}
=\Lambda e  J^{-1}
=\Lambda (e' J)  J^{-1}
=\Lambda e'
=\Lambda.
\ee
Now that the matrix $\Lambda$ has been found, we can find $h{^a}_\mu$. According to \eqref{HhTr}, we have 
\be
h{^a}_\mu={ (\Lambda^{-1})^a}_c\, H{^c}_\mu.
\label{b transformation}
\ee
Substituting \eqref{tetrad Kerr} and \eqref{Lambda matrix} into \eqref{b transformation}, we derive \eqref{b matrix} of Sec.~\ref{KN}. One can check that the Lorentz connection of $h{^a}_\mu$ indeed vanishes, by inserting \eqref{b matrix} into \eqref{A formula}.

%%%%%%%
\section{Formulas for the phase differences\label{formula derivation}}
Here we derive the formulas \eqref{delta phi0g} and \eqref{delta phi0EM}. We would like to mention that \eqref{delta phi0g} was derived in Ref.~\cite{Mo} for Kerr spacetime. But for safety, we re-derive it for a more general function. 

Consider an integral $\Phi=\int f_\mu (r,\theta, u^\alpha)\, \dd x^\mu$ along the sides of the rectangle in FIG.~\ref{experimentKN}. Recalling the assumption (a) in Sec.~\ref{interference}, we have
\bea
\Phi_{AB}
&\approx& \bigl[f_0^A (t_{B}-t_{A} ) +f_3^A (\varphi_{B}-\varphi_{A})\bigr]_{AB},
\label{phaseAB}
\\
\Phi_{DC}
&\approx& \bigl[f_0^D (t_{C}-t_{D} ) +f_3^D (\varphi_{C}-\varphi_{D}) \bigr]_{DC},
\label{phaseDC}
\\
\Phi_{AD}
&\approx&
\bigl[ f_0^{p_0} (t_{D}-t_{A})
+f_1^{p_1} (r_2-r_1) +f_2^{p_2} (\theta_2-\theta_1) +f_3^{p_3} (\varphi_{D}-\varphi_{A})
\bigr]_{AD}
\nonumber\\
&\approx& 
\bigl[ f_0^A (t_{D}-t_{A})
+f_1^A (r_2-r_1) +f_2^A (\theta_2-\theta_1) +f_3^A (\varphi_{D}-\varphi_{A})
\bigr]_{AD},
\label{phaseAD}
\\
\phi_{BC}
&\approx& 
\bigl[ f_0^{q_0}  (t_{C}-t_{B})
+f_1^{q_1} (r_2-r_1) +f_2^{q_2} (\theta_2-\theta_1) +f_3^{q_3} (\varphi_{C}-\varphi_{B} )
\bigr]_{BC}
\nonumber\\
&\approx& 
\bigl[ f_0^B  (t_{C}-t_{B})
+f_1^B (r_2-r_1) +f_2^B (\theta_2-\theta_1) +f_3^B (\varphi_{C}-\varphi_{B} )
\bigr]_{BC},
\label{phaseBC}
\eea 
where the superscripts denote the points, and we used the mean value theorem for $\Phi_{AD}$ and $\Phi_{BC}$. The last step of $\Phi_{AD}$ holds because $(f_\mu^{p}-f_\mu^A)$ are smaller than or equal to $O(l/r_1)$ and $x^\mu_D-x^\mu_A$ are in the order $O(l/r_1)$ (recall that we neglect $O(l^2/r_1^2)$ in this paper). Therefore we get 
\bea
\Phi_{DC}-\Phi_{AB}
&=&\bigl[f_0^D t_{AB} +f_3^D \varphi_{AB}\bigr]_{DC}
-\Phi_{AB}
+\bigl[f_0^D (t_{DC}-t_{AB}) +f_3^D (\varphi_{DC}-\varphi_{AB})\bigr]_{DC},
\label{DCminusAB}
\\
\Phi_{AD}-\Phi_{BC}
&=&
\bigl[f_0^A (t_{D}-t_{A})+ f_1^A (r_2-r_1)+f_2^A (\theta_2-\theta_1)+f_3^A (\varphi_{D}-\varphi_{A})
\bigr]_{AD}
\nonumber\\
&&-\bigl[f_0^B (t_{C}-t_{B})+ f_1^B (r_2-r_1)+f_2^B (\theta_2-\theta_1)+f_3^B (\varphi_{C}-\varphi_{B} )
\bigr]_{BC},
\label{ADminusBC}
\eea
where $x^\mu_{p_1 p_2}$ denotes $x^\mu_{p_2}-x^\mu_{p_1}$. Combining the formulas \eqref{DCminusAB} and \eqref{ADminusBC}, we get
\bea
\Phi_{ADC}-\Phi_{ABC}
&=&\Phi_{DC}-\Phi_{AB} +\Phi_{AD}-\Phi_{BC}
\nonumber\\
&=&\delta\Phi_a +\delta\Phi_b +\delta\Phi_c,
\label{ADCminusABC}
\eea
where 
\bea
\delta\Phi_a
&=&\bigl[f_0^D t_{AB} +f_3^D \varphi_{AB}\bigr]_{DC}
-\Phi_{AB},
\label{deltaPhia}
\\
\delta\Phi_b
&=&
(f_0^D)_{DC} (t_{DC}-t_{AB})
+(f_0^A)_{AD} t_{AD}
-(f_0^B)_{BC} t_{BC},
\label{deltaPhib}
\\
\delta\Phi_c
&=&
(f_3^D)_{DC}(\varphi_{DC}-\varphi_{AB})
+\bigl[ f_1^A (r_2-r_1)+f_2^A (\theta_2-\theta_1)+f_3^A (\varphi_{D}-\varphi_{A})
\bigr]_{AD}
\nonumber\\
&&
-\bigl[ f_1^B (r_2-r_1)+f_2^B (\theta_2-\theta_1)+f_3^B (\varphi_{C}-\varphi_{B} )
\bigr]_{BC}.
\label{deltaPhic}
\eea

The formula \eqref{ADCminusABC} is applicable to the function $S_\mu$ and $F_\mu$, because they can be written as $S_\mu (r,\theta, u^\alpha)$ and $F_\mu (r,\theta, u^\alpha)$ in Kerr-Newman spacetime. The formula \eqref{ADCminusABC} can be simplified if $f=f(r,\theta)$. We will apply the latter to the electromagnetic potential. Now apply \eqref{ADCminusABC} to $S_\mu$. 
Firstly, we know $S_0^B=S_0^A$ (recall \eqref{S beta2}), by which we can simplify \eqref{deltaPhib}. Then, we know $S_2=0$ (recall \eqref{S beta2}). Furthermore, for the rectangle ABCD, we know
\be
(v^\varphi)^A_{AD}\approx 0,
\qquad
(v^\varphi)^B_{BC}\approx 0,
\label{assumption}
\ee
hence we have $\mathcal{L}^A_{AD}=\mathcal{L}^B_{BC}$ (recall \eqref{L given}), such that $(S_1^A)_{AD}=(S_1^B)_{BC}$ and $(S_3^A)_{AD}=(S_3^B)_{BC}$ (recall \eqref{S beta2} and \eqref{P1}), by which we can simplify \eqref{deltaPhic}. With the above simplifications, we find from \eqref{ADCminusABC} the formula for $S_\mu$: 
\be
\Phi^g_{ADC}-\Phi^g_{ABC}
=\delta\Phi_a^g +\delta\Phi_b^g +\delta\Phi_c^g,
\label{ADCminusABC S}
\ee
where 
\bea
\delta\Phi_a^g
&=&\bigl[S_0^D t_{AB} +S_3^D \varphi_{AB}\bigr]_{DC}
-\Phi_{AB}^g,
\label{deltaPhiaS}
\\
\delta\Phi_b^g
&=&
[(S_0^D)_{DC}-(S_0^B)_{BC}] (t_{DC}-t_{AB}),
\label{deltaPhibS}
\\
\delta\Phi_c^g
&=&
[(S_3^D)_{DC}-(S_3^B)_{BC}] (\varphi_{DC}-\varphi_{AB}).
\label{deltaPhicS}
\eea
Therefore, the formula \eqref{delta phi0g} for the gravitational phase difference holds.

Then we apply \eqref{ADCminusABC} to the electromagnetic potential $A_\mu$. Firstly, for a function $f_\mu (r,\theta)$, we have $f_\mu^A=f_\mu^B$, such that \eqref{ADCminusABC} is simplified to
\be
\Phi_{ADC}-\Phi_{ABC}
=\delta\Phi_a +\delta\Phi_b +\delta\Phi_c,
\label{ADCminusABC EM2}
\ee
where 
\bea
\delta\Phi_a
&=&\bigl[f_0^D t_{AB} +f_3^D \varphi_{AB}\bigr]_{DC}
-\Phi_{AB},
\label{deltaPhiaEM2}
\\
\delta\Phi_b
&=&
[(f_0^D)_{DC}-(f_0^B)_{BC}] (t_{DC}-t_{AB}),
\label{deltaPhibEM2}
\\
\delta\Phi_c
&=&
[(f_3^D)_{DC}-(f_3^B)_{BC}] (\varphi_{DC}-\varphi_{AB}).
\label{deltaPhicEM2}
\eea
Obviously, the electromagnetic potential is independent of the four-velocity $u^\alpha$. Besides, in Kerr-Newman spacetime or dyonic Kerr-Newman spacetime, it is independent of $t$ and $\phi$ (see \eqref{dmudxmu} and \eqref{dmudxmu2}). In short, we can write the electromagnetic potential as $A_\mu (r,\theta)$. Therefore, the formula \eqref{ADCminusABC EM2} applying to $A_\mu (r,\theta)$ yields
\be
\Phi^{\rm EM}_{ADC}-\Phi^{\rm EM}_{ABC}
=\delta\Phi_a^{\rm EM} +\delta\Phi_b^{\rm EM} +\delta\Phi_c^{\rm EM},
\label{ADCminusABC EM}
\ee
where 
\bea
\delta\Phi_a^{\rm EM}
&=&\bigl[A_0^D t_{AB} +A_3^D \varphi_{AB}\bigr]_{DC}
-\Phi_{AB}^{\rm EM},
\label{deltaPhiaEM}
\\
\delta\Phi_b^{\rm EM}
&=&
[(A_0^D)_{DC}-(A_0^B)_{BC}] (t_{DC}-t_{AB}),
\label{deltaPhibEM}
\\
\delta\Phi_c^{\rm EM}
&=&
[(A_3^D)_{DC}-(A_3^B)_{BC}] (\varphi_{DC}-\varphi_{AB}).
\label{deltaPhicEM}
\eea
Thus, the formula \eqref{delta phi0EM} of the electromagnetic phase difference holds.

Finally, we apply the formula \eqref{ADCminusABC} to $F_\mu$. Similar to the case of $S_\mu$, in terms of \eqref{F beta020} we know $F_0^A=F_0^B$, by which we can simplify \eqref{deltaPhib}. Then according to \eqref{F beta020}, \eqref{P1}, \eqref{PP2}, and $\mathcal{L}^A_{AD}=\mathcal{L}^B_{BC}$ (recall the sentence after \eqref{assumption}), we have $(F_1^A)_{AD}=(F_1^B)_{BC}$, $(F_2^A)_{AD}=(F_2^B)_{BC}$, and $(F_3^A)_{AD}=(F_3^B)_{BC}$. With these conditions, we can simplify \eqref{deltaPhic}. Finally, we find the formula for $F_\mu$:
\be
\Phi^{\rm free}_{ADC}-\Phi^{\rm free}_{ABC}
=\delta\Phi_a^{\rm free} +\delta\Phi_b^{\rm free} +\delta\Phi_c^{\rm free},
\label{ADCminusABC free}
\ee
where 
\bea
\delta\Phi_a^{\rm free}
&=&\bigl[F_0^D t_{AB} +F_3^D \varphi_{AB}\bigr]_{DC}
-\Phi_{AB}^{\rm free},
\label{deltaPhia free}
\\
\delta\Phi_b^{\rm free}
&=&
[(F_0^D)_{DC}-(F_0^B)_{BC}] (t_{DC}-t_{AB}),
\label{deltaPhib free}
\\
\delta\Phi_c^{\rm free}
&=&
[(F_3^D)_{DC}-(F_3^B)_{BC}] (\varphi_{DC}-\varphi_{AB}).
\label{deltaPhic free}
\eea
Consequently, we obtain the formula \eqref{ADCminusABCf} of the free phase difference.

%%%%%%%
\section{The energy and momentum in constant gravitational field\label{observation}}
In this appendix, we derive the expressions of $t_{AB}$, $\varphi_{AB}$, $\mathcal{E}$, and $\mathcal{L}$ given in Sec.~\ref{interference}. As for the first two quantities, we can take directly the result (73) in Ref.~\cite{Mo}. But here we would like to discuss them for a generic constant gravitational field (a constant gravitational field means that we can always find a frame in which the metric does not depend on the time coordinate \cite{Landau:1975pou}), then apply the results to Kerr-Newman spacetime to find these quantities. 

As Ref.~\cite{Landau:1975pou} shows, in a constant gravitational field, the four-velocity of a massive particle is
\be
u^0=\frac{1}{\sqrt{1-v^2} } \Bigl(\frac{1}{\sqrt{g_{00} }} -\frac{g_{0k}}{g_{00}}v^k\Bigr),
\qquad
u^j=\frac{v^j}{\sqrt{1-v^2} },
\label{u0uj}
\ee
where $v^j$ is defined in the first equation of \eqref{v components}, with the velocity $v$ defined by
\be
v^2=\Gamma_{ij}v^i v^j,
\label{vL}
\ee
and the proper length $L$ is defined by (as shown in Ref.~\cite{Landau:1975pou}, only when $g_{\mu\nu}$ is independent of time, the integral $\int \dd L$ is valid to define a length)
\be
\dd L^2=\Gamma_{ij}\dd x^i \dd x^j,
\label{vL2}
\ee
where $\Gamma_{ij}$ is defined in the second equation of \eqref{v components}. To find the accumulated time along a path, let us seek for the relation between $\dd x^0$ and $\dd L$. For this purpose, dividing $\dd L^2$ by $\dd s^2$, we obtain
\be
\frac{\dd L^2}{\dd s^2}=\Gamma_{ij} u^i u^j
=\Gamma_{ij}\frac{v^i v^j}{1-v^2}
=\frac{v^2}{1-v^2},
\label{dLds}
\ee
where \eqref{vL2}, \eqref{u0uj}, and \eqref{vL} have been used. Then according to the first equation of \eqref{u0uj} and \eqref{dLds}, we get
\be
\frac{\dd t}{\dd L}
=\frac{u^0}{\dd L/\dd s}
=\frac{1}{v} \Bigl(\frac{1}{\sqrt{g_{00} }} -\frac{g_{0k}}{g_{00}}v^k\Bigr)
\label{dtdL}.
\ee
Now we discuss the covariant four-momentum for a particle in the constant gravitational field. As for $P_0$, plugging \eqref{u0uj} into $P_0=g_{0\nu} P^\nu$ results in~\cite{Landau:1975pou}
\be
P_0=\frac{m\sqrt{g_{00}}}{\sqrt{1-v^2}}.
\label{P00}
\ee
While for $P_j$, inserting \eqref{u0uj} into $P_j=g_{j\nu} P^\nu$, and using the second equation of \eqref{v components} and \eqref{P00}, we derive
\bea
P_j
&=&\frac{m}{\sqrt{g_{00}(1-v^2)}}
(g_{0j}-\Gamma_{jk}v^k\sqrt{g_{00}})
\nonumber\\
&=&
\frac{P_0}{g_{00}} (g_{0j}-\Gamma_{jk}v^k\sqrt{g_{00}}).
\label{Pjj}
\eea

Now we apply these results to Kerr-Newman spacetime to derive \eqref{t varphi}, \eqref{t varphi2}, \eqref{mathcalE}, and \eqref{L given} in Sec.~\ref{interference}. Discuss $t_{AB}$ firstly. For a path of the particle, we rewrite \eqref{dtdL} in Kerr-Newman spacetime as
\be
\frac{\dd t}{\dd L}
=\frac{1}{v(\lambda) } \Bigl(\frac{1}{\sqrt{g_{00}(r(\lambda),\theta(\lambda) ) }} -\frac{g_{03}(r(\lambda),\theta(\lambda) ) }{g_{00}(r(\lambda),\theta(\lambda) ) }v^\varphi (\lambda) \Bigr),
\label{IntL}
\ee
where $\lambda$ is the affine parameter along the path. In what follows we prove that the right hand side of \eqref{IntL} is a constant along the path AB. Firstly, along AB, the coordinates $r$ and $\theta$ keep constant (the assumption (a) in Sec.~\ref{interference}). Secondly, because of the conserved energy \eqref{mathcalE}, the velocity $v$ also keeps constant on AB. As for $v^\varphi$, with the same way in Ref.~\cite{Mo} (i.e., applying \eqref{vL} on this path), one can find the $\varphi$ component of the velocity at a point of AB, as follows
\be
(v^\varphi)_1=\sqrt{\frac{v^2}{\Gamma_{33}} },
\label{vphiGamma}
\ee
which is also constant. Therefore, the right hand side of \eqref{IntL} is constant along AB. Hence on this path we get
\be
\dd t=  \frac{1}{v} \Bigl(\frac{1}{\sqrt{g_{00} }} -\frac{g_{03}}{g_{00}}(v^\varphi)_1 \Bigr) \dd L.
\label{vdtdL}
\ee
Integrating \eqref{vdtdL} we obtain
\be
t_{AB} = \frac{1}{v} \Bigl(\frac{1}{\sqrt{g_{00} }} -\frac{g_{03}}{g_{00}}(v^\varphi)_1\Bigr) s .
\label{vts}
\ee
After that, substituting \eqref{vphiGamma} into \eqref{vts}, the expression of $t_{AB}$ in \eqref{t varphi} is found. While for $\varphi_{AB}$ in \eqref{t varphi2}, as Ref.~\cite{Mo} shows, it is derived by using \eqref{vL2} on the path AB. The conserved energy \eqref{mathcalE} is obtained by combining \eqref{conserved quantities1} with \eqref{P00}. As for the conserved momentum,  inserting \eqref{Pjj} into \eqref{conserved quantities}, we get
\be
\mathcal{L}=\frac{P_0}{g_{00}}\bigl(
-g_{03} +\Gamma_{33}\, v^\varphi  \sqrt{g_{00}}
\bigr) 
-q A_3,
\label{mathcalLL}
\ee
where $\Gamma_{31}=\Gamma_{32}=0$ has been used (as for $\Gamma_{ij}$, see the second equation of \eqref{v components}). Then combining \eqref{conserved quantities1} with \eqref{mathcalLL}, the expression~\eqref{L given} is found.

\section{Phase differences in gravitation without universality\label{without}}

In this appendix we derive the equation of motion \eqref{EOMwithoutUn}, the gravitational  phase difference \eqref{delta phi gW0}, the electromagnetic phase difference \eqref{deltaPhi ee}, and the free phase difference \eqref{deltaphiFF1} in the case that the universality of gravitation breaks down (namely $m_{\rm g} \ne m_{\rm i}$, where $m_{\rm g}$ is the gravitational mass and $m_{\rm i}$ is the inertial mass). In such case, the action for a charged particle in the presence of gravity and electromagnetic field is given by~\cite{Aldrovandi:2013wha}
\be
S=-\int \Bigl[ m_{\rm i}  u_a (\partial_\beta x^a+\dot{A}{^a}_{b\beta}x^b ) +m_{\rm g} u_a B{^a}_\beta
+q A_\beta \Bigr] \dd x^\beta.
\label{S action}
\ee

We derive the equation \eqref{EOMwithoutUn} following a way as the one in Sec.~11.3.2 of Ref.~\cite{Aldrovandi:2013wha}. Firstly, rewrite \eqref{S action} as
\be
S=-m_{\rm i}\int \Bigl[   u_a {h^a}_\beta +\Bigl( \frac{m_{\rm g}}{m_{\rm i}}-1\Bigr) u_a B{^a}_\beta
+\frac{q}{m_{\rm i}} A_\beta \Bigr] \dd x^\beta.
\label{S action2}
\ee
The variation of \eqref{S action2} produces
\be
\frac{\dd u_a}{\dd s} -\dot{A}{^b}_{a\rho} u_b u^\rho 
-\frac{q}{m_{\rm i} } h{_a}^\mu F_{\mu\nu} u^\nu
=-\dot{K}{^b}_{a\rho} u_b u^\rho +F_a,
\label{duds0}
\ee
where $\dot{K}{^b}_{a\rho}$ is the contortion (see (1.63) of Ref.~\cite{Aldrovandi:2013wha}), and $F_a$ is
\be
F_a=\Bigl(1-\frac{m_{\rm g} }{m_{\rm i} } \Bigr) h{_a}^\sigma \Bigl[ ( {\delta^\rho}_\sigma-u^\rho u_\sigma) {B^b}_\rho \frac{\dd u_b}{\dd s} -(\partial_\sigma {B^b}_\rho -\partial_\rho {B^b}_\sigma ) u_b u^\rho
\Bigr].
\ee
Raising the index of $u_a$ in \eqref{duds0}, one gets
\be
\frac{\dd u^a}{\dd s} +\dot{A}{^a}_{b\rho} u^b u^\rho 
-\frac{q}{m_{\rm i} } \eta^{ab}  h{_b}^\rho F_{\rho\nu} u^\nu
=\dot{K}{^a}_{b\rho} u^b u^\rho +\eta^{ab} F_b.
\label{duds}
\ee
Then combining \eqref{duds} with the identity (4.66) of Ref.~\cite{Aldrovandi:2013wha}, yields
\be
\frac{\dd u^\mu}{\dd s}
+( \dot{\Gamma}{^\mu}_{\nu\rho} -\dot{K}{^\mu}_{\nu\rho} ) u^\nu u^\rho
=
\frac{q}{m_{\rm i}} \eta^{ab}h{_a}^\mu h{_b}^\rho F_{\rho\nu} u^\nu
+h{_c}^\mu \eta^{ca}F_a,
\label{duGamma}
\ee
where $\dot{\Gamma}{^\mu}_{\nu\rho}$ is the Weitzenb\"ock connection defined in (4.65) of Ref.~\cite{Aldrovandi:2013wha}. Finally, using the relation $\Gamma{^\mu}_{\nu\rho}=\dot{\Gamma}{^\mu}_{\nu\rho}-\dot{K}{^\mu}_{\nu\rho}$ of Ref.~\cite{Aldrovandi:2013wha} and the identity $g^{\mu\sigma}=\eta^{ca}h{_c}^\mu h{_a}^\sigma$ (which follows from $g_{\alpha\beta}=\eta_{ab}h{^a}_\alpha h{^b}_\beta$, by raising the spacetime indices of the tetrad using the spacetime metric~\cite{Carroll:2004st}) for \eqref{duGamma},  we get \eqref{EOMwithoutUn}. 

In what follows we derive the phase differences \eqref{delta phi gW0} and \eqref{deltaPhi ee}. Similar to \eqref{phi B}, the gravitational phase corresponding to the action \eqref{S action} is
\be
\phi_{\rm g}=\frac{m_{\rm g} }{\hbar}\int  u_a B{^a}_\beta\dd x^\beta,
\label{phi B2}
\ee
while the electromagnetic phase is still given by \eqref{EMphase}, i.e.
\be
\phi_{\rm e}=\frac{q}{\hbar} \int  A_\beta \dd x^\beta.
\label{EMphase2}
\ee
Comparing \eqref{phi B2} with \eqref{phi B}, it seems that the former is not different from the latter, except needing to replace $m$ with $m_{\rm g}$. However, considering that the equation of motion for a particle in non-universal gravitation (see \eqref{EOMwithoutUn}) is different from the one in general relativity, these phase differences may be different.

Before studying the phase differences, let us study the gravitational phase \eqref{phi B2} firstly. It can be written as
\be
\phi_{\rm g}
=\frac{1}{\hbar} \int  S_\beta \dd x^\beta,
\qquad
S_\beta=m_{\rm g}  u_\nu B{^\nu}_\beta,
\label{phi B3}
\ee
where $B{^\nu}_\beta={h_a}^\nu B{^a}_\beta$. Recall that we always choose a proper frame, as we did in Sec.~\ref{GEphase}. Same as the derivation for \eqref{S beta}, plugging \eqref{B expression} into the second equation of \eqref{phi B3}, yields
\be
(S_\beta)
=m_{\rm g}\begin{pmatrix}
u_0(1-\frac{\rho}{\chi})\\
u_1(1-\frac{\chi_0}{\rho_0})\\
0\\
u_0 \frac{a(2Mr-Q^2)\rho_0}{\rho\chi\chi_0} \sin^2\theta +u_3(1-\frac{\rho_0\chi}{\rho\chi_0})
\end{pmatrix}.
\label{S beta02}
\ee
Like \eqref{conserved quantities1} and \eqref{conserved quantities} in Sec.~\ref{KN}, we want to know whether there are conserved quantities related to $u_0$ and $u_3$, such that we can express them according to the conserved quantities. Let us start from the Euler-Lagrange equation
\be
\frac{\dd }{\dd \lambda} \frac{\partial L}{\partial \dot{x}^\mu }
=\frac{\partial L }{\partial x^\mu},
\label{Euler-Lagrange}
\ee
where $\dot{x^\mu}=dx^\mu /d\lambda$ and $\lambda$ is the affine parameter. In terms of \eqref{Euler-Lagrange}, if the Lagrangian $L$ is independent of a coordinate component $x^\epsilon$, the component $\partial L/\partial \dot{x}^\epsilon$ is conserved. For this purpose, we rewrite \eqref{S action2} as
\be
S=\int L \dd \lambda,
\ee
where the Lagrangian is
\bea
L&=&-m_{\rm i}\Bigl[
\frac{\dd s}{\dd \lambda}
+\Bigl(\frac{m_{\rm g}}{m_{\rm i}}-1 \Bigr) g_{\nu\sigma} \dot{x}^\sigma\dot{x}^\beta {B^\nu}_\beta \frac{\dd \lambda}{\dd s} 
+\frac{q }{m_{\rm i} } A_\beta\dot{x}^\beta
\Bigr]
\nonumber\\
&=&
-m_{\rm i}\Bigl[
(g_{\rho\gamma}\dot{x}^\rho \dot{x}^\gamma)^{1/2}
 \! + \! \Bigl(\frac{m_{\rm g}}{m_{\rm i}} \! - \! 1 \Bigr) g_{\nu\sigma} \dot{x}^\sigma\dot{x}^\beta {B^\nu}_\beta 
(g_{\rho\gamma}\dot{x}^\rho \dot{x}^\gamma)^{-1/2}
 \! + \! \frac{q }{m_{\rm i} } A_\beta \dot{x}^\beta
\Bigr].
\label{Lagrange}
\eea
Taking the derivative of \eqref{Lagrange} we obtain the generalized momentum
\be
\pi_\mu=\frac{\partial L}{\partial \dot{x}^\mu}
= -m_{\rm i}\Bigl[
u_\mu
+\Bigl(\frac{m_{\rm g}}{m_{\rm i}}-1 \Bigr)
( g_{\nu\mu} {B^\nu}_\beta u^\beta
+{B^\nu}_\mu u_\nu
-{B^\nu}_\beta u_\nu u^\beta u_\mu)
+\frac{q }{m_{\rm i} } A_\mu
\Bigr].
\label{pimu}
\ee
Because $g_{\mu\nu}$, ${B^\nu}_\beta$, and $A_\mu$ do not depend on $t$ and $\varphi$ (see \eqref{metric}, \eqref{B expression}, and \eqref{dmudxmu}), the Lagrangian \eqref{Lagrange} is also independent of $t$ and $\varphi$, namely
\be
\frac{\partial L}{\partial t}=0,
\qquad
\frac{\partial L}{\partial \varphi}=0.
\ee 
Therefore, according to \eqref{Euler-Lagrange}, both the generalized momentums $\pi_t$ and $\pi_\varphi$ are constant. Denote them as
\be
\pi_t=-\mathcal{E},
\qquad
\pi_\varphi=\mathcal{L}.
\label{EL2}
\ee
Up to now, we have found two conserved quantities related with the four-velocity. However, according to \eqref{pimu}, the relation between the generalized momentum and the four-velocity is non-linear, which makes it very difficult to express $u_0$ and $u_3$ by the conserved quantities. To eliminate the nonlinearity, we assume $v\ll 1$ so that \eqref{u0uj} reduces to
\be
u^0\approx \frac{1}{\sqrt{g_{00} }},
\qquad
u^j\approx 0,
\label{u0uj2}
\ee
and we implement the approximations \eqref{u0uj2} to the second term of the square brackets of \eqref{pimu}, such that the generalized momentum simplifies to
\be
\pi_\mu
\approx
 -\Bigl(m_{\rm i}
u_\mu
+W_\mu+q A_\mu
\Bigr),
\label{pimu2}
\ee
where
\be
W_\mu 
= (m_{\rm g}-m_{\rm i} )
\Bigl[ g_{\nu\mu} {B^\nu}_0 \frac{1}{\sqrt{g_{00}} }
+ \frac{1}{\sqrt{g_{00}} } g_{\nu 0}{B^\nu}_\mu 
-\Bigl(\frac{1}{\sqrt{g_{00}} }\Bigr)^3 g_{\nu 0} g_{\mu 0} {B^\nu}_0 
\Bigr].
\label{Wmu}
\ee
Combining \eqref{pimu2} with \eqref{EL2}, we find
\bea
m_{\rm i} u_0 &=& \mathcal{E}-q A_0-W_0,
\label{miu0}\\
m_{\rm i} u_3 &=& -\mathcal{L}-q A_3-W_3.
\label{miu3}
\eea 
Then substituting \eqref{miu0} and \eqref{miu3} into \eqref{S beta02}, we get
\be
(S_\beta)
=\frac{m_{\rm g}}{m_{\rm i}} 
\begin{pmatrix}
(\mathcal{E}-q A_0-W_0)(1-\frac{\rho}{\chi})\\
m_{\rm i} u_1(1-\frac{\chi_0}{\rho_0})\\
0\\
(\mathcal{E}-q A_0-W_0) \frac{a(2Mr-Q^2)\rho_0}{\rho\chi\chi_0} \sin^2\theta 
-(\mathcal{L}+ q A_3 + W_3)(1-\frac{\rho_0\chi}{\rho\chi_0})
\end{pmatrix}.
\label{S beta03}
\ee
Comparing \eqref{S beta03} with \eqref{S beta2}, we can find that there is a factor $m_{\rm g}/m_{\rm i}$ in front of the matrix, and that $(\mathcal{E}-qA_0)$ and $(\mathcal{L}+qA_3)$ are replaced by $(\mathcal{E}-qA_0-W_0)$ and $(\mathcal{L}+qA_3+W_3)$ respectively. Therefore, we expect that the new gravitational phase difference has a form as $(m_{\rm g}/m_{\rm i}) \delta\phi^g +\delta\phi^{ad}$, where $\delta\phi^g$ is the gravitational phase difference \eqref{delta phi g},  and $\delta\phi^{ad}$ is an additional term proportional to $(m_{\rm g}/m_{\rm i})(m_{\rm g}-m_{\rm i})$.

Before proceeding, let us check whether some formulas we previously used to calculate the phase differences still hold in the case $m_{\rm g}\ne m_{\rm i}$. Firstly, because both $S_0$ and $S_3$ are independent of $t$ and $\varphi$ (recall \eqref{S beta03}), the formula \eqref{delta phi0g} is still valid. So does \eqref{phiAB0}. Additionally, we need to know whether the quantities $t_{AB}$, $t_{DC}$, $\varphi_{AB}$, and $\varphi_{DC}$ are changed (they appear in \eqref{delta phi ag}, \eqref{delta phi bg}, and \eqref{delta phi cg}). Previously, these quantities were derived in the Appendix~\ref{observation} which is closely related to the equations (84.4) to (84.7), (84.14), and (88.10) to (88.14) of Ref.~\cite{Landau:1975pou}. We emphasize that even in the case $m_{\rm g}\ne m_{\rm i}$, the above equations of Ref.~\cite{Landau:1975pou} still hold, if we still take the way of its Chap.~84 to define the distance and simultaneity, and still take the way of its Chap.~88 to define the velocity $v$. Accordingly, with a careful analysis, we find that even in the case $m_{\rm g}\ne m_{\rm i}$, the equations \eqref{u0uj} to \eqref{vts} still hold (here we need to replace $m$ with $m_{\rm i}$ in \eqref{P00} and \eqref{Pjj}). So does \eqref{v components}. Therefore, as for $t_{AB}$ and $\varphi_{AB}$, the formulas \eqref{t varphi} and \eqref{t varphi2} still hold. These formulas also apply to $t_{DC}$ and $\varphi_{DC}$ (except needing to replace $r_1$ with $r_2$, $\theta_1$ with $\theta_2$, and $v$ with $v_2$). While for $\mathcal{E}$ and $\mathcal{L}$, they are derived by combining \eqref{miu0} and \eqref{miu3} with \eqref{P00} and \eqref{Pjj}. The results are
\bea
&&\mathcal{E}=
\frac{m_{\rm i} \sqrt{g_{00}} }{\sqrt{1-v^2}} +q A_0 +W_0,
\label{mathcalEE0}
\\
&&\mathcal{L}=\frac{1 }{g_{00} }(\mathcal{E}-q A_0-W_0)\bigl(
-g_{03} +v^\varphi \Gamma_{33}  \sqrt{g_{00}}
\bigr) 
-q A_3-W_3.
\label{mathcalLL0}
\eea
By the way, the two relations in \eqref{r2 theta2} still hold (see (A4) and (A5) of Ref.~\cite{Mo} for the derivations).

In summary, even in the case $m_{\rm g}\ne m_{\rm i}$, the equations \eqref{delta phi0g} to \eqref{phiAB0}, \eqref{t varphi}, \eqref{t varphi2}, \eqref{v components}, \eqref{r2 theta2}, and \eqref{u0uj} to \eqref{vts} do not change. While the equations \eqref{mathcalE} and \eqref{L given} need to be modified as \eqref{mathcalEE0} and \eqref{mathcalLL0} respectively. And the expression \eqref{S beta2} for $S_\beta$ should be changed to \eqref{S beta03}. With these modifications, repeating the calculations in Sec.~\ref{interference}, the gravitational phase difference \eqref{delta phi gW0} is derived.

Now we study the electromagnetic phase difference. Firstly, the formula \eqref{delta phi0EM} is still valid. Nonetheless, we still need to know whether the quantities in this formula change. According to the above discussions, we know that the quantities $t_{AB}$, $\varphi_{AB}$, $t_{DC}$, and $\varphi_{DC}$ are not changed. Evidently, the electromagnetic potential is still given by \eqref{dmudxmu}. Therefore, the term $\delta\phi^e_a$ given by \eqref{delta phi agEM} also does not change. The modifications may appear in $\delta\phi^e_b$ (see \eqref{delta phi bgEM}), because $t_{DC}$ depends on $v_2$ which can be expressed as a function containing the factor $(m_{\rm g}-m_{\rm i})/m_{\rm i}$ (similar to the expression in \eqref{v2}). However, the value of $\delta\phi^e_b$ is negligible because it is of the order of $O(l^2/r_1^2)$. As for $\delta\phi^e_c$,  the modification does not appear in it, and it is also of the order of $O(l^2/r_1^2)$ (see \eqref{delta phi cgEM}). In summary, in our approximations ($\delta r\approx 0$ and $\delta \theta\approx 0$ along the
paths AB and DC, and $O(l^2/r_1^2)\approx 0$), the electromagnetic phase difference does not change. It is still given by \eqref{delta e}.

Finally, we deal with the free phase. According to \eqref{S action}, we have
\bea
\phi_{\rm f}+\phi_{\rm i}
\, =\frac{m_{\rm i}}{\hbar}\int  u_a (\partial_\beta x^a+\dot{A}{^a}_{b\beta}x^b ) \dd x^\beta
\,  =\frac{m_{\rm i}}{\hbar}\int  u_a ({h^a}_\beta-{B^a}_\beta ) \dd x^\beta
\, =\frac{m_{\rm i}}{\hbar}\int  u_\beta \dd x^\beta
-\frac{m_{\rm i}}{ m_{\rm g}}\phi_{\rm g},
\label{free phase2}
\eea
where $\phi_{\rm g}=(m_{\rm g}/\hbar)\int  u_a {B^a}_\beta \dd x^\beta$ is the gravitational phase (recall \eqref{phi B2}). Recall that in this paper we have chosen a proper frame such that $\dot{A}{^a}_{b\beta}=0$ (see Sec.~\ref{QMphase}). Accordingly, the inertial phase $\phi_{\rm i}$ vanishes. Of course, \eqref{free phase2} also holds in such case. Plugging \eqref{S beta02} into this equation, we get
\be
\phi_{\rm f}
=\frac{1}{\hbar} \int F_\beta \dd x^\beta,
\ee
where
\be
(F_\beta)
=m_{\rm i}\begin{pmatrix}
u_0 \frac{\rho}{\chi}\\
u_1 \frac{\chi_0}{\rho_0} \\
u_2\\
u_0 \frac{a(Q^2-2Mr)\rho_0}{\rho\chi\chi_0} \sin^2\theta +u_3 \frac{\rho_0\chi}{\rho\chi_0}
\end{pmatrix}
=\begin{pmatrix}
(\mathcal{E}-q A_0-W_0)\frac{\rho}{\chi}\\
m_{\rm i} u_1 \frac{\chi_0}{\rho_0}\\
m_{\rm i} u_2\\
(\mathcal{E}-q A_0-W_0) \frac{a(Q^2-2Mr)\rho_0}{\rho\chi\chi_0} \sin^2\theta 
-(\mathcal{L}+ q A_3 + W_3)\frac{\rho_0\chi}{\rho\chi_0}
\end{pmatrix},
\label{F beta02}
\ee
where $W_\mu$ is given by \eqref{Wmu}. With this expression, repeating the calculations in Sec.~\ref{interference}, we can find the free phase difference \eqref{deltaphiFF1}.

\section{The electromagnetic field for dyonic black hole\label{EMpotential}}
A usual way to derive the electromagnetic field is using the equation (here we adopt the convention in Ref.~\cite{Landau:1975pou})
\be
F_{\mu\nu}=\partial_\mu A_\nu-\partial_\nu A_\mu.
\label{FmunuAA}
\ee
However, as mentioned in Sec.~\ref{massive particles}, the potential in \eqref{dmudxmu2} is singular at $\theta=\pi$, such that \eqref{FmunuAA} fails on the negative semi-axis $z$, and a way to resolve such difficulty is introducing Dirac's string and modifying \eqref{FmunuAA} as \eqref{field string}. 

Nonetheless, \eqref{FmunuAA} is still valid in the regions which avoid the singularities. Let us find the electromagnetic field in such regions. Using \eqref{FmunuAA} and the potential in \eqref{dmudxmu2}, we find
\be
F_{\mu\nu}
=
\begin{pmatrix}
0&\Omega_1 &\Omega_2 a&0\\
-\Omega_1&0&0& \Omega_1 a\sin^2\theta\\
-\Omega_2 a&0&0&\Omega_2\rho_0^2\\
0&- \Omega_1 a\sin^2\theta&-\Omega_2\rho_0^2  &0
\label{Fmunu1}
\end{pmatrix},
\ee
where
\bea
\Omega_1&=&\frac{Q r^2 - 2 a P r \cos\theta - 
 a^2 Q \cos^2\theta}{\rho^4},
\\
\Omega_2&=&\frac{\sin\theta\, (-P r^2 - 2 a Q r \cos\theta + a^2 P \cos^2\theta)}{\rho^4},
\eea
with $\rho$ and $\rho_0$ given in \eqref{rgrho} and \eqref{rho0} respectively. We would like to mention that the result \eqref{Fmunu1} can be extended to including the negative semi-axis $z$, because we can use \eqref{dmudxmu3} to replace \eqref{dmudxmu2} such that the singularities are absent and the Dirac's strings are unnecessary. 

Using $F^{\mu\nu}=g^{\mu\alpha}g^{\nu\beta}F_{\alpha\beta}$ we get
\be
F^{\mu\nu}
=
\begin{pmatrix}
0&-\frac{\Omega_1\rho_0^2}{\rho^2}&-\frac{\Omega_2 a}{\rho^2}&0\\
\frac{\Omega_1\rho_0^2}{\rho^2}&0&0& \frac{\Omega_1 a}{\rho^2}\\
\frac{\Omega_2 a}{\rho^2}&0&0&\frac{\Omega_2}{\rho^2\sin^2\theta}\\
0&-\frac{\Omega_1 a}{\rho^2}&-\frac{\Omega_2}{\rho^2\sin^2\theta} &0
\label{Fmunu2}
\end{pmatrix}.
\ee
For a non-rotating black hole ($a=0$), the expression \eqref{Fmunu2} reduces to
\be
F^{\mu\nu}
=
\begin{pmatrix}
0&-\frac{Q}{r^2}&0&0\\
\frac{Q}{r^2}&0&0& 0\\
0&0&0&-\frac{P}{r^4\sin\theta}\\
0&0&\frac{P}{r^4\sin\theta} &0
\label{Fmunu3}
\end{pmatrix}.
\ee
Transforming \eqref{Fmunu3} to the expression in cartesian coordinates $K'(t,x,y,z)$, we get
\be
F'^{\mu\nu}
=
\begin{pmatrix}
0&-\frac{Qx}{r^3}&-\frac{Qy}{r^3}&-\frac{Qz}{r^3}\\
\frac{Qx}{r^3}&0&-\frac{Pz}{r^3}&\frac{Py}{r^3}\\
\frac{Qy}{r^3}&\frac{Pz}{r^3}&0&-\frac{Px}{r^3}\\
\frac{Qz}{r^3}&-\frac{Py}{r^3}&\frac{Px}{r^3}&0
\end{pmatrix}.
\label{fieldF}
\ee
Comparing \eqref{fieldF} with the definition~\cite{Landau:1975pou} 
\be
F'^{\mu\nu}
=
\begin{pmatrix}
0&-E_x&-E_y&-E_z\\
E_x&0&-B_z&B_y\\
E_y&B_z&0&-B_x\\
E_z&-B_y&B_x&0
\end{pmatrix},
\label{definitionLandau}
\ee
the electric field and the magnetic field are respectively
\be
\vec{E}=\frac{Q\vec{r}}{r^3},\qquad
\vec{B}=\frac{P\vec{r}}{r^3}.
\label{EMfieldCartesian}
\ee

%%%%%%%%%
%%%
\newpage
\bibliography{sn-bibliography}

\end{document}